\documentclass[11pt,a4paper]{article}%
\usepackage{a4}
\usepackage{amsmath}
\usepackage{graphicx}
\usepackage{amsfonts}
\usepackage{amssymb}%
\newtheorem{theorem}{Theorem}
\newtheorem{lemma}[theorem]{Lemma}
\newenvironment{proof}[1][Proof]{\textbf{#1.} }{\ \rule{0.5em}{0.5em}}
\begin{document}

\title{The nested $SU(N)$ off-shell Bethe ansatz\\and exact form factors}
\author{Hratchia M. Babujian\thanks{Address: Alikhanian Brothers 2, Yerevan,
375036 Armenia} \thanks{E-mail: babujian@physik.fu-berlin.de} ,
Angela Foerster\thanks{Address: Instituto de F\'{\i}sica da UFRGS, Av. Bento
Gon\c{c}alves 9500, Porto Alegre, RS - Brazil} \thanks{E-mail:
angela@if.ufrgs.br} ,
and Michael Karowski\thanks{E-mail: karowski@physik.fu-berlin.de}\\[1em]
Institut f\"{u}r Theoretische Physik, Freie Universit\"{a}t Berlin,\\
Arnimallee 14, 14195 Berlin, Germany}
\date{November 1, 2006\\[1em]
{\it This work is dedicated to the 75th anniversary of H. Bethe's foundational
work on the Heisenberg chain}}
\maketitle

\begin{abstract}
The form factor equations are solved for an $SU(N)$ invariant S-matrix under
the assumption that the anti-particle is identified with the bound state of
$N-1$ particles. The solution is obtained explicitly in terms of the nested
off-shell Bethe ansatz where the contribution from each level is written in
terms of multiple contour integrals. \\[8pt]
PACS: 11.10.-z; 11.10.Kk; 11.55.Ds
\newline Keywords: Integrable quantum field theory, Form factors

\end{abstract}

\section{Introduction}

The Bethe ansatz \cite{Bethe}, was initially formulated by Bethe 75 years ago
to solve the eigenvalue problem for the isotropic Heisenberg model. The
approach has found applications in the context of several integrable systems
in different areas, such as Statistical Mechanics, Quantum Field Theory,
Condensed Matter Physics, Atom and Molecular Physics, among others. The
original techniques have been refined into several directions: Lieb and
Lininger \cite{LL} solved the one-dimensional bose gas problem with $\delta
$-function potential using the Bethe ansatz. The 6-vertex model was solved by
Lieb \cite{Li,Li1} with the same technique. C.N.~Yang and C.P.~Yang \cite{YY}
proved `Bethe's hypothesis'\footnote{Yang and Yang decided to honor Bethe's
insight by calling his assumption \textquotedblleft Bethe's
hypothesis\textquotedblright\ \cite{Yangbook}, now usually called
\textquotedblleft Bethe ansatz\textquotedblright.} for the ground state of the
anisotropic Heisenberg spin chain. Due to Yang \cite{Yang1} and Baxter
\cite{Baxter} we have the fundamental Yang-Baxter equation for the
two-particle S-matrix or for the matrix of the Boltzmann weights, which
guarantees exact integrability of the system. Subsequently, Faddeev and
collaborators \cite{FST,TF} formulated these ideas in an elegant algebraic
way, known as the \textquotedblleft algebraic Bethe ansatz\textquotedblright.
Yang \cite{Yang1} and Sutherland \cite{sut} generalized the technique of the
Bethe ansatz for those cases where the underlying symmetry group is larger
than $SU(2)$. This method is now called the \textquotedblleft
nested\textquotedblright\ Bethe ansatz. This technique was applied in
\cite{AL2} to derive the spectrum of the chiral $SU(N)$ Gross-Neveu model
\cite{GN}. The algebraic nested Bethe ansatz was formulated in \cite{BdVV} for
the $SU(N)$ and in \cite{dVK} for the $O(2N)$ symmetric case, respectively.
Another generalization of the Bethe ansatz is the \textquotedblleft
off-shell\textquotedblright\ Bethe ansatz, which was originally formulated by
one of the authors (H.B.) \cite{B1,B2,B3,BF} to calculate correlation function
in WZNW models (see also \cite{FR,SV}). This version of the Bethe ansatz paves
the way to an analysis of off-shell quantities and opens up the intriguing
possibility to merge the Bethe ansatz and the form factor approach. In this
context we point out that recently the form factor program has received
renewed interest in connection with condensed matter physics \cite{EK,Ts,GNT}
and atomic physics \cite{LZMGo}. In particular, applications to Mott
insulators and carbon nanotubes \cite{CE,EK} doped two-leg ladders
\cite{EK1} and in the field of Bose-Einstein condensates of ultracold
atomic and molecular gases \cite{LZ,LZMGo} have been discussed and in some
instances correlation functions have been computed.

For integrable quantum field theories in 1+1 dimensions one of the authors
(M.K.) et al.~formulated the on-shell program \cite{KTTW} i.e.~the exact
determination of the scattering matrix using the Yang-Baxter equations and the
off-shell program \cite{KW} i.e.~the exact determination of form factors which
are matrix elements of local operators. This approach was developed further
and studied in the context of several explicit models by Smirnov \cite{Sm} who
proposed the form factor equations (i) -- (v) (see below) as extensions of
similar formulae in the original article \cite{KW}. The formulae were proven
by two of the authors (H.B. and M.K.) et al.~\cite{BFKZ}. In this article the
techniques of the \textquotedblleft off-shell\textquotedblright\ Bethe ansatz
was used to determine the form factors for the sine-Gordon model. There,
however, the underlying group structure is simple and there was no need to use
a nested version of the off-shell Bethe ansatz. In the present article we will
focus on the determination of the form factors for an $SU(N)$ model. The
procedure is similar as for the scaling $Z(N)$ Ising and affine $A(N-1)$ Toda
models \cite{BK04,BFK} because the bound state structures of these models are
similar. However, the algebraic structure of the form factors for the $SU(N)$
model is more complicated, because the S-matrix possesses backward scattering.
Therefore we have to apply a nontrivial algebraic off-shell Bethe ansatz. For
$N>2$ we have to develop the nested version of this technique (see also
\cite{BKZ2}).

It is expected that the results of this paper apply to the chiral $SU(N)$
Gross-Neveu model \cite{GN,BW,ABW,KKS}. In a separate article \cite{BFK3} we
will investigate these physical applications and compare our exact results
with two different $1/N$-expansions of the chiral Gross-Neveu model \cite{BW}
and \cite{KKS}. We note that $SU(N)$ form factors were also calculated in
\cite{Sm,NT,Ta} using other techniques, see also the related paper \cite{MTV}.

\subsection{The $SU(N)$ S-matrix}

The general solutions of the Yang-Baxter equations, unitarity and crossing
relations for a $U(N)$ invariant S-matrix have been obtained in \cite{BKKW}.
The S-matrix for the scattering of two particles belonging to the vector
representation of $SU(N)$ can be written as
\begin{equation}
S_{\alpha\beta}^{\delta\gamma}(\theta)=\delta_{\alpha\gamma}\delta
_{\beta\delta}\,b(\theta)+\delta_{\alpha\delta}\delta_{\beta\gamma}%
\,c(\theta)\,. \label{1.2}%
\end{equation}
Unitarity reads as $S_{i}(-\theta)S_{i}(\theta)=1\,$for the S-matrix
eigenvalues
\[
S_{+}(\theta)=b(\theta)+c(\theta)\,,~~~S_{-}(\theta)=b(\theta)-c(\theta)\,.
\]
The amplitude $S_{+}(\theta)=a(\theta)$ is the highest weight $w=(2,0,\dots
,0)$ S-matrix eigenvalue for the two particle scattering. It will be
essential for the Bethe ansatz below.

As usual in this context we use in the notation
\[
v^{1\dots n}\in V^{1\dots n}=V^{1}\otimes\cdots\otimes V^{n}%
\]
for a vector in a tensor product space. The vector components are denoted by
$v^{\underline{\alpha}}=v^{\alpha_{1}\dots\alpha_{n}}$. Below we will also use
co-vectors $v_{1\dots n}\in\left(  V^{1\dots n}\right)  ^{\dag}$ (the dual of
$V^{1\dots n}$) with components $v_{\underline{\alpha}}$. A linear operator
connecting two such spaces with matrix elements $A_{\alpha_{1}\dots\alpha_{n}%
}^{\alpha_{1}^{\prime}\dots\alpha_{n^{\prime}}^{\prime}}$ is denoted by
\[
A_{1\dots n}^{1^{\prime}\dots n^{\prime}}:~V^{1\dots n}\rightarrow
V^{1^{\prime}\dots n^{\prime}}%
\]
where we omit the upper indices if they are obvious. All vector spaces $V^{i}$
are isomorphic to a space $V$ whose basis vectors label all kinds of particles
(e.g. $V\cong\mathbb{C}^{N}$ for the vector representation of $SU(N)$). The
vector spaces $V^{i}$ is associated to a rapidity variable $\theta_{i}$. An
S-matrix such as $S_{ij}(\theta_{ij})=S_{ij}^{ji}(\theta_{i}-\theta_{j})$ acts
nontrivially only on the factors $V_{i}\otimes V_{j}$ and exchanges these
factors. Using this notation, the Yang-Baxter relation writes as%
\begin{equation}
S_{12}(\theta_{12})S_{13}(\theta_{13})S_{23}(\theta_{23})=S_{23}(\theta
_{23})S_{13}(\theta_{13})S_{12}(\theta_{12}) \label{1.3}%
\end{equation}
and implies here the relation between the amplitudes \cite{BKKW}%
\[
c(\theta)=-\frac{i\eta}{\theta}b(\theta)\,,~~\eta=\frac{2\pi}{N}.
\]
A solution \cite{BKKW,BW,KS} of all these equations writes as
\begin{equation}
a(\theta)=b(\theta)+c(\theta)=-\frac{\Gamma\left(  1-\frac{\theta}{2\pi
i}\right)  \Gamma\left(  1-\frac{1}{N}+\frac{\theta}{2\pi i}\right)  }%
{\Gamma\left(  1+\frac{\theta}{2\pi i}\right)  \Gamma\left(  1-\frac{1}%
{N}-\frac{\theta}{2\pi i}\right)  }\,. \label{1.4}%
\end{equation}
This S-matrix possesses a bound state pole in $S_{-}(\theta)$ i.e. in the
anti-symmetric tensor channel. It is consistent with Swieca's
\cite{KuS,KS,KKS} picture that the anti-particle is a bound state of $N-1$
particles (see also \cite{BK04,BFK}).

For later convenience and in order to simplify the formulae we introduce%
\begin{equation}
\tilde{S}(\theta)=\frac{S(\theta)}{a(\theta)}=\frac{\mathbf{1}\theta
-\mathbf{P}i\eta}{\theta-i\eta} \label{1.5}%
\end{equation}
where $\mathbf{1}$ is the unit, $\mathbf{P}$ the permutation operator. We
depict this matrix as
\[
\tilde{S}_{\alpha\beta}^{\delta\gamma}(\theta_{12})=%
\begin{array}
[c]{c}%
\unitlength3mm\begin{picture}(6,6) \put(1,1){\line(1,1){4}}
\put(5,1){\line(-1,1){4}} \put(.5,0){$\alpha$} \put(5,0){$\beta$}
\put(5,5.4){$\gamma$} \put(.5,5.4){$\delta$} \put(1.8,.7){$\theta_1$}
\put(3.5,.7){$\theta_2$} \end{picture}
\end{array}
=\delta_{\alpha\gamma}\delta_{\beta\delta}\,\tilde{b}(\theta_{12}%
)+\delta_{\alpha\delta}\delta_{\beta\gamma}\,\tilde{c}(\theta_{12})
\]
and the amplitudes are explicitly%
\[
\tilde{b}(\theta)=\frac{\theta}{\theta-i\eta}\,,~~\tilde{c}(\theta
)=\frac{-i\eta}{\theta-i\eta}\,.
\]

\subsection{Generalized Form factors}

For a state of $n$ particles of kind $\alpha_{i}$ with rapidities $\theta_{i}$
and a local operator $\mathcal{O}(x)$ we define the form factor functions
$F_{\alpha_{1}\dots\alpha_{n}}^{\mathcal{O}}(\theta_{1},\dots,\theta_{n})$, or
using a short hand notation $F_{\underline{\alpha}}^{\mathcal{O}}%
(\underline{\theta})$, by
\begin{equation}
\langle\,0\,|\,\mathcal{O}(x)\,|\,\theta_{1},\dots,\theta_{n}\,\rangle
_{\underline{\alpha}}^{in}=e^{-ix(p_{1}+\cdots+p_{n})}F_{\underline{\alpha}%
}^{\mathcal{O}}(\underline{\theta})~,~~\text{for}~\theta_{1}>\dots>\theta_{n}.
\label{1.8}%
\end{equation}
where $\underline{\alpha}=(\alpha_{1},\dots,\alpha_{n})$ and $\underline
{\theta}=(\theta_{1},\dots,\theta_{n})$. For all other arrangements of the
rapidities the functions $F_{\underline{\alpha}}^{\mathcal{O}}(\underline
{\theta})$ are given by analytic continuation. Note that the physical value of
the form factor, i.e. the left hand side of (\ref{1.8}), is given for ordered
rapidities as indicated above and the statistics of the particles. The
$F_{\underline{\alpha}}^{\mathcal{O}}(\underline{\theta})$ are considered as
the components of a co-vector valued function $F_{1\dots n}^{\mathcal{O}%
}(\underline{\theta})\in V_{1\dots n}=\left(  V^{1\dots n}\right)  ^{\dagger}$
which may be depicted as
\begin{equation}
F_{1\dots n}^{\mathcal{O}}(\underline{\theta})=%
\begin{array}
[c]{c}%
\unitlength3.5mm\begin{picture}(4,4) \put(2,2){\oval(4,2)}
\put(2,2){\makebox(0,0){${\cal O}$}} \put(1,0){\line(0,1){1}}
\put(3,0){\line(0,1){1}} \put(-.1,0){$\theta_1$} \put(3.4,0){$\theta_n$}
\put(1.5,.3){$\dots$} \end{picture}
\end{array}
\,. \label{1.9}%
\end{equation}

Now we formulate the main properties of form factors in terms of the functions
$F_{1\dots n}^{\mathcal{O}}$. They follow from general LSZ-assumptions and
\textquotedblleft maximal analyticity\textquotedblright, which means that
$F_{1\dots n}^{\mathcal{O}}(\underline{\theta})$ is a meromorphic function
with respect to all $\theta$'s and in the `physical' strips
$0<\operatorname{Im}\theta_{ij}<\pi$~$(\theta_{ij}=\theta_{i}-\theta
_{j}\,i<j)$ there are only poles of physical origin as for example bound state
poles. The generalized form factor functions satisfy the following

\paragraph{Form factor equations:}

The co-vector valued auxiliary function $F_{1\dots n}^{\mathcal{O}%
}({\underline{\theta}})$ is meromorphic in all variables $\theta_{1}%
,\dots,\theta_{n}$ and satisfies the following relations:

\begin{itemize}
\item[(i)] The Watson's equations describe the symmetry property under the
permutation of both, the variables $\theta_{i},\theta_{j}$ and the spaces
$i,j=i+1$ at the same time
\begin{equation}
F_{\dots ij\dots}^{\mathcal{O}}(\dots,\theta_{i},\theta_{j},\dots)=F_{\dots
ji\dots}^{\mathcal{O}}(\dots,\theta_{j},\theta_{i},\dots)\,S_{ij}(\theta_{ij})
\label{1.10}%
\end{equation}
for all possible arrangements of the $\theta$'s.

\item[(ii)] The crossing relation implies a periodicity property under the
cyclic permutation of the rapidity variables and spaces
\begin{multline}
^{~\text{out,}\bar{1}}\langle\,p_{1}\,|\,\mathcal{O}(0)\,|\,p_{2},\dots
,p_{n}\,\rangle_{2\dots n}^{\text{in,conn.}}\\
=F_{1\ldots n}^{\mathcal{O}}(\theta_{1}+i\pi,\theta_{2},\dots,\theta
_{n})\sigma_{1}^{\mathcal{O}}\mathbf{C}^{\bar{1}1}=F_{2\ldots n1}%
^{\mathcal{O}}(\theta_{2},\dots,\theta_{n},\theta_{1}-i\pi)\mathbf{C}%
^{1\bar{1}} \label{1.12}%
\end{multline}
where $\sigma_{\alpha}^{\mathcal{O}}$ takes into account the statistics of the
particle $\alpha$ with respect to $\mathcal{O}$. The charge conjugation matrix
$\mathbf{C}^{\bar{1}1}$ will be discussed below.

\item[(iii)] There are poles determined by one-particle states in each
sub-channel. In particular the function $F_{\underline{\alpha}}^{\mathcal{O}%
}({\underline{\theta}})$ has a pole at $\theta_{12}=i\pi$ such that
\begin{equation}
\operatorname*{Res}_{\theta_{12}=i\pi}F_{1\dots n}^{\mathcal{O}}(\theta
_{1},\dots,\theta_{n})=2i\,\mathbf{C}_{12}\,F_{3\dots n}^{\mathcal{O}}%
(\theta_{3},\dots,\theta_{n})\left(  \mathbf{1}-\sigma_{2}^{\mathcal{O}}%
S_{2n}\dots S_{23}\right)  \,. \label{1.14}%
\end{equation}

\item[(iv)] If there are also bound states in the model the function
$F_{\underline{\alpha}}^{\mathcal{O}}({\underline{\theta}})$ has additional
poles. If for instance the particles 1 and 2 form a bound state (12), there is
a pole at $\theta_{12}=i\eta,~(0<\eta<\pi)$ such that
\begin{equation}
\operatorname*{Res}_{\theta_{12}=i\eta}F_{12\dots n}^{\mathcal{O}}(\theta
_{1},\theta_{2},\dots,\theta_{n})\,=F_{(12)\dots n}^{\mathcal{O}}%
(\theta_{(12)},\dots,\theta_{n})\,\sqrt{2}\Gamma_{12}^{(12)} \label{1.16}%
\end{equation}
where the bound state intertwiner $\Gamma_{12}^{(12)}$ and the values of
$\theta_{1},~\theta_{2},~\theta_{(12)}$ and $\eta$ are given in \cite{K1,BK}.

\item[(v)] Naturally, since we are dealing with relativistic quantum field
theories we finally have
\begin{equation}
F_{1\dots n}^{\mathcal{O}}(\theta_{1}+\mu,\dots,\theta_{n}+\mu)=e^{s\mu
}\,F_{1\dots n}^{\mathcal{O}}(\theta_{1},\dots,\theta_{n}) \label{1.18}%
\end{equation}
if the local operator transforms under Lorentz transformations as
$\mathcal{O}\rightarrow e^{s\mu}\mathcal{O}$ where $s$ is the
\textquotedblleft spin\textquotedblright\ of $\mathcal{O}$.
\end{itemize}

The property (i) - (iv) may be depicted as
\[%
\begin{array}
[c]{rrcl}%
\text{(i)} &
\begin{array}
[c]{c}%
\unitlength3.2mm\begin{picture}(7,3) \put(3.5,2){\oval(7,2)}
\put(3.5,2){\makebox(0,0){${\cal O}$}} \put(1,0){\line(0,1){1}}
\put(3,0){\line(0,1){1}} \put(4,0){\line(0,1){1}} \put(6,0){\line(0,1){1}}
\put(1.4,.5){$\dots$} \put(4.4,.5){$\dots$} \end{picture}
\end{array}
& = &
\begin{array}
[c]{c}%
\unitlength3.2mm\begin{picture}(7,4) \put(3.5,3){\oval(7,2)}
\put(3.5,3){\makebox(0,0){${\cal O}$}} \put(1,0){\line(0,1){2}}
\put(3,0){\line(1,2){1}} \put(4,0){\line(-1,2){1}} \put(6,0){\line(0,1){2}}
\put(1.4,1){$\dots$} \put(4.4,1){$\dots$} \end{picture}
\end{array}
\\
\text{(ii)} & ~~~%
\begin{array}
[c]{c}%
\unitlength3.2mm\begin{picture}(5,4)(2,0) \put(2,0){\line(0,1){1}}
\put(3.5,3){\line(0,1){1}} \put(3.5,2){\oval(5,2)}
\put(3.5,2){\makebox(0,0){${\cal O}$}} \put(5,0){\line(0,1){1}}
\put(6,3){\scriptsize conn.} \put(2.8,.3){$\dots$} \end{picture}
\end{array}
~~=~~%
\begin{array}
[c]{c}%
\unitlength3.2mm\begin{picture}(6,4) \put(0,1){\line(0,1){3}} \put
(1,1){\oval(2,2.5)[b]} \put (1.49,.2){$\times$} \put(3.5,2){\oval(5,2)}
\put(3.5,2){\makebox(0,0){${\cal O}$}} \put(5,0){\line(0,1){1}}
\put(2.8,.3){$\dots$} \end{picture}
\end{array}
~~ & = &
\begin{array}
[c]{c}%
\unitlength3.2mm\begin{picture}(7,4) \put(7,1){\line(0,1){3}} \put
(6,1){\oval(2,2)[b]} \put(3.5,2){\oval(5,2)} \put(3.5,2){\makebox(0,0){${\cal
O}$}} \put(2,0){\line(0,1){1}} \put(2.6,.3){$\dots$} \end{picture}
\end{array}
\\
\text{(iii)} & \dfrac{1}{2i}\,\operatorname*{Res}\limits_{\theta_{12}=i\pi}~~~%
\begin{array}
[c]{c}%
\unitlength3.2mm\begin{picture}(6,4) \put(3,2){\oval(6,2)}
\put(3,2){\makebox(0,0){${\cal O}$}} \put(1,0){\line(0,1){1}}
\put(2,0){\line(0,1){1}} \put(3,0){\line(0,1){1}} \put(5,0){\line(0,1){1}}
\put(3.4,.5){$\dots$} \end{picture}
\end{array}
& = &
\begin{array}
[c]{c}%
\unitlength3.2mm\begin{picture}(5,4) \put(.5,0){\oval(1,2)[t]}
\put(3,2){\oval(4,2)} \put(3,2){\makebox(0,0){${\cal O}$}}
\put(2,0){\line(0,1){1}} \put(4,0){\line(0,1){1}} \put(2.4,.5){$\dots$}
\end{picture}
\end{array}
-
\begin{array}
[c]{c}%
\unitlength3.2mm\begin{picture}(6,5) \put(0,0){\line(0,1){3}}
\put(3,3){\oval(6,4)[t]} \put (5.45,2.3){$\times$} \put(3,3){\oval(6,4)[br]}
\put(3,0){\oval(4,2)[tl]} \put(3,3){\oval(4,2)} \put(3,3){\makebox(0,0){${\cal
O}$}} \put(2,0){\line(0,1){2}} \put(4,0){\line(0,1){2}} \put(2.4,1.5){$\dots$}
\end{picture}
\end{array}
\\
\text{(iv)} & \dfrac{1}{\sqrt{2}}\,\operatorname*{Res}\limits_{\theta
_{12}=i\eta}~%
\begin{array}
[c]{c}%
\unitlength3.2mm\begin{picture}(5,3) \put(2.5,2){\oval(5,2)}
\put(2.5,2){\makebox(0,0){${\cal O}$}} \put(1,0){\line(0,1){1}}
\put(2,0){\line(0,1){1}} \put(4,0){\line(0,1){1}} \put(2.4,.5){$\dots$}
\end{picture}
\end{array}
& = &
\begin{array}
[c]{c}%
\unitlength3.2mm\begin{picture}(5,4) \put(2.5,3){\oval(5,2)}
\put(2.5,3){\makebox(0,0){${\cal O}$}}
\put(1.5,0){\oval(1,2)[t]}
\put(1.5,1){\line(0,1){1}} \put(4,0){\line(0,1){2}} \put(2.2,1.1){$\dots$}
\end{picture}
\end{array}
\end{array}
\]
where $\times$ denotes the statistics factor $\sigma^{\mathcal{O}}$. As was
shown in \cite{BFKZ} the properties (i) -- (iii) follow from general
LSZ-assumptions and \textquotedblleft maximal analyticity\textquotedblright.

We will now provide a constructive and systematic way of how to solve the form
factor equations (i) -- (v) for the co-vector valued function $F_{1\dots
n}^{\mathcal{O}}$, once the scattering matrix is given.

\subparagraph{Minimal form factor:}

The solutions of Watson's and the crossing equations (i) and (ii) for two
particles with no poles in the physical strip $0\leq\operatorname{Im}%
\theta\leq\pi$ and at most a simple zero at $\theta=0$ are the minimal form
factors. In particular those for highest weight states are essential for the
construction of the off-shell Bethe ansatz. One easily finds the minimal
solution of%
\[
F\left(  \theta\right)  =a\left(  \theta\right)  F\left(  -\theta\right)
=F\left(  2\pi i-\theta\right)
\]
using (\ref{1.4}) as%
\begin{equation}
F\left(  \theta\right)  =c\exp\int\limits_{0}^{\infty}\frac{dt}{t\sinh^{2}%
t}e^{\frac{t}{N}}\sinh t\left(  1-\frac{1}{N}\right)  \left(  1-\cosh t\left(
1-\frac{\theta}{i\pi}\right)  \right)  \,. \label{1.22}%
\end{equation}
It belongs to the highest weight $w=(2,0,\dots,0)$. We define the
corresponding `Jost-function' as for the $Z(N)$ models \cite{BK04,BFK} by the
equation%
\begin{equation}
\prod_{k=0}^{N-2}\phi\left(  \theta+ki\eta\right)  \prod_{k=0}^{N-1}F\left(
\theta+ki\eta\right)  =1~,~~\eta=\frac{2\pi}{N} \label{1.24}%
\end{equation}
which is typical for models where the bound state of $N-1$ particles is the
anti-particle \cite{BFK}. The solution is
\begin{equation}
\phi(\theta)=\Gamma\left(  \frac{\theta}{2\pi i}\right)  \Gamma\left(
1-\frac{1}{N}-\frac{\theta}{2\pi i}\right)  \label{1.26}%
\end{equation}
and satisfies the relations%
\begin{multline}
\phi(\theta)=\phi(-\theta)a\left(  -\theta\right)  =\phi((N-1)i\eta
-\theta)\label{1.28}\\
=\frac{1}{-b(\theta)}\phi(2\pi i-\theta)=\frac{a(\theta-2\pi i)}{-b(\theta
)}\phi(\theta-2\pi i)\,.
\end{multline}
Notice that the equations (\ref{1.26}) and (\ref{1.24}) also determine the
normalization constant $c$ in (\ref{1.22}) as%
\[
c=\Gamma^{-2(1-1/N)}\left(  \frac{1}{2}-\frac{1}{2N}\right)  \exp\left(
-\int_{0}^{\infty}e^{\frac{1}{N}t}\left(  \frac{\sinh\left(  1-\frac{1}%
{N}\right)  t}{t\sinh^{2}t}-\frac{\left(  1-\frac{1}{N}\right)  }{t\sinh
t}\right)  dt\right)  \,.
\]

\subparagraph{Generalized form factors:}

The co-vector valued function (\ref{1.9}) for n-particles can be written as
\cite{KW}
\begin{equation}
F_{1\dots n}^{\mathcal{O}}(\underline{\theta})=K_{1\dots n}^{\mathcal{O}%
}(\underline{\theta})\prod_{1\leq i<j\leq n}F(\theta_{ij}) \label{2.10}%
\end{equation}
where $F(\theta)$ is the minimal form factor function (\ref{1.22}). The
K-function $K_{1\dots n}^{\mathcal{O}}(\underline{\theta})$ contains the
entire pole structure and its symmetry is determined by the form factor
equations (i) and (ii) where the S-matrix is replaced by $\tilde{S}%
(\theta)=S(\theta)/a(\theta)$%
\begin{align}
K_{\dots ij\dots}^{\mathcal{O}}(\dots,\theta_{i},\theta_{j},\dots)  &
=K_{\dots ji\dots}^{\mathcal{O}}(\dots,\theta_{j},\theta_{i},\dots)\,\tilde
{S}_{ij}(\theta_{ij})\label{2.12}\\
K_{1\ldots n}^{\mathcal{O}}(\theta_{1}+i\pi,\theta_{2},\dots,\theta_{n}%
)\sigma_{1}^{\mathcal{O}}\mathbf{C}^{\bar{1}1}  &  =K_{2\ldots n1}%
^{\mathcal{O}}(\theta_{2},\dots,\theta_{n},\theta_{1}-i\pi)\mathbf{C}%
^{1\bar{1}} \label{2.14}%
\end{align}
for all possible arrangements of the $\theta$'s.

\subsection{Nested \textquotedblleft off-shell\textquotedblright\ Bethe ansatz
for $SU(N)$}

We consider a state with $n$ particles and define as usual in the context of
the algebraic Bethe ansatz \cite{FST,TF} the monodromy matrix
\begin{equation}
\tilde{T}_{1\dots n,0}(\underline{\theta},\theta_{0})=\tilde{S}_{10}%
(\theta_{10})\,\cdots\tilde{S}_{n0}(\theta_{n0})=%
\begin{array}
[c]{c}%
\unitlength3mm\begin{picture}(9,4.5) \put(0,2){\line(1,0){9}}
\put(2,0){\line(0,1){4}} \put(7,0){\line(0,1){4}} \put(1,0){$1$} \put(5.8,0){$
n$} \put(8.2,.7){$ 0$} \put(3.6,3){$\dots$} \end{picture}
\end{array}
\,. \label{1.30}%
\end{equation}
It is a matrix acting in the tensor product of the \textquotedblleft quantum
space\textquotedblright\ $V^{1\dots n}=V^{1}\otimes\cdots\otimes V^{n}$ and
the \textquotedblleft auxiliary space\textquotedblright\ $V^{0}$. All vector
spaces $V^{i}$ are isomorphic to a space $V$ whose basis vectors label all
kinds of particles. Here we consider $V\cong\mathbb{C}^{N}$ as the space of
the vector representation of $SU(N)$. The Yang-Baxter algebra relation for the
S-matrix (\ref{1.3}) yields
\begin{gather}
\tilde{T}_{1\dots n,a}(\underline{\theta},\theta_{a})\,\tilde{T}_{1\dots
n,b}(\underline{\theta},\theta_{b})\,\tilde{S}_{ab}(\theta_{a}-\theta
_{b})=\tilde{S}_{ab}(\theta_{a}-\theta_{b})\,\tilde{T}_{1\dots n,b}%
(\underline{\theta},\theta_{b})\,\tilde{T}_{1\dots n,a}(\underline{\theta
},\theta_{a})\label{1.32}\\
\unitlength4mm\begin{picture}(22,5.5) \put(.3,0){$1$} \put(4.3,0){$n$}
\put(0,2.3){$b$} \put(0,4.3){$a$} \put(7.5,2.3){$b$} \put(7.5,.5){$a$}
\put(2.5,3){$\dots$} \put(1,0){\line(0,1){5}} \put(0,2){\line(1,0){8}}
\put(5,0){\line(0,1){5}} \put(0,2){\oval(14,4)[rt]} \put(8,2){\oval(2,4)[lb]}
\put(9.3,2.3){$=$} \put(13.3,0){$1$} \put(17.3,0){$n$} \put(11,2){$b$}
\put(11,4.2){$a$} \put(18.5,3.4){$b$} \put(18.5,1.4){$a$}
\put(15.5,2){$\dots$} \put(14,0){\line(0,1){5}} \put(18,0){\line(0,1){5}}
\put(11,3){\line(1,0){8}} \put(19,3){\oval(14,4)[lb]}
\put(11,3){\oval(2,4)[rt]} \end{picture}\nonumber
\end{gather}
which implies the basic algebraic properties of the sub-matrices
$A,B,C,D$ with respect to the auxiliary space defined by
\begin{equation}
\tilde{T}_{1\dots n,0}(\underline{\theta},z)\equiv\left(
\begin{array}
[c]{cc}%
\tilde{A}_{1\dots n}(\underline{\theta},z) & \tilde{B}_{1\dots n,\beta
}(\underline{\theta},z)\\
\tilde{C}_{1\dots n}^{\beta}(\underline{\theta},z) & \tilde{D}_{1\dots
n,\beta}^{\beta^{\prime}}(\underline{\theta},z)
\end{array}
\,\right)  ,~~~~2\leq\beta,\beta^{\prime}\leq N\,. \label{1.34}%
\end{equation}
We propose the following ansatz for the general form factor $F_{1\dots
n}^{\mathcal{O}}(\underline{\theta})$ or the K-function defined by
(\ref{2.10}) in terms of a nested `off-shell' Bethe ansatz and written as a
multiple contour integral%
\begin{equation}
\fbox{$\rule[-0.2in]{0in}{0.5in}\displaystyle~K_{\underline{\alpha}%
}^{\mathcal{O}}(\underline{\theta})=\frac{N_{n}}{m!}\int_{\mathcal{C}%
_{\underline{\theta}}}\frac{dz_{1}}{R}\cdots\int_{\mathcal{C}_{\underline
{\theta}}}\frac{dz_{m}}{R}\,\tilde{h}(\underline{\theta},\underline
{z})\,p^{\mathcal{O}}(\underline{\theta},\underline{z})\,\,\tilde{\Psi
}_{\underline{\alpha}}(\underline{\theta},\underline{z})$~} \label{2.16}%
\end{equation}
where $\tilde{h}(\underline{\theta},\underline{z})$ is a scalar function which
depends only on the S-matrix and not on the specific operator $\mathcal{O}(x)$%
\begin{align}
\tilde{h}(\underline{\theta},\underline{z})  &  =\prod_{i=1}^{n}\prod
_{j=1}^{m}\tilde{\phi}(\theta_{i}-z_{j})\prod_{1\leq i<j\leq m}\tau
(z_{i}-z_{j})\label{2.18}\\
\tau(z)  &  =\frac{1}{\phi(\theta)\phi(-\theta)}\,,~~\tilde{\phi}(\theta
)=\phi(\theta)a(\theta)=\phi(-\theta)\,. \label{2.19}%
\end{align}
For the $SU(N)$ S-matrix the function $\phi(\theta)$ is given by (\ref{1.24})
with the solution (\ref{1.26}). The integration contour $\mathcal{C}%
_{\underline{\theta}}$ is depicted in Fig. \ref{f5.1}. The constant $R$ is
defined by $R=\oint_{\theta}dz\tilde{\phi}(\theta-z)$ where the integration
contour is a small circle around $z=\theta$ as part of $\mathcal{C}%
_{\underline{\theta}}$.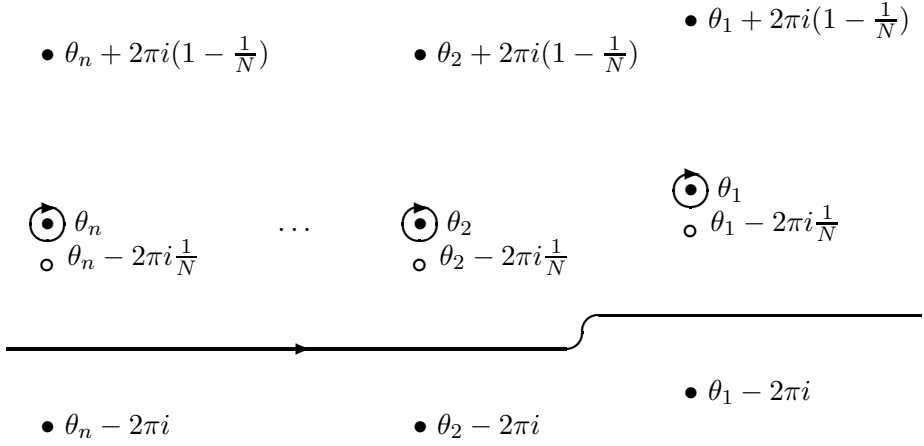
\begin{figure}[tbh]%
\[
\unitlength4.5mm\begin{picture}(27,13)
\thicklines
\put(1,0){
\put(0,0){$\bullet~\theta_n-2\pi i$}
\put(.19,5){\circle{.3}~$\theta_n-2\pi i\frac1N$}
\put(0,6){$\bullet~~\theta_n$}
\put(.2,6.2){\oval(1,1)}
\put(.5,6.68){\vector(1,0){0}}
\put(0,11){$\bullet~\theta_n+2\pi i(1-\frac1N)$}
}
\put(8,6){\dots}
\put(12,0){
\put(0,0){$\bullet~\theta_2-2\pi i$}
\put(.19,5){\circle{.3}~$\theta_2-2\pi i\frac1N$}
\put(0,6){$\bullet~~\theta_2$}
\put(.2,6.2){\oval(1,1)}
\put(.5,6.68){\vector(1,0){0}}
\put(0,11){$\bullet~\theta_2+2\pi i(1-\frac1N)$}
}
\put(20,1){
\put(0,0){$\bullet~\theta_1-2\pi i$}
\put(.19,5){\circle{.3}~$\theta_1-2\pi i\frac1N$}
\put(0,6){$\bullet~~\theta_1$}
\put(.2,6.2){\oval(1,1)}
\put(.5,6.68){\vector(1,0){0}}
\put(0,11){$\bullet~\theta_1+2\pi i(1-\frac1N)$}
}
\put(9,2.5){\vector(1,0){0}}
\put(0,3){\oval(34,1)[br]}
\put(27,3){\oval(20,1)[tl]}
\end{picture}
\]
\caption{The integration contour $\mathcal{C}_{\underline{\theta}}$. The
bullets refer to poles of the integrand resulting from $a(\theta_{i}%
-z_{j})\,\phi(\theta_{i}-z_{j})$ and the small open circles refer to poles
originating from $b(\theta_{i}-z_{j})$ and $c(\theta_{i}-z_{j})$.}%
\label{f5.1}%
\end{figure}

The dependence on the specific operator $\mathcal{O}(x)$ is encoded in the
scalar p-function $p^{\mathcal{O}}(\underline{\theta},\underline{z})$ which is
in general a simple function of $e^{\theta_{i}}$ and $e^{z_{j}}$ (see below).
By means of the ansatz (\ref{2.10}) and (\ref{2.16}) we have transformed the
complicated form factor equations (i) - (v) (which are in general matrix
equations) into much simpler scalar equations for the p-function (see below).
The K-function is in general a linear combination of the \emph{fundamental
building blocks} \cite{BK2,BK04,BFK} given by (\ref{2.16}) - (\ref{2.19}). We
consider here cases where the sum consists only of one term.

If in (\ref{1.34}) the range of $\beta$'s is non-trivial, i.e. if $N>2$ the
Bethe ansatz co-vectors are of the form%
\begin{equation}
\tilde{\Psi}_{\underline{\alpha}}(\underline{\theta},\underline{z}%
)=L_{\underline{\beta}}(\underline{z})\,\tilde{\Phi}_{\underline{\alpha}%
}^{\underline{\beta}}(\underline{\theta},\underline{z}) \label{1.37}%
\end{equation}
where summation over all $\underline{\beta}=(\beta_{1},\dots,\beta_{m})$ with
$\beta_{i}>1$ is assumed. The basic Bethe ansatz co-vectors $\tilde{\Phi
}_{1\dots n}^{\underline{\beta}}\in\left(  V^{1\dots n}\right)  ^{\dag}$ are
defined as%
\begin{equation}%
\begin{array}
[c]{rcl}%
\tilde{\Phi}_{1\dots n}^{\underline{\beta}}(\underline{\theta},\underline
{z}) & = & \Omega_{1\dots n}\tilde{C}_{1\dots n}^{\beta_{m}}(\underline
{\theta},z_{m})\cdots\tilde{C}_{1\dots n}^{\beta_{1}}(\underline{\theta}%
,z_{1})\\
\tilde{\Phi}_{\underline{\alpha}}^{\underline{\beta}}(\underline{\theta
},\underline{z}) & = &
\begin{array}
[c]{c}%
\unitlength4mm\begin{picture}(9,7) \put(9,5){\oval(14,2)[lb]}
\put(9,5){\oval(18,6)[lb]} \put(4,1){\line(0,1){4}} \put(8,1){\line(0,1){4}}
\put(-.2,5.4){$\beta_1$} \put(1.8,5.4){$\beta_m$} \put(3.8,.1){$\alpha_1$}
\put(7.8,.1){$\alpha_n$} \put(3.8,5.4){$1$} \put(7.8,5.4){$1$}
\put(9.2,1.8){$1$} \put(9.2,3.8){$1$} \put(3,2.5){$\theta_1$}
\put(6.8,2.5){$\theta_{n}$} \put(.8,2.5){$z_1$} \put(1.7,3.6){$z_m$}
\put(5.4,4.5){$\dots$} \put(8.5,2.6){$\vdots$} \end{picture}
\end{array}
~~~,~~~
\begin{array}
[c]{l}%
2\leq\beta_{i}\leq N\\
1\leq\alpha_{i}\leq N~.
\end{array}
\end{array}
\label{1.38}%
\end{equation}
Here the \textquotedblleft pseudo-vacuum\textquotedblright\ is the highest
weight co-vector (with weight $w=(n,0,\dots,0)$)
\[
\Omega_{1\dots n}=e(1)\otimes\cdots\otimes e(1)
\]
where the unit vectors $e(\alpha)~(\alpha=1,\dots,N)$ correspond to the
particle of type $\alpha$ which belong to the vector representation of
$SU(N)$. The pseudo-vacuum vector satisfies
\begin{equation}%
\begin{array}
[c]{rcl}%
\Omega_{1\dots n}\,\tilde{B}_{1\dots n}^{\beta}(\underline{\theta},z) & = &
0\\
\Omega_{1\dots n}\,\tilde{A}_{1\dots n}(\underline{\theta},z) & = &
\Omega_{1\dots n}\\
\Omega_{1\dots n}\,\tilde{D}_{1\dots n,\beta}^{\beta^{\prime}}(\underline
{\theta},z) & = & \delta_{\beta}^{\beta^{\prime}}\prod\limits_{i=1}^{n}%
\tilde{b}(\theta_{i}-z)\Omega_{1\dots n}\,.
\end{array}
\label{1.40}%
\end{equation}
The amplitudes of the scattering matrices are given by eqs. (\ref{1.2}) and
(\ref{1.4}). The technique of the \textbf{`nested Bethe ansatz'} means that
one makes for the coefficients $L_{\underline{\beta}}(\underline{z})$ in
(\ref{1.37}) the analogous construction (\ref{2.16}) - (\ref{2.19}) as for
$K_{\underline{\alpha}}(\underline{\theta})$ where now the indices
$\underline{\beta}$ take only the values $2\leq\beta_{i}\leq N$. This nesting
is repeated until the space of the coefficients becomes one dimensional.

In this article we will focus on the determination of the form factors for an
$SU(N)$ S-matrix. The paper is organized as follows: In Section 2 we construct
the general form factor formula for the simplest $SU(2)$ case. In Section 3 we
construct the general form factor formula for the $SU(3)$ case, which is more
complex due to the presence of the nesting procedure (more explicitly, we have
here two levels). We extend these results in Section 4, where the general form
factors for $SU(N)$ are constructed and discussed in detail.

\section{$SU(2)$ form factors}

In this section we start with the simplest case. We perform the form factor
program for the $SU(2)$ S-matrix. The results should apply to the well-known
$SU(2)$ Gross-Neveu model \cite{GN}, investigated by Bethe ansatz methods in
\cite{AL1,Be}. From the technical point of view the calculation of the form
factors is very similar to the one done for the sine-Gordon model in
\cite{BFKZ,BK}.

\paragraph{S-matrix:}

The $SU(2)$ S matrix\footnote{It is related to the sine-Gordon S-matrix for
$\beta^{2}=8\pi$ by $(a,b,c)^{SU(2)}=(a,-b,c)^{S.G.}$ (see e.g. \cite{BFKZ}).}
is given by (\ref{1.2}) and (\ref{1.4}) for $N=2$. It turns out that the
amplitudes satisfy the relations $b(\theta)=-a(i\pi-\theta)$ and
$c(\theta)=c(i\pi-\theta)$ which may be written as%
\begin{equation}
S_{\alpha\beta}^{\delta\gamma}(\theta)=-\left(  \delta_{\alpha}^{\gamma}%
\delta_{\beta}^{\delta}\,b(i\pi-\theta)+\epsilon^{\delta\gamma}\epsilon
_{\alpha\beta}\,c(i\pi-\theta)\right)  \,. \label{2.1}%
\end{equation}
We have introduced
\[
\epsilon_{\alpha\beta}=%
\begin{array}
[c]{c}%
\unitlength3.4mm\begin{picture}(2,2) \put(1,1){\oval(2,2)[t]}
\put(1,2){\makebox(0,0){$\bullet$}} \put(-.2,0){$\alpha$} \put(1.7,0){$\beta$}
\end{picture}
\end{array}
~~,~~~\epsilon^{\alpha\beta}=%
\begin{array}
[c]{c}%
\unitlength3.4mm\begin{picture}(2,2) \put(1,1){\oval(2,2)[b]}
\put(1,0){\makebox(0,0){$\bullet$}} \put(-.2,1.3){$\alpha$}
\put(1.7,1.3){$\beta$} \end{picture}
\end{array}
\]
which are antisymmetric and $\epsilon_{12}=\epsilon^{21}=1$ such that
\begin{equation}
\epsilon_{\alpha\beta}\epsilon^{\beta\gamma}=\delta_{\alpha}^{\gamma}:\text{
}~~%
\begin{array}
[c]{c}%
\unitlength3mm\begin{picture}(8,4) \put(1,2){\oval(2,2)[t]}
\put(1,3){\makebox(0,0){$\bullet$}} \put(3,2){\oval(2,2)[b]}
\put(3,1){\makebox(0,0){$\bullet$}} \put(0,0){\line(0,1){2}}
\put(4,2){\line(0,1){2}} \put(5.5,2){$=$} \put(8,0){\line(0,1){4}}
\end{picture}
\end{array}
~~\text{and}~~\epsilon_{\alpha\beta}A_{\gamma}^{\alpha}\epsilon^{\gamma\beta
}=~%
\begin{array}
[c]{c}%
\unitlength3mm\begin{picture}(3,4) \put(1,2){\oval(2,2)}
\put(1,2){\makebox(0,0){$A$}} \put(2,3){\oval(2,2)[t]}
\put(2,4){\makebox(0,0){$\bullet$}} \put(2,1){\oval(2,2)[b]}
\put(2,0){\makebox(0,0){$\bullet$}} \put(3,1){\line(0,1){2}} \end{picture}
\end{array}
~=-\operatorname*{tr}A\, \label{2.6}%
\end{equation}
for a matrix $A$. Formula (\ref{2.1}) may be understood as an unusual crossing
relation (c.f. \cite{KuS})
\begin{gather}
S_{\alpha\beta}^{\delta\gamma}(\theta)=-\mathbf{C}_{\alpha\alpha^{\prime}%
}S_{\beta\gamma^{\prime}}^{\alpha^{\prime}\delta}(i\pi-\theta)\mathbf{C}%
^{\gamma^{\prime}\gamma}\label{2.2}\\
\unitlength3mm\begin{picture}(16,4) \put(2,0){\line(-1,2){2}}
\put(0,0){\line(1,2){2}} \put(4,1.8){$= ~-$} \put(10,2){\oval(2,2)[t]}
\put(10,3){\makebox(0,0){$\bullet$}} \put(12,2){\oval(2,2)[b]}
\put(12,1){\makebox(0,0){$\bullet$}} \put(10.5,0){\line(1,4){1}}
\put(9,0){\line(0,1){2}} \put(13,2){\line(0,1){2}} \end{picture}\nonumber
\end{gather}
if we define the \textquotedblleft charge conjugation
matrices\textquotedblright\ as $\mathbf{C}_{\alpha\beta}=\epsilon_{\alpha
\beta}$ and $\mathbf{C}^{\alpha\beta}=\epsilon^{\alpha\beta}$. This means that
particle $1$ is the anti-particle of $2$ and vice versa.

For $\theta\rightarrow0$ and $i\pi$ the S-matrix yield the permutation matrix
and the annihilation-creation operator, respectively, or in terms of
$\tilde{S}=S/a$
\[
\tilde{S}_{\alpha\beta}^{\delta\gamma}(0)=\delta_{\alpha}^{\delta}%
\delta_{\beta}^{\gamma}=~%
\begin{array}
[c]{c}%
\unitlength3.4mm\begin{picture}(2,5) \put(0,1){\line(1,3){.5}}
\put(0,4){\line(1,-3){.5}} \put(2,1){\line(-1,3){.5}}
\put(2,4){\line(-1,-3){.5}} \put(-.2,0){$\alpha$} \put(1.7,0){$\beta$}
\put(-.2,4.5){$\delta$} \put(1.7,4.5){$\gamma$} \end{picture}
\end{array}
~,~~\operatorname*{Res}_{\theta=i\pi}\tilde{S}_{\alpha\beta}^{\delta\gamma
}(\theta)=i\pi\epsilon^{\delta\gamma}\epsilon_{\alpha\beta}=i\pi~%
\begin{array}
[c]{c}%
\unitlength3.4mm\begin{picture}(2,5) \put(1,1){\oval(2,2)[t]}
\put(1,4){\oval(2,2)[b]} \put(1,2){\makebox(0,0){$\bullet$}}
\put(1,3){\makebox(0,0){$\bullet$}} \put(-.2,0){$\alpha$} \put(1.7,0){$\beta$}
\put(-.2,4.3){$\delta$} \put(1.7,4.3){$\gamma$} \end{picture}
\end{array}
\,.
\]

\paragraph{Ansatz for form factors:}

Because of (\ref{2.6}) the crossing equation (\ref{2.14}) writes in components
as
\begin{equation}
K_{\alpha_{1}\dots\alpha_{n}}^{\mathcal{O}}(\theta_{1},\theta_{2},\dots
,\theta_{n})\sigma^{\mathcal{O}}=-K_{\alpha_{2}\dots\alpha_{n}\alpha_{1}%
}^{\mathcal{O}}(\theta_{2},\dots,\theta_{n},\theta_{1}-2\pi i) \label{2.14a}%
\end{equation}
The form factor equation (iii) here reads as%
\begin{equation}
\operatorname*{Res}_{\theta_{12}=i\pi}F_{1\dots n}^{\mathcal{O}}(\theta
_{1},\dots,\theta_{n})=2i\,\epsilon_{12}\,F_{3\dots n}^{\mathcal{O}}%
(\theta_{3},\dots,\theta_{n})\left(  \mathbf{1}-\sigma_{2}^{\mathcal{O}}%
S_{2n}\dots S_{23}\right)  \,. \label{2.15}%
\end{equation}
Using the crossing relation (\ref{2.2}) it turns out that the statistics
factor in (\ref{2.14}) and (\ref{2.15}) has to satisfy $\left(  \sigma
^{\mathcal{O}}\right)  ^{2}=(-1)^{n}$ with the solutions
\begin{equation}
\sigma^{\mathcal{O}}=i^{Q_{\mathcal{O}}}\text{~~~for ~}Q_{\mathcal{O}%
}=n\operatorname{mod}2\,. \label{2.15a}%
\end{equation}

We define the $SU(2)$ `Jost-function' $\phi(\theta)$ and $\tau(z)$ by the same
equations as for the sine-Gordon \cite{BFKZ} and the $Z(2)$ models
\cite{BK04,BFK}
\begin{align}
\phi(\theta)F(\theta)F(\theta+i\pi)  &  =1\label{2.22}\\
\tau(z)\phi(z)\phi(-z)  &  =1 \label{2.23}%
\end{align}
with the solution for $\phi(\theta)$ given by (\ref{1.26}). The functions
$\phi\left(  z\right)  $ and $\tau(z)$ are for $SU(2)$ explicitly given as%
\[
\phi(z)=\tilde{\phi}(-z)=\Gamma\left(  \frac{z}{2\pi i}\right)  \Gamma\left(
\frac{1}{2}-\frac{z}{2\pi i}\right)  \,,~~\tau(z)=\frac{z\sinh z}{4\pi^{3}}.
\]

The scalar function $\tilde{h}(\underline{\theta},\underline{z})$ in
(\ref{2.16}) encodes only data from the scattering matrix. The
\textbf{p-function} $p^{\mathcal{O}}(\underline{\theta},\underline{z})$ on the
other hand depends on the explicit nature of the local operator $\mathcal{O}$.
It is analytic in all variables and in order that the form factors satisfy
(i), (ii) and (iii) of eqs. (\ref{1.10}) - (\ref{1.14}) $p_{nm}(\underline
{\theta},\underline{z})$ has to satisfy%
\begin{equation}%
\begin{array}
[c]{lll}%
(\mathrm{i}^{\prime})_{2} &  & p_{nm}(\underline{\theta},\underline
{z})~\text{is symmetric under }\theta_{i}\leftrightarrow\theta_{j}\text{ and
}z_{i}\leftrightarrow z_{j}\\[1mm]%
(\mathrm{ii}^{\prime})_{2} &  & \left\{
\begin{array}
[c]{l}%
\sigma p_{nm}(\theta_{1}+2\pi i,\theta_{2},\dots,\underline{z})=(-1)^{m-1}%
p_{nm}(\theta_{1},\theta_{2},\dots,\underline{z})\\
p_{nm}(\underline{\theta},z_{1}+2\pi i,z_{2},\dots)=(-1)^{n}p_{nm}%
(\underline{\theta},z_{1},z_{2},\dots)
\end{array}
\right. \\[1mm]%
(\mathrm{iii}^{\prime})_{2} &  & \text{if }\theta_{12}=i\pi:p_{nm}%
(\underline{\theta},\underline{z})|_{z_{1}=\theta_{1}}=(-1)^{m-1}\sigma
p_{nm}(\underline{\theta},\underline{z})|_{z_{1}=\theta_{2}}\\[1mm]
&  & ~~~~~~~~~~~~~~~~~~~=\sigma p_{n-2m-1}(\theta_{3},\ldots,\theta_{n}%
,z_{2},\ldots,z_{m})+\tilde{p}%
\end{array}
\label{2.30}%
\end{equation}
where $\sigma$ is the statistics factor of (\ref{2.14a}) and (\ref{2.15}). In
order to simplify the notation we have suppressed the dependence of the
p-function $p^{\mathcal{O}}$ and the statistics factor $\sigma^{\mathcal{O}}$
on the operator $\mathcal{O}(x)$. By means of the ansatz (\ref{2.10}) and
(\ref{2.16}) - (\ref{2.19}) we have transformed the complicated form factor
equations (i) - (iii) to the simple ones $(\mathrm{i}^{\prime})_{2}$ -
$(\mathrm{iii}^{\prime})_{2}$ for the p-function.

\begin{theorem}
\label{T2}The co-vector valued function $F_{\underline{\alpha}}^{\mathcal{O}%
}(\underline{\theta})$ given by the ansatz (\ref{2.10}) and (\ref{2.16})
satisfies the form factor equations $(\mathrm{i}),\ (\mathrm{ii})$ and
$(\mathrm{iii})$ (see (\ref{1.10}) - (\ref{1.14})) if the p-function
$p^{\mathcal{O}}(\underline{\theta},\underline{z})$ satisfies the equations
$(\mathrm{i}^{\prime})_{2}$, $(\mathrm{ii}^{\prime})_{2}$, $(\mathrm{iii}%
^{\prime})_{2}$ of (\ref{2.30}) and the normalization constants satisfies the
recursion relation
\[
i\pi\tilde{\phi}(i\pi)F(i\pi)N_{n}=2iN_{n-2}\,.
\]

\end{theorem}

\noindent The proof of this theorem for $SU(2)$ is similar to that for the
sine-Gordon model in \cite{BFKZ} and may easily obtained from the ones for
$SU(3)$ or $SU(N)$ established below.

\section{$SU(3)$ form factors}

In this section we construct the form factors for the $SU(3)$ model, which
corresponds to the simplest example where the nested Bethe ansatz technique
has to be applied together with the off-shell Bethe ansatz. In comparison to
the previous $SU(2)$ case, there are additional properties to be obeyed by the
second-level Bethe ansatz function (See lemma \ref{l6} below for details).

\subsection{S-matrix}

The $SU(3)$ S-matrix is given by (\ref{1.2}) and (\ref{1.4}) for $N=3$. The
eigenvalue $S_{-}$ has a pole at $\theta=i\eta=\frac{2}{3}i\pi$ which means
that there exist bound states of two fundamental particles $\alpha
+\beta\rightarrow(\rho\sigma)\,$(with $1\leq\rho<\sigma\leq3$) which transform
as the anti-symmetric $SU(3)$ tensor representation. The general bound state
S-matrix formula \cite{K1,BK} for the scattering of a bound state with another
particle reads in particular for the $SU(3)$ case as%
\begin{align}
S_{(\rho\sigma)\gamma}^{\gamma^{\prime}(\rho^{\prime}\sigma^{\prime})}%
(\theta_{(12)3})\Gamma_{\alpha\beta}^{(\rho\sigma)}  &  =\Gamma_{\alpha
^{\prime}\beta^{\prime}}^{(\rho^{\prime}\sigma^{\prime})}S_{\alpha
\gamma^{\prime\prime}}^{\gamma^{\prime}\alpha^{\prime}}(\theta_{13}%
)S_{\beta\gamma}^{\gamma^{\prime\prime}\beta^{\prime}}(\theta_{23}%
)\,\Big|_{\theta_{12}=i\eta}\label{3.2}\\%
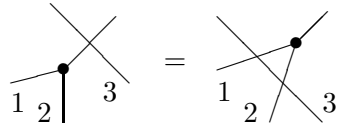
\begin{array}
[c]{c}%
\unitlength3.5mm\begin{picture}(5,5) \put(2,2){\line(1,1){2}}
\put(4.5,1.5){\line(-1,1){3}} \put(2,2){\line(-4,-1){2}}
\put(2,0){\line(0,1){2}} \put(0,.4){$1$} \put(1,0){$2$} \put(3.5,.8){$3$}
\put(2,2){\makebox(0,0){$\bullet$}} \end{picture}
\end{array}
&  =~%
\begin{array}
[c]{c}%
\unitlength3.5mm\begin{picture}(5,5) \put(3,3){\line(1,1){1.2}}
\put(4,0){\line(-1,1){4}} \put(0,2){\line(3,1){3}} \put(2,0){\line(1,3){1}}
\put(0,.8){$1$} \put(1,0){$2$} \put(4,.4){$3$}
\put(3,3){\makebox(0,0){$\bullet$}} \end{picture}
\end{array}
\nonumber
\end{align}
where $\theta_{(12)}$ is the bound state rapidity and $\eta$ the bound state
fusion angle. The bound state fusion intertwiner $\Gamma_{\alpha\beta}%
^{(\rho\sigma)}$ is defined by%
\[
i\operatorname*{Res}_{\theta=i\eta}S_{\alpha\beta}^{\beta^{\prime}%
\alpha^{\prime}}(\theta)=\sum_{\rho<\sigma}\Gamma_{(\rho\sigma)}%
^{\beta^{\prime}\alpha^{\prime}}\Gamma_{\alpha\beta}^{(\rho\sigma)}=~~%
\begin{array}
[c]{c}%
\unitlength3.4mm\begin{picture}(3,6) \put(1,1){\oval(2,2)[t]}
\put(1,5){\oval(2,2)[b]} \put(1,2){\makebox(0,0){$\bullet$}}
\put(1,4){\makebox(0,0){$\bullet$}} \put(-.2,0){$\alpha$} \put(1.7,0){$\beta$}
\put(1.3,2.7){$(\rho\sigma)$} \put(-.2,5.3){$\beta'$} \put(1.7,5.3){$\alpha'$}
\put(1,2){\line(0,1){2}} \end{picture}
\end{array}
~.
\]
With a convenient choice of an undetermined phase factor one obtains
\begin{equation}
\Gamma_{\alpha\beta}^{(\rho\sigma)}=\Gamma\left(  \delta_{\alpha}^{\rho}%
\delta_{\beta}^{\sigma}-\delta_{\alpha}^{\sigma}\delta_{\beta}^{\rho}\right)
\,,~~\Gamma_{(\rho\sigma)}^{\beta^{\prime}\alpha^{\prime}}=\Gamma\left(
\delta_{\rho}^{\beta^{\prime}}\delta_{\sigma}^{\alpha^{\prime}}-\delta
_{\sigma}^{\beta^{\prime}}\delta_{\rho}^{\alpha^{\prime}}\right)  \label{3.4}%
\end{equation}
where $\Gamma=i\sqrt{ia(i\eta)i\eta}$ is a number. Choosing in (\ref{3.2})
special cases for the external particles we calculate
\begin{align*}
S_{(12)3}^{3(12)}(\theta)  &  =b(\theta+\tfrac{1}{3}i\pi)b(\theta-\tfrac{1}%
{3}i\pi)=a(i\pi-\theta)\\
S_{(12)2}^{2(12)}(\theta)  &  =b(\theta+\tfrac{1}{3}i\pi)a(\theta-\tfrac{1}%
{3}i\pi)=b(\pi i-\theta)\\
S_{(23)1}^{3(12)}(\theta)  &  =-b(\theta+\tfrac{1}{3}i\pi)c(\theta-\tfrac
{1}{3}i\pi)=c(\pi i-\theta)
\end{align*}
which may be written as%
\[
S_{(\alpha\beta)\gamma}^{\gamma^{\prime}(\alpha^{\prime}\beta^{\prime}%
)}(\theta)=\delta_{\gamma}^{\gamma^{\prime}}\delta_{(\alpha\beta)}%
^{(\alpha^{\prime}\beta^{\prime})}b(\pi i-\theta)+\epsilon^{\gamma^{\prime
}\alpha^{\prime}\beta^{\prime}}\epsilon_{\alpha\beta\gamma}c(\pi i-\theta)
\]
with the total anti-symmetric tensors $\epsilon_{\alpha\beta\gamma}$ and
$\epsilon^{\alpha\gamma\beta}$ ($\epsilon_{123}=\epsilon^{321}=1$). These
formulae may be again understood as a crossing relation (c.f. \cite{KuS})
\begin{gather}
\epsilon_{\alpha^{\prime}\beta^{\prime}\delta}S_{\alpha\gamma^{\prime\prime}%
}^{\gamma^{\prime}\alpha^{\prime}}(\theta+\tfrac{1}{2}i\eta)S_{\beta\gamma
}^{\gamma^{\prime\prime}\beta^{\prime}}(\theta-\tfrac{1}{2}i\eta
)=\epsilon_{\alpha\beta\delta^{\prime}}S_{\gamma\delta}^{\delta^{\prime}%
\gamma^{\prime}}(i\pi-\theta)\label{3.6}\\%
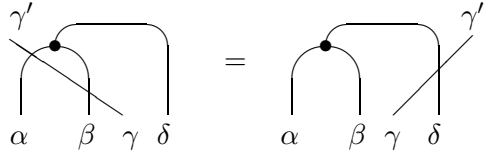
\begin{array}
[c]{c}%
\unitlength3mm\begin{picture}(21,6) \put(2,1){\oval(3,6)[t]}
\put(4.5,4){\oval(5,2)[t]} \put(2,4){\makebox(0,0){$\bullet$}}
\put(5,1){\line(-3,2){5}} \put(7,1){\line(0,1){3}} \put(0,-.3){$\alpha$}
\put(3,-.3){$\beta$} \put(5,-.3){$\gamma$} \put(0,5){$\gamma'$}
\put(6.5,-.3){$\delta$} \put(9.5,3){$=$} \put(14,1){\oval(3,6)[t]}
\put(16.5,4){\oval(5,2)[t]} \put(14,4){\makebox(0,0){$\bullet$}}
\put(17,1){\line(1,1){3.5}} \put(19,1){\line(0,1){3}} \put(12,-.3){$\alpha$}
\put(15,-.3){$\beta$} \put(16.6,-.3){$\gamma$} \put(20,5){$\gamma'$}
\put(18.5,-.3){$\delta$} \end{picture}
\end{array}
\nonumber
\end{gather}
or $S_{(\alpha\beta)\gamma}^{\gamma^{\prime}(\alpha^{\prime}\beta^{\prime}%
)}(\theta)=\mathbf{C}_{(\alpha\beta)\delta^{\prime}}S_{\gamma\delta}%
^{\delta^{\prime}\gamma^{\prime}}(i\pi-\theta)\mathbf{C}^{\delta
(\alpha^{\prime}\beta^{\prime})}$ if we write the charge conjugation matrices
as
\begin{equation}
\mathbf{C}_{(\alpha\beta)\gamma}=\mathbf{C}_{\alpha(\beta\gamma)}%
=\epsilon_{\alpha\beta\gamma}\,,~~\mathbf{C}^{\alpha(\beta\gamma)}%
=\mathbf{C}^{(\alpha\beta)\gamma}=\epsilon^{\alpha\beta\gamma}\,. \label{3.8}%
\end{equation}
Therefore we have the relations (c.f. (\ref{2.6}))%
\begin{equation}
\mathbf{C}_{\alpha(\rho\sigma)}\mathbf{C}^{(\rho\sigma)\beta}=\delta_{\alpha
}^{\beta}~,~~\mathbf{C}_{\alpha(\rho\sigma)}A_{\beta}^{\alpha}\mathbf{C}%
^{\beta(\rho\sigma)}=\operatorname*{tr}A \label{3.10}%
\end{equation}
and%
\begin{equation}
\mathbf{C}_{\alpha(\rho\sigma)}\Gamma_{\beta\gamma}^{(\rho\sigma)}%
=\mathbf{C}_{(\rho\sigma)\gamma}\Gamma_{\alpha\beta}^{(\rho\sigma)}%
=\epsilon_{\alpha\beta\gamma}\Gamma\,. \label{3.9}%
\end{equation}
These results\footnote{The physical aspects of these facts will we discussed
elsewhere \cite{BFK3}.} are consistent with the picture that the bound state
of particles 1 and 2 is to be identified with the anti-particle of 3. For
later convenience we consider the total 3-particle S-matrix in the
neighborhood of its poles at $\theta_{12},\theta_{23}=i\eta$
\begin{align}
\tilde{S}_{\alpha\beta\gamma}^{\gamma^{\prime}\beta^{\prime}\alpha^{\prime}%
}(\theta_{1},\theta_{2},\theta_{3})  & =\tilde{S}_{\alpha^{\prime\prime}%
\beta^{\prime\prime}}^{\beta^{\prime}\alpha^{\prime}}(\theta_{12})\tilde
{S}_{\alpha\gamma^{\prime\prime}}^{\gamma^{\prime}\alpha^{\prime\prime}%
}(\theta_{13})\tilde{S}_{\beta\gamma}^{\gamma^{\prime\prime}\beta
^{\prime\prime}}(\theta_{23})\nonumber\\
& \approx2\frac{i\eta}{\theta_{12}-i\eta}\frac{i\eta}{\theta_{23}-i\eta
}\epsilon^{\gamma^{\prime}\beta^{\prime}\alpha^{\prime}}\epsilon_{\alpha
\beta\gamma}\,.\label{3.12}%
\end{align}

\subsection{Form factors}

In order to obtain a recursion relation where only form factors for the
fundamental particles of type $\alpha=1,2,3$ (which transform as the $SU(3)$
vector representation) are involved, we have to apply the bound state relation
(iv) to get the anti-particle and then the creation annihilation equation
(iii)%
\begin{align*}
\operatorname*{Res}_{\theta_{12}=i\eta}F_{123\dots n}^{\mathcal{O}}(\theta
_{1},\theta_{2},\theta_{3},\dots,\theta_{n}) &  =F_{(12)3\dots n}%
^{\mathcal{O}}(\theta_{(12)},\theta_{3},\dots,\theta_{n})\sqrt{2}\Gamma
_{12}^{(12)}\\
\operatorname*{Res}_{\theta_{(12)3}=i\pi}F_{(12)3\dots n}^{\mathcal{O}}%
(\theta_{(12)},\theta_{3},\dots,\theta_{n}) &  =2i\mathbf{C}_{(12)3}F_{4\dots
n}^{\mathcal{O}}(\theta_{4},\dots,\theta_{n})\\
&~~~\times\left(\mathbf{1}-\sigma_{3}^{\mathcal{O}}S_{3n}\dots S_{34}\right)
\end{align*}
where $\Gamma_{12}^{(12)}$ is the bound state intertwiner (\ref{3.4}) (see
\cite{K1,BK}), $\theta_{(12)}=\frac{1}{2}(\theta_{1}+\theta_{2})$ is the bound
state rapidity, $\eta=\frac{2}{3}\pi$ is the bound state fusion angle and
$\mathbf{C}_{(12)3}$ is the charge conjugation matrix (\ref{3.8}). We obtain
with the short notation $\underline{\check{\theta}}=(\theta_{4},\dots
,\theta_{n})$%
\begin{equation}
\operatorname*{Res}_{\theta_{23}=i\eta}\operatorname*{Res}_{\theta_{12}=i\eta
}F_{123\dots n}^{\mathcal{O}}(\underline{\theta})=2i\varepsilon_{123}\sqrt
{2}\Gamma F_{4\dots n}^{\mathcal{O}}(\underline{\check{\theta}})\left(
\mathbf{1}-\sigma_{3}^{\mathcal{O}}S_{3n}\dots S_{34}\right)  \label{3.14}%
\end{equation}
where (\ref{3.9}) has been used.

\paragraph{Ansatz for form factors:}

We propose that the $n$-particle form factor of an operator $\mathcal{O}(x)$
is given by the same formula (\ref{2.10}) as for $SU(2)$ where the form factor
equations (i) and (ii) for K-function write again as (\ref{2.12}) and
(\ref{2.14}). Consistency of (\ref{3.14}), (\ref{2.14}) and the crossing
relation (\ref{3.6}) means that the statistics factors are of the form
$\sigma_{\alpha}^{\mathcal{O}}=\sigma^{\mathcal{O}}(r_{\alpha})$ if the
particle of type $\alpha$ belongs to a $SU(3)$ representation of rank
$r_{\alpha}=1,2$ and%
\begin{equation}
\sigma^{\mathcal{O}}(r)=e^{i\pi\frac{2}{3}rQ_{\mathcal{O}}}~~~\text{for
~}Q_{\mathcal{O}}=n\operatorname{mod}3 \label{3.15}%
\end{equation}
as an extension of (\ref{2.15a}). As for $SU(2)$, we propose again for the
K-function the ansatz in form of the integral representation (\ref{2.16}) with
(\ref{2.18}). However, the Bethe ansatz co-vector is here for $SU(3)$ of the
form
\begin{equation}
\tilde{\Psi}_{\underline{\alpha}}(\underline{\theta},\underline{z}%
)=L_{\underline{\beta}}(\underline{z}){\tilde{\Phi}}_{\underline{\alpha}%
}^{\underline{\beta}}(\underline{\theta},\underline{z}) \label{3.16}%
\end{equation}
where the basic Bethe states ${\tilde{\Phi}}_{\underline{\alpha}}%
^{\underline{\beta}}(\underline{\theta},\underline{z})$ are given by
(\ref{1.38}) and the indices $\alpha_{i}$ run over $1\leq\alpha_{i}\leq3$ and
the $\beta_{i}$ over $2\leq\beta_{i}\leq3$. For the function $L_{\underline
{\beta}}(\underline{z})$ one makes an analogous ansatz (\ref{2.16}) as for
$K_{\underline{\alpha}}(\underline{\theta})$ where the indices run here over a
set with one element less. By this procedure one obtains the nested Bethe
ansatz. The next level function $L_{\underline{\beta}}(\underline{z})$ is
assumed to satisfy

\begin{itemize}
\item[(i)$^{(1)}$]
\begin{equation}
L_{\dots ij\dots}(\dots,z_{i},z_{j},\dots)=L_{\dots ji\dots}(\dots,z_{j}%
,z_{i},\dots)\,\tilde{S}_{ij}(z_{ij}) \label{3.22}%
\end{equation}

\item[(ii)$^{(1)}$]
\begin{equation}
\mathbf{C}^{\bar{1}1}L_{1\ldots m}(z_{1}+i\pi,z_{2},\dots,z_{m})=L_{2\ldots
m1}(z_{2},\dots,z_{m},z_{1}-i\pi)\mathbf{C}^{1\bar{1}} \label{3.24}%
\end{equation}

\item[(iii)$^{(1)}$] there is a pole at $z_{12}=i\eta$ such that%
\begin{equation}
\operatorname*{Res}_{z_{12}=i\eta}L_{\underline{\beta}}(\underline{z}%
)=c_{1}\prod_{i=3}^{m}\tilde{\phi}(z_{i2})\operatorname*{Res}_{z_{12}=i\eta
}\tilde{S}_{\beta_{1}\beta_{2}}^{32}(z_{12})L_{\underline{\check{\beta}}%
}(\underline{\check{z}}) \label{3.26}%
\end{equation}
with $c_{1}=\tilde{\phi}(i\eta).$
\end{itemize}

\noindent These properties of the next level Bethe ansatz function
$L_{\underline{\beta}}(\underline{z})$ are discussed in lemma \ref{l6}.

The minimal two particle form factor function%
\begin{equation}
F\left(  \theta\right)  =c\exp\int\limits_{0}^{\infty}\frac{dt}{t\sinh^{2}%
t}e^{\frac{t}{3}}\sinh\frac{2}{3}t\left(  1-\cosh t\left(  1-\frac{\theta
}{i\pi}\right)  \right)  \label{3.28}%
\end{equation}
belongs to the highest weight $w=(2,0,0)$. The $SU(3)$ and the $Z(3)$ model
\cite{BK04,BFK} possess the same bound state structure, namely the
anti-particle is to be identified with the bound state of two particles.
Therefore we define the $SU(3)$ `Jost-function' $\phi\left(  z\right)  $ by
the same equation as for the $Z(3)$ model
\begin{equation}
\phi(\theta)\phi(\theta+i\eta)F(\theta)F(\theta+i\eta)F(\theta+2i\eta)=1
\label{3.29}%
\end{equation}
with the solution
\[
\phi(z)=\Gamma\left(  \frac{z}{2\pi i}\right)  \Gamma\left(  \frac{2}{3}%
-\frac{z}{2\pi i}\right)  \,.
\]
These equations determine also the constant $c$ in (\ref{3.28}). The function
$\tau(z)$ is again defined by (\ref{2.19}).

The function $\tilde{h}(\underline{\theta},\underline{z})$ is scalar and
encodes only data from the scattering matrix. The \textbf{p-function}
$p^{\mathcal{O}}(\underline{\theta},\underline{z})$ on the other hand depends
on the explicit nature of the local operator $\mathcal{O}$. It is analytic in
all variables and, in order that the form factors satisfy (i) - (iii) of eqs.
(\ref{1.10}) - (\ref{1.14}), we assume%
\begin{equation}%
\begin{array}
[c]{lll}%
(\mathrm{i}^{\prime})_{3} &  & p(\underline{\theta},\underline{z})~\text{is
symmetric under }\theta_{i}\leftrightarrow\theta_{j}\\[2mm]%
(\mathrm{ii}^{\prime})_{3} &  & \left\{
\begin{array}
[c]{l}%
\sigma p(\theta_{1}+2\pi i,\theta_{2},\dots,\underline{z})=(-1)^{m}%
p(\theta_{1},\theta_{2},\dots,\underline{z})\\
p(\underline{\theta},z_{1}+2\pi i,z_{2},\dots)=(-1)^{n}p(\underline{\theta
},z_{1},z_{2},\dots)
\end{array}
\right. \\[1mm]%
(\mathrm{iii}^{\prime})_{3} &  & \text{for }\theta_{12}=\theta_{23}%
=i\eta~\left\{
\begin{array}
[c]{l}%
p(\underline{\theta};\theta_{2},\theta_{3},\underline{\check{z}}%
)=(-1)^{m}\,p(\underline{\check{\theta}},\underline{\check{z}})\\
p(\underline{\theta};\theta_{1},\theta_{2},\underline{\check{z}}%
)=\sigma\,p(\underline{\check{\theta}},\underline{\check{z}})
\end{array}
\right.
\end{array}
\label{3.30}%
\end{equation}
with $\underline{\check{\theta}}=(\theta_{4},\dots,\theta_{n}),\,\underline
{\check{z}}=(z_{3},\dots,z_{m})$. In $(\mathrm{ii}^{\prime})_{3}$ and
$(\mathrm{iii}^{\prime})_{3}$ $\sigma$ is the statistics factor of the
operator ${\mathcal{O}}$ with respect to the fundamental particle belonging to
the vector representation of $SU(3)$.

\begin{theorem}
\label{T13}Let the co-vector valued function $F_{\underline{\alpha}%
}^{\mathcal{O}}(\underline{\theta})$ be defined by the ansatz (\ref{2.10}),
(\ref{2.16}), (\ref{2.18}) and (\ref{3.16}). Let the p-function satisfy
$(\mathrm{i}^{\prime})_{3},\ (\mathrm{ii}^{\prime})_{3}$ and $(\mathrm{iii}%
^{\prime})_{3}$ of (\ref{3.30}) and let the function $L_{\underline{\beta}%
}(\underline{z})$ satisfy $(\mathrm{i})^{(1)},\ (\mathrm{ii})^{(1)}$ and
$(\mathrm{iii})^{(1)}$ of (\ref{3.22}) - (\ref{3.26}). Let the normalization
constants satisfy the recursion relation
\begin{equation}
2(i\eta)^{2}\tilde{\phi}^{2}(i\eta)\tilde{\phi}(2i\eta)F^{2}(i\eta
)F(2i\eta)N_{n}=2i\sqrt{2}\Gamma N_{n-3}\,. \label{norm3}%
\end{equation}
Then the function $F_{\underline{\alpha}}^{\mathcal{O}}(\underline{\theta})$
satisfies the form factor equations $(\mathrm{i}),\ (\mathrm{ii})$ and
$(\mathrm{iii})$ (see (\ref{1.10}) - (\ref{1.14})). In particular (\ref{3.14})
is satisfied.
\end{theorem}

\begin{proof}
Property (i) in the form of (\ref{2.12}) follows directly from $(\mathrm{i}%
^{\prime})_{3}$, the Yang-Baxter equations and the action of the S-matrix on
the pseudo-ground state $\Omega$
\begin{align*}
\,\Omega_{\dots ji\dots}\,\tilde{C}_{\dots ji\dots}(\dots\theta_{j},\theta
_{i}\dots)\,\tilde{S}_{ij}(\theta_{ij})  &  =\Omega_{\dots ji\dots}\,\tilde
{S}_{ij}(\theta_{ij})\,\tilde{C}_{\dots ij\dots}(\dots\theta_{i},\theta
_{j}\dots)\\
&  =\,\Omega_{\dots ij\dots}\,\tilde{C}_{\dots ij\dots}({\dots}\theta
_{i},\theta_{j}{\dots})\,.
\end{align*}
because $\,\tilde{S}_{11}^{11}(\theta)=S_{11}^{11}(\theta)/a(\theta)=1$ and
$F(\theta)=F(-\theta)a(\theta)$.

Using (i) and (\ref{3.10}) the property (ii) in the form of (\ref{2.14}) may
be rewritten as a matrix difference equation \cite{BKZ,BKZ1,BKZ2}
\begin{equation}
K_{1\dots n}(\underline{\theta}^{\prime})\sigma=K_{1\dots n}(\underline
{\theta})\,Q_{1\dots n}(\underline{\theta}) \label{3.34}%
\end{equation}
where $\underline{\theta}^{\prime}=(\theta_{1}+2\pi i,\theta_{2}\dots
,\theta_{n})$ and $\sigma$ is a statistics factor. The matrix $Q(\underline
{\theta})=Q(\underline{\theta},1)$ is a special case (for $i=1$) of the trace
\[
Q_{1\dots n}(\underline{\theta},i)=\operatorname*{tr}\nolimits_{0}%
\mathrm{\,}\tilde{T}_{Q,1\dots n,0}(\underline{\theta},i)=%
\begin{array}
[c]{c}%
\unitlength6mm\begin{picture}(10,4)(0,0) \put(1,1){\oval(8,2)[rt]}
\put(9,3){\oval(8,2)[lb]} \put(2,1){\line(0,1){2}} \put(4,1){\line(0,1){2}}
\put(6,1){\line(0,1){2}} \put(8,1){\line(0,1){2}} \put(1.9,.3){$\alpha_1$}
\put(1.9,3.3){$\alpha'_1$} \put(4.9,.3){$\alpha_i$} \put(4.9,3.3){$\alpha'_i$}
\put(7.9,.3){$\alpha_n$} \put(7.9,3.3){$\alpha'_n$} \put(1.2,1.1){$\theta_1$}
\put(5.2,1.1){$\theta'_i$} \put(4.4,2.4){$\theta_i$} \put(8.2,1.1){$\theta_n$}
\put(2.6,2.5){\dots} \put(6.6,2.5){\dots} \put(5,3){\oval(10,2)[t]}
\put(1,3){\oval(2,2)[lb]} \put(9,3){\oval(2,2)[rb]} \end{picture}
\end{array}
\]
of a modified monodromy matrix%
\[
\tilde{T}_{Q,1\dots n,0}(\underline{\theta},i)=\tilde{S}_{10}(\theta
_{1}-\theta_{i}^{\prime})\,\cdots\mathbf{P}_{i0}\cdots\tilde{S}_{n0}%
(\theta_{n}-\theta_{i})
\]
where $\mathbf{P}$ is the permutation matrix. For the special case $i=1$ the
matrix $Q(\underline{\theta},1)=Q(\underline{\theta})$ may be written as a
trace of the ordinary monodromy matrix (\ref{1.30}) over the auxiliary space
for the specific value of the spectral parameter $\theta_{0}=\theta_{1}$
\begin{equation}
Q_{1\dots n}(\underline{\theta})=\operatorname*{tr}\nolimits_{0}\tilde
{T}_{1\dots n,0}(\underline{\theta},\theta_{1}) \label{3.36}%
\end{equation}
because $\tilde{S}(0)=\mathbf{P}$. The difference equation (\ref{3.34}) may be
depicted as
\[%
\begin{array}
[c]{c}%
\unitlength4mm\begin{picture}(5,4) \put(2.5,2){\oval(5,2)}
\put(2.5,2){\makebox(0,0){$K$}} \put(1,0){\line(0,1){1}}
\put(2,0){\line(0,1){1}} \put(4,0){\line(0,1){1}} \put(2.4,.5){$\dots$}
\end{picture}
\end{array}
~\sigma~~=~~%
\begin{array}
[c]{c}%
\unitlength4mm\begin{picture}(7,5) \put(4.5,2){\oval(5,2)[b]}
\put(3.5,2){\oval(7,6)[t]} \put(1,2){\oval(2,2)[lb]} \put(1,0){\oval(2,2)[rt]}
\put(3.5,3){\oval(5,2)} \put(3.5,3){\makebox(0,0){$K$}}
\put(3,0){\line(0,1){2}} \put(5,0){\line(0,1){2}} \put(3.4,.5){$\dots$}
\end{picture}
\end{array}
\]
where we use the rule that the rapidity of a line changes by $2\pi i$ if the
line bends by $360^{0}$ in the positive sense.

In the following we will suppress the indices $1\dots n$. The Yang-Baxter
relations (\ref{1.32}) imply the typical commutation rules for the matrices
$\tilde{A},\tilde{C},\tilde{D}$ defined in eq.~(\ref{1.34})
\begin{align}
\tilde{C}^{\beta}(\underline{\theta},z)\tilde{A}(\underline{\theta},\theta) &
=\frac{1}{\tilde{b}(\theta-z)}\tilde{A}(\underline{\theta},\theta)\tilde
{C}^{\beta}(\underline{\theta},z)-\frac{\tilde{c}(\theta-z)}{\tilde{b}%
(\theta-z)}\tilde{A}(\underline{\theta},z)\tilde{C}^{\beta}(\underline{\theta
},\theta)\label{3.38}\\
\tilde{C}^{\beta}(\underline{\theta},z)\tilde{D}_{\gamma}^{\gamma^{\prime}%
}(\underline{\theta},\theta) &  =\frac{1}{\tilde{b}(z-\theta)}\tilde{S}%
_{\beta^{\prime}\gamma^{\prime\prime}}^{\gamma^{\prime}\beta}(z-\theta
)\tilde{D}_{\gamma}^{\gamma^{\prime\prime}}(\underline{\theta},\theta
)\tilde{C}^{\beta^{\prime}}(\underline{\theta},z)\nonumber\\
&  -\frac{\tilde{c}(z-\theta)}{\tilde{b}(z-\theta)}\tilde{D}_{\gamma}^{\beta
}(\underline{\theta},z)\tilde{C}^{\gamma^{\prime}}(\underline{\theta}%
,\theta)\nonumber
\end{align}
where $\beta,\beta^{\prime},\gamma,\gamma^{\prime},\gamma^{\prime\prime}%
\in\left\{  2,3\right\}  $. In addition there are the Zapletal commutation
rules \cite{BKZ,BKZ1,BKZ2} where also the matrices $A_{Q},C_{Q},D_{Q}$ defined
by
\[
\tilde{T}_{Q}({\underline{\theta}})=\left(
\begin{array}
[c]{cc}%
\tilde{A}_{Q}(\underline{\theta}) & \tilde{B}_{Q}^{\beta}(\underline{\theta
})\\
\tilde{C}_{Q}^{\beta}(\underline{\theta}) & \tilde{D}_{Q}{{}}_{\beta}%
^{\beta^{\prime}}(\underline{\theta})
\end{array}
\right)
\]
are involved \cite{BKZ}
\begin{align}
\tilde{C}^{\beta}(\underline{\theta},z)\tilde{A}_{Q}(\underline{\theta}) &
=\frac{1}{\tilde{b}(\theta_{1}^{\prime}-z)}\tilde{A}_{Q}(\underline{\theta
})\tilde{C}^{\beta}(\underline{\theta}^{\prime},z)-\frac{\tilde{c}(\theta
_{1}^{\prime}-z)}{\tilde{b}(\theta_{1}^{\prime}-z)}\tilde{A}(\underline
{\theta},z)\tilde{C}_{Q}^{\beta}(\underline{\theta})\label{3.40}\\
\tilde{C}^{\beta}(\underline{\theta},z)\tilde{D}{_{Q}{}}_{\gamma}%
^{\gamma^{\prime}}(\underline{\theta}) &  =\frac{1}{\tilde{b}(z-\theta
_{1}^{\prime})}\tilde{S}_{\beta^{\prime}\gamma^{\prime\prime}}^{\gamma
^{\prime}\beta}(z-\theta_{1}^{\prime})\tilde{D}_{Q}{}_{\gamma}^{\gamma
^{\prime\prime}}(\underline{\theta})\tilde{C}^{\beta^{\prime}}(\underline
{\theta}^{\prime},z)\nonumber\\
&  -\frac{\tilde{c}(z-\theta_{1})}{\tilde{b}(z-\theta_{1})}\tilde{D}_{\gamma
}^{\beta}(\underline{\theta},z)\tilde{C}_{Q}^{\gamma^{\prime}}(\underline
{\theta})\,.\label{3.41}%
\end{align}
Note that we assign to the auxiliary space of $\tilde{T}_{Q}(\underline
{\theta})$ corresponding to the horizontal line the spectral parameter
$\theta_{1}$ on the right hand side and $\theta_{1}^{\prime}=$ $\theta
_{1}+2\pi i$ on the left hand side.

We are now going to prove (\ref{3.34}) in the form%
\begin{equation}
K(\underline{\theta})\left(  \tilde{A}_{Q}(\underline{\theta})+\sum_{\beta
=2}^{3}\tilde{D}{_{Q}{}}_{\beta}^{\beta}(\underline{\theta})\right)
=K(\underline{\theta})\,Q(\underline{\theta})=K(\underline{\theta}^{\prime
})\sigma\label{3.42}%
\end{equation}
where $K(\underline{\theta})$ is a co-vector valued function as given by
eq.~(\ref{2.16}) and the Bethe ansatz state (\ref{1.38}). To analyze the left
hand side of eq.~(\ref{3.42}) we proceed as follows: We apply the trace of
$\tilde{T}_{Q}$ which is $\tilde{A}_{Q}+\sum_{\beta=2}^{3}{\tilde{D}_{Q}{}%
}_{\beta}^{\beta}$ to the co-vector $K(\underline{\theta})$
\[
\Omega\,\tilde{C}^{\beta_{m}}(\underline{\theta},z_{m})\cdots\tilde{C}%
^{\beta_{1}}(\underline{\theta},z_{1})\tilde{T}{_{Q}{}}_{\gamma}%
^{\gamma^{\prime}}\,(\underline{\theta})=~%
\begin{array}
[c]{c}%
\unitlength5mm\begin{picture}(11,5.5) \put(10,5){\oval(17,2)[lb]}
\put(10,5){\oval(20,6)[lb]} \put(4,0){\line(0,1){5}} \put(9,0){\line(0,1){5}}
\put(10,5){\oval(14,8)[lb]} \put(2,0){\oval(2,2)[rt]} \put(-.5,5.3){$\beta_1$}
\put(1,5.3){$\beta_m$} \put(2.8,5.3){1} \put(3.8,5.3){1} \put(8.8,5.3){1}
\put(1.4,.8){$\gamma'$} \put(10.2,.8){$\gamma$} \put(10.2,1.8){1}
\put(10.2,3.8){1} \put(4.3,0){$\theta_{2}$} \put(9.2,0){$\theta_{n}$}
\put(7.5,2.3){$z_1$} \put(7.5,4.3){$z_m$} \put(5.5,3){$\dots$}
\put(9.5,2.7){$\vdots$} \put(7.5,1.2){$\theta_1$} \put(2,0){$\theta_1'$}
\end{picture}
\end{array}
.
\]
In the contribution from $\tilde{A}_{Q}(\underline{\theta})$ which means
$\gamma^{\prime}=\gamma=1$ one may use Yang-Baxter relations to observe that
only the amplitudes $\tilde{S}_{11}^{11}(\theta_{1}-z_{j})=1$ appear in the
S-matrices $\tilde{S}(\theta_{1}-z_{j})$ which are constituents of the
C-operators. Therefore we may shift all $z_{j}$-integration contours
$\mathcal{C}_{\underline{\theta}}$ to $\mathcal{C}_{\underline{\theta}%
^{\prime}}$ without changing the values of the integrals, because there are no
singularities inside $\mathcal{C}_{\underline{\theta}}\cup-\mathcal{C}%
_{\underline{\theta}^{\prime}}$ (c.f. Fig. \ref{f5.1}). Using a short notation
we have
\[
K(\underline{\theta})\,\tilde{A}_{Q}(\underline{\theta})=\int_{\mathcal{C}%
_{{\underline{\theta}}^{\prime}}}d\underline{z}\,\tilde{h}(\underline{\theta
},\underline{z})p(\underline{\theta},\underline{z})\,\tilde{\Psi}%
(\underline{\theta},\underline{z})\,\tilde{A}_{Q}(\underline{\theta})
\]
(with $\int_{\mathcal{C}_{\underline{\theta}}}d\underline{z}=\frac{1}{m!}%
\int_{\mathcal{C}_{\underline{\theta}}}\frac{dz_{1}}{R}\cdots\int
_{\mathcal{C}_{\underline{\theta}}}\frac{dz_{m}}{R}$). We now proceed as usual
in the algebraic Bethe ansatz and push the $\tilde{A}_{Q}(\underline{\theta})$
and $\tilde{D}_{Q}(\underline{\theta})$ through all the $\tilde{C}$-operators
using the commutation rules (\ref{3.40}) and (\ref{3.41}) and obtain
\begin{align*}
\tilde{C}^{\beta_{m}}(\underline{\theta},z_{m})\cdots\tilde{C}^{\beta_{1}%
}(\underline{\theta},z_{1})\,\tilde{A}_{Q}(\underline{\theta}) &  =\prod
_{j=1}^{m}\frac{1}{\tilde{b}(\theta_{1}^{\prime}-z_{j})}\,\tilde{A}%
_{Q}(\underline{\theta})\\
&  \times\tilde{C}^{\beta_{m}}(\underline{\theta}^{\prime
},z_{m})\cdots\tilde{C}^{\beta_{1}}(\underline{\theta}^{\prime},z_{1})
+\sum uw_{A}\,,
\end{align*}%
\begin{align*}
\tilde{C}^{\beta_{m}}(\underline{\theta},z_{m})\cdots\tilde{C}^{\beta_{1}%
}(\underline{\theta},z_{1})\,\tilde{D}{_{Q}{}}_{\beta}^{\beta^{\prime}%
}(\underline{\theta}) &  =\prod_{j=1}^{m}\frac{1}{\tilde{b}(z_{j}-\theta
_{1}^{\prime})}{\tilde{T}^{(1)}\,}_{\underline{\beta}\;\beta^{\prime\prime}%
}^{\beta^{\prime}\underline{\beta}^{\prime}}(\underline{z},\theta_{1}^{\prime
})\tilde{D}{_{Q}{}}_{\beta}^{\beta^{\prime\prime}}(\underline{\theta})\\
&  \times\tilde{C}^{\beta_{m}^{\prime}}({\underline{\theta}},z_{m}%
)\cdots\tilde{C}^{\beta_{1}^{\prime}}({\underline{\theta}},z_{1})+\sum
uw_{D}\,.
\end{align*}
The \textquotedblleft wanted terms\textquotedblright\ written out explicitly
originate from the first term in the commutations rules (\ref{3.40}); all
other contributions yield the so-called \textquotedblleft unwanted
terms\textquotedblright. The next level monodromy matrix is
\[
{\tilde{T}^{(1)}\,}_{\underline{\beta}\;\beta}^{\beta^{\prime}\underline
{\beta}^{\prime}}(\underline{z},\theta)=\left(  \tilde{S}_{10}(z_{1}%
-\theta)\cdots\tilde{S}_{m0}(z_{m}-\theta)\right)  _{\underline{\beta}\;\beta
}^{\beta^{\prime}\;\underline{\beta}^{\prime}}%
\]
where the $\beta$'s and also the internal summation indices take the values
$2,3$. If we insert these equations into the representation (\ref{2.16}) of
$K({\underline{\theta}})$ we first find that the wanted contribution from
$\tilde{A}_{Q}$ already gives the result we are looking for. Secondly the
wanted contribution from $\tilde{D}_{Q}$ applied to $\Omega$ gives zero.
Thirdly the unwanted contributions from $\tilde{A}_{Q}$ and $\tilde{D}_{Q}$
cancel after integration over the $z_{j}$. All these three facts can be seen
as follows. We have
\begin{equation}
\Omega\,\tilde{A}_{Q}(\underline{\theta})=\Omega~,~~~\Omega\,\tilde{D}{_{Q}{}%
}_{\beta}^{\beta^{\prime}}(\underline{\theta})=\delta_{\beta}^{\beta^{\prime}%
}\prod_{i=1}^{n}\tilde{b}(\theta_{i}-\theta_{1})\Omega=0\label{3.48}%
\end{equation}
which follow from eq.~(\ref{1.40}). Therefore the wanted term from $\tilde
{A}_{Q}$ yields
\begin{align*}
\left[  K(\underline{\theta})\,\tilde{A}_{Q}(\underline{\theta})\right]  ^{w}
&  =\int_{\mathcal{C}_{{\underline{\theta}}^{\prime}}}d\underline{z}%
\,\prod_{j=1}^{m}\frac{1}{\tilde{b}(\theta_{1}^{\prime}-z_{j})}\,\tilde
{h}(\underline{\theta},\underline{z})p(\underline{\theta},\underline{z}%
)\tilde{\Psi}(\underline{\theta}^{\prime},\underline{z})\\
&  =\sigma\int_{\mathcal{C}_{{\underline{\theta}}^{\prime}}}d\underline
{z}\,\tilde{h}(\underline{\theta}^{\prime},\underline{z})p(\underline{\theta
}^{\prime},\underline{z})\tilde{\Psi}(\underline{\theta}^{\prime}%
,\underline{z})\\
&  =\sigma K(\underline{\theta}^{\prime})\,.
\end{align*}
It has been used that the `shift relation' of the function $\phi(\theta)$ in
(\ref{1.28}) and (ii') of (\ref{3.30}) for the function $p$ imply that%
\[
\prod_{j=1}^{m}\frac{1}{\tilde{b}(\theta_{1}^{\prime}-z_{j})}\,\tilde
{h}(\underline{\theta},\underline{z})p(\underline{\theta},\underline
{z})=\sigma\tilde{h}(\underline{\theta}^{\prime},\underline{z})p(\underline
{\theta}^{\prime},\underline{z})\,.
\]
The wanted contribution from $\tilde{D}{_{Q}}$ vanish, since $\tilde{b}(0)=0$.
Therefore it remains to prove that the unwanted terms cancel. The commutation
relations (\ref{3.40}) and (\ref{3.41}) imply that the unwanted terms are
proportional to a product of $\tilde{C}$-operators, where one $\tilde
{C}(\underline{\theta},z_{j})$ is replaced by $\tilde{C}_{Q}(\underline
{\theta})$. Because of the symmetry (i)$^{1}$ of $L_{\underline{\beta}%
}(\underline{z})$ it is sufficient to consider only the unwanted terms for
$j=1$ which are denoted by $uw_{A}^{1}$ and $uw_{D}^{1}$. They originate from
the second term in the commutation rules (\ref{3.40}) when $\tilde{A}%
_{Q}(\underline{\theta})$ is commuted with $\tilde{C}(\underline{\theta}%
,z_{1})$. Then the resulting $\tilde{A}(\underline{\theta},z_{1})$ pushed
through the other $C$'s. Taking into account only the first terms in
(\ref{3.38}) we arrive at
\[
uw_{A}^{1}(\underline{z})=-\frac{\tilde{c}(\theta_{1}^{\prime}-z_{1})}%
{\tilde{b}(\theta_{1}^{\prime}-z_{1})}\,\prod_{j=2}^{m}\frac{1}{\tilde
{b}(z_{1}-z_{j})}\,\Omega\,\tilde{A}(\underline{\theta},z_{1})\,\tilde
{C}^{\beta_{m}}(\underline{\theta},z_{m})\dots\tilde{C}_{Q}^{\beta_{1}%
}(\underline{\theta})
\]
Using (\ref{1.40}) and Yang-Baxter relations (always taking into account the
`$SU(N)$ ice rule' which means `color conservation') and $\tilde{a}(\theta)=1$
one obtains%
\begin{align*}
\left[  \Omega\,\tilde{A}({\underline{\theta}},z_{1})\,\tilde{C}^{\beta_{m}%
}({\underline{\theta}},z_{m})\dots\tilde{C}_{Q}^{\beta_{1}}({\underline
{\theta}})\right]  _{\underline{\alpha}}  &  =\,\delta_{\alpha_{1}}^{\beta
_{1}}{\check{\Phi}}_{\underline{\tilde{\alpha}}}^{\underline{\tilde{\beta}}%
}(\underline{\check{\theta}},\underline{\check{z}})\\%
\begin{array}
[c]{l}%
\unitlength4mm\begin{picture}(12,8) \put(11,7){\oval(19,4)[lb]}
\put(11,7){\oval(21,7)[lb]} \put(3,7){\oval(6,10)[lb]}
\put(3,6){\line(1,0){8}} \put(5,1){\line(0,1){6}} \put(10,1){\line(0,1){6}}
\put(11,7){\oval(14,10)[lb]} \put(3,1){\oval(2,2)[rt]}
\put(-.7,7.4){$\beta_1$} \put(.2,7.4){$\beta_2$} \put(1.2,7.4){$\beta_m$}
\put(3.8,7.3){1} \put(4.8,7.3){1} \put(9.8,7.3){1} \put(3.7,0){$\alpha_1$}
\put(4.9,0){$\alpha_2$} \put(9.8,0){$\alpha_n$} \put(2.4,5.8){1}
\put(11.2,1.8){1} \put(11.2,3.3){1} \put(11.2,4.8){1} \put(11.2,5.8){1}
\put(5.3,1){$\theta_{2}$} \put(10.3,1){$\theta_{n}$} \put(8.5,3.8){$z_2$}
\put(8.5,6.3){$z_1$} \put(8.5,5.3){$z_m$} \put(6.8,4.3){$\dots$}
\put(10.5,3.9){$\vdots$} \put(8.5,2.3){$\theta_1$} \end{picture}
\end{array}
&  =\,%
\begin{array}
[c]{l}%
\unitlength4mm\begin{picture}(12,8) \put(11,6){\oval(19,4)[lb]}
\put(11,6){\oval(21,7)[lb]} \put(2,6){\oval(4,10)[lb]}
\put(9,6){\oval(12,2)[lb]} \put(9,2){\oval(2,6)[rt]}
\put(11,2){\oval(2,2)[lb]} \put(4,0){\line(0,1){6}} \put(9,0){\line(0,1){6}}
\put(2,0){\oval(2,2)[rt]} \put(-.7,6.4){$\beta_1$} \put(.2,6.4){$\beta_2$}
\put(1.2,6.4){$\beta_m$} \put(2.8,6.3){1} \put(3.8,6.3){1} \put(8.8,6.3){1}
\put(2.7,-1){$\alpha_1$} \put(3.9,-1){$\alpha_2$} \put(8.8,-1){$\alpha_n$}
\put(11.2,.8){1} \put(11.2,2.3){1} \put(11.2,3.8){1} \put(4.3,0){$\theta_{2}$}
\put(9.3,0){$\theta_{n}$} \put(7.5,2.8){$z_2$} \put(7.5,4.3){$z_m$}
\put(5.8,3.3){$\dots$} \put(9.5,2.9){$\vdots$} \put(7.5,5.3){$\theta_1$}
\end{picture}
\end{array}
\end{align*}
where ${\check{\Phi}}_{\underline{\tilde{\alpha}}}^{\underline{\tilde{\beta}}%
}(\underline{\check{\theta}},\underline{\check{z}})$ does not depend on
$\alpha_{1},\beta_{1},\theta_{1},z_{1}$. Similarly, we obtain the first
unwanted contribution from $\sum_{\beta}\tilde{D}{_{Q}{}}_{\beta}^{\beta}$ as
\begin{multline*}
uw_{D}^{1}=-\frac{\tilde{c}(z_{1}-\theta_{1})}{\tilde{b}(z_{1}-\theta_{1}%
)}\prod_{j=2}^{m}\frac{1}{\tilde{b}(z_{j}-z_{1})}\,\left(  {\tilde{T}%
^{(1)}{\,}}_{\underline{\beta}^{\prime}\beta^{\prime\prime}}^{\beta
\;\underline{\beta}}(\underline{z},z_{1})\right)  \\
\times\Omega\tilde{D}_{\beta}^{\beta^{\prime\prime}}(\underline{\theta}%
,z_{1})\tilde{C}^{\beta_{m}^{\prime}}(\underline{\theta},z_{m})\cdots\tilde
{C}_{Q}^{\beta_{1}^{\prime}}(\underline{\theta})
\end{multline*}
which may be depicted as
\begin{multline*}
\left(  {\tilde{T}^{(1)}{}}_{\underline{\beta}^{\prime}\;\beta^{\prime\prime}%
}^{\beta\;\underline{\beta}}(\underline{z},z_{1})\right)  \Omega\tilde
{D}_{\beta}^{\beta^{\prime\prime}}(\underline{\theta},z_{1})\tilde{C}%
^{\beta_{m}^{\prime}}(\underline{\theta},z_{m})\cdots\tilde{C}_{Q}^{\beta
_{1}^{\prime}}(\underline{\theta})\\
=~~~~%
\begin{array}
[c]{l}%
\unitlength4mm\begin{picture}(12,8) \put(11,7){\oval(17,4)[lb]}
\put(11,7){\oval(20,7)[lb]} \put(3,5){\oval(6,6)[lb]}
\put(1,7){\oval(2,2)[lb]} \put(-1,5){\oval(2,2)[rt]} \put(1,6){\line(1,0){10}}
\put(5,1){\line(0,1){6}} \put(10,1){\line(0,1){6}}
\put(11,7){\oval(14,10)[lb]} \put(3,1){\oval(2,2)[rt]}
\put(-.5,7.4){$\beta_1$} \put(.8,7.4){$\beta_2$} \put(2.2,7.4){$\beta_m$}
\put(3.8,7.3){1} \put(4.8,7.3){1} \put(9.8,7.3){1} \put(3.8,0){$\alpha_1$}
\put(4.8,0){$\alpha_2$} \put(9.8,0){$\alpha_n$} \put(11.2,1.8){1}
\put(11.2,3.3){1} \put(11.2,4.8){1} \put(11.2,5.8){$\beta$}
\put(-2,5.8){$\beta$} \put(5.3,1){$\theta_{2}$} \put(10.2,1){$\theta_{n}$}
\put(8.5,3.8){$z_2$} \put(8.5,6.3){$z_1$} \put(8.5,5.3){$z_m$}
\put(7,4.3){$\dots$} \put(10.5,3.9){$\vdots$} \put(8.5,2.2){$\theta_1$}
\end{picture}
\end{array}
.
\end{multline*}
Again (\ref{1.40}) and Yang-Baxter relations yield
\[
\left[  \Omega\,\tilde{D}_{\beta}^{\beta^{\prime\prime}}(\underline{\theta
},z_{1})\tilde{C}^{\beta_{m}^{\prime}}(\underline{\theta},z_{m})\cdots
\tilde{C}_{Q}^{\beta_{1}^{\prime}}(\underline{\theta})\right]  _{\underline
{\alpha}}=\delta_{\beta}^{\beta^{\prime\prime}}\prod_{i=1}^{n}\tilde{b}%
(\theta_{i}-z_{1})\,\delta_{\alpha_{1}}^{\beta_{1}^{\prime}}{\check{\Phi}%
}_{\underline{\tilde{\alpha}}}^{\underline{\tilde{\beta}}^{\prime}}%
(\underline{\check{\theta}},\underline{\check{z}})
\]
while assumption (ii)$^{(1)}$ for $L_{\underline{\beta}}(\underline{z})$
means
\[
L_{\underline{\beta}}(\underline{z}){Q^{(1)}}_{\underline{\beta}^{\prime}%
}^{\underline{\beta}}(\underline{z})=L_{\underline{\beta}}(\underline
{z}^{\prime})
\]
where analogously to eq. (\ref{3.36}) ${Q^{(1)}}_{\underline{\beta}^{\prime}%
}^{\underline{\beta}}(\underline{z})={\tilde{T}^{(1)}{}}_{\underline{\beta
}^{\prime}\;\beta}^{\beta\;\underline{\beta}}(\underline{z},z_{1})$ is the
next level $Q$-matrix and $\underline{z}^{\prime}=(z_{1}^{\prime}=z_{1}+2\pi
i,\dots,z_{m})$. Therefore we finally obtain%
\begin{align*}
&  \left[  K(\underline{\theta})\,\tilde{A}_{Q}(\underline{\theta})\right]
_{\underline{\alpha}}^{uw}\\
&  =-\int_{\mathcal{C}_{{\underline{\theta}}^{\prime}}}d\underline{z}%
\,\tilde{h}(\underline{\theta},\underline{z})\,p(\underline{\theta}%
,\underline{z})\frac{\tilde{c}(\theta_{1}^{\prime}-z_{1})}{\tilde{b}%
(\theta_{1}^{\prime}-z_{1})}\,\prod_{j=2}^{m}\frac{1}{\tilde{b}(z_{1}-z_{j}%
)}\,L_{\underline{\beta}}(\underline{z})\delta_{\alpha_{1}}^{\beta_{1}}%
{\check{\Phi}}_{\underline{\tilde{\alpha}}}^{\underline{\tilde{\beta}}%
}(\underline{\check{\theta}},\underline{\check{z}})\\
&  \left[  K(\underline{\theta})\,\tilde{D}{_{Q}{}}_{\beta}^{\beta}\right]
_{\underline{\alpha}}^{uw}\\
&  =-\int_{\mathcal{C}_{{\underline{\theta}}}}d\underline{z}\,\tilde
{h}(\underline{\theta},\underline{z}^{\prime})\,p(\underline{\theta
},\underline{z}^{\prime})\frac{\tilde{c}(z_{1}-\theta_{1})}{\tilde{b}%
(z_{1}-\theta_{1})}\prod_{j=2}^{m}\frac{1}{\tilde{b}(z_{1}^{\prime}-z_{j}%
)}\,L_{\underline{\beta}}(\underline{z}^{\prime})\delta_{\alpha_{1}}%
^{\beta_{1}}{\check{\Phi}}_{\underline{\tilde{\alpha}}}^{\underline
{\tilde{\beta}}}(\underline{\check{\theta}},\underline{\check{z}})\,.
\end{align*}
It has been used that the `shift relation' of the function $\phi(\theta)$ in
(\ref{1.28}) and (ii') of (\ref{3.30}) for the function $p$ imply%
\[
\prod_{i=1}^{n}\tilde{b}(\theta_{i}-z_{1})\prod_{j=2}^{m}\frac{1}{\tilde
{b}(z_{j}-z_{1})}\,\tilde{h}(\underline{\theta},\underline{z})p(\underline
{\theta},\underline{z})=\prod_{j=2}^{m}\frac{1}{\tilde{b}(z_{1}^{\prime}%
-z_{j})}\,\tilde{h}(\underline{\theta},\underline{z}^{\prime})p(\underline
{\theta},\underline{z}^{\prime})\,.
\]
For the $\tilde{D}_{Q}$-term we rewrite the $z_{1}$-integral by replacing
$z_{1}\rightarrow z_{1}-2\pi i$ (such that $z_{1}^{\prime}\rightarrow z_{1}$
and $\mathcal{C}_{{\underline{\theta}}}\rightarrow\mathcal{C}_{{\underline
{\theta}}}+2\pi i$) and obtain for the sum of the unwanted term from
$\tilde{A}_{Q}$ and $\tilde{D}_{Q}$ (using $\tilde{c}(z)/\tilde{b}%
(z)=-\tilde{c}(-z)/\tilde{b}(-z)$)
\begin{multline*}
\left[  K(\underline{\theta})\,\left(  \tilde{A}_{Q}(\underline{\theta
})+\tilde{D}_{Q}(\underline{\theta})\right)  \right]  _{\underline{\alpha}%
}^{uw}=\left(  \int_{\mathcal{C}_{{\underline{\theta}}^{\prime}}}%
d\underline{z}-\int_{\mathcal{C}_{{\underline{\theta}}}+2\pi i}dz_{1}%
\int_{\mathcal{C}_{{\underline{\theta}}}}d\underline{\check{z}}\,\right) \\
\times\frac{\tilde{c}(z_{1}-\theta_{1}^{\prime})}{\tilde{b}(z_{1}-\theta
_{1}^{\prime})}\tilde{h}(\underline{\theta},\underline{z})\,p^{\mathcal{O}%
}(\underline{\theta},\underline{z})L_{\underline{\beta}}(\underline{z}%
)\prod_{j=2}^{m}\frac{1}{\tilde{b}(z_{1}-z_{j})}\,\delta_{\alpha_{1}}%
^{\beta_{1}}{\check{\Phi}}_{\underline{\tilde{\alpha}}}^{\underline
{\tilde{\beta}}}(\underline{\tilde{\theta}},\underline{\tilde{z}})\,.
\end{multline*}
Taking into account the pole structure of $\frac{\tilde{c}(z_{1}-\theta
_{1}^{\prime})}{\tilde{b}(z_{1}-\theta_{1}^{\prime})}\tilde{h}(\underline
{\theta},\underline{z})$ we may replace for the $\tilde{A}_{Q}$ -term
\[
\int_{\mathcal{C}_{{\underline{\theta}}^{\prime}}}d\underline{z}\cdots=\left(
-2\pi i\operatorname*{Res}_{z_{1}=\theta_{1}^{\prime}}+\int_{\mathcal{C}%
_{{\underline{\theta}}}}dz_{1}\right)  \int_{\mathcal{C}_{{\underline{\theta}%
}^{\prime}}}d\underline{\tilde{z}}\cdots
\]
where again $\left(  \int_{\mathcal{C}_{{\underline{\theta}}^{\prime}}}%
-\int_{\mathcal{C}_{{\underline{\theta}}}}\right)  d\underline{\tilde{z}%
}\cdots=0$ and for the $\tilde{D}_{Q}$-term
\[
\int_{\mathcal{C}_{{\underline{\theta}}}+2\pi i}dz_{1}\int_{\mathcal{C}%
_{{\underline{\theta}}}}d\underline{\tilde{z}}\,\cdots=\left(  -2\pi
i\operatorname*{Res}_{z_{1}=\theta_{1}^{\prime}}+\int_{\mathcal{C}%
_{{\underline{\theta}}}}dz_{1}\right)  \int_{\mathcal{C}_{{\underline{\theta}%
}}}d\underline{\tilde{z}}\cdots
\]
which proves the cancellation of the unwanted terms.

The proof of (iii) is similar to that for the $Z(3)$ model in \cite{BFK}. We
use the short notations%
\begin{align*}
\underline{\theta}  &  =(\theta_{1},\dots,\theta_{n}),~\underline{\hat{\theta
}}=(\theta_{1},\theta_{2},\theta_{3}),~\underline{\check{\theta}}=(\theta
_{4},\dots,\theta_{n}),\\
\underline{\alpha}  &  =(\alpha_{1},\dots,\alpha_{n}),~\underline{\hat{\alpha
}}=(\alpha_{1},\alpha_{2},\alpha_{3}),~\underline{\check{\alpha}}=(\alpha
_{4},\dots,\alpha_{n}),\\
\underline{z}  &  =(z_{1},\dots,z_{m}),~\underline{\hat{z}}=(z_{1}%
,z_{2}),~\underline{\check{z}}=(z_{3},\dots,z_{m}),\\
\underline{\beta}  &  =(\beta_{1},\dots,\beta_{m}),~\underline{\hat{\beta}%
}=(\beta_{1},\beta_{2}),~\underline{\check{\beta}}=(\beta_{3},\dots,\beta_{m})
\end{align*}
and prove (\ref{3.14}) which may be depicted as%
\[
\operatorname*{Res}_{\theta_{23}=i\eta}\operatorname*{Res}_{\theta_{12}=i\eta
}~~~%
\begin{array}
[c]{c}%
\unitlength3.2mm\begin{picture}(7,4) \put(3.5,2){\oval(7,2)}
\put(3,2){\makebox(0,0){${\cal O}$}} \put(1,0){\line(0,1){1}}
\put(2,0){\line(0,1){1}} \put(3,0){\line(0,1){1}} \put(4,0){\line(0,1){1}}
\put(6,0){\line(0,1){1}} \put(4.4,.5){$\dots$} \end{picture}
\end{array}
=2i\sqrt{2}\Gamma\left(
\begin{array}
[c]{c}%
\unitlength3.2mm\begin{picture}(6,4) \put(.5,0){\oval(1,2)[t]}
\put(1.1,1){\oval(1.2,1)[t]} \put(1.7,0){\line(0,1){1}} \put(4,2){\oval(4,2)}
\put(4,2){\makebox(0,0){${\cal O}$}} \put(3,0){\line(0,1){1}}
\put(5,0){\line(0,1){1}} \put(3.4,.5){$\dots$} \end{picture}
\end{array}
~~-~~%
\begin{array}
[c]{c}%
\unitlength3.2mm\begin{picture}(6,5) \put(0,0){\oval(1,2)[t]}
\put(0,1){\line(0,1){2}} \put(3,3){\oval(6,4)[t]} \put(3,3){\oval(6,4)[br]}
\put(3,0){\oval(3.5,2)[tl]} \put(3,3){\oval(4,2)}
\put(3,3){\makebox(0,0){${\cal O}$}} \put(2,0){\line(0,1){2}}
\put(4,0){\line(0,1){2}} \put(2.4,1.5){$\dots$} \put(5.5,2.5){$\times$}
\end{picture}
\end{array}
\right)  .
\]
We will show that the K-function given by the integral representation
(\ref{2.16}) with (\ref{2.18}) satisfies%
\begin{align}
\operatorname*{Res}_{\theta_{23}=i\eta}\operatorname*{Res}_{\theta_{12}=i\eta
}K_{1\dots n}(\underline{\theta})  &  =c_{0}\prod_{i=4}^{n}\prod_{j=2}%
^{3}\tilde{\phi}(\theta_{ij})\,\varepsilon_{123}\,K_{4\dots n}(\underline
{\check{\theta}})\left(  \mathbf{1}-\sigma_{3}S_{3n}\dots S_{34}\right)
\label{3.54}\\
c_{0}  &  =2i\sqrt{2}\Gamma F^{-2}(i\eta)F^{-1}(2i\eta)\,.\nonumber
\end{align}
This is equivalent to (\ref{3.14}) because of the ansatz (\ref{2.10}) for
$F_{1\dots n}(\underline{\theta})$ and the relation of $F(z)$ and $\phi(z)$
given by (\ref{3.29}). The residues of $K_{1\dots n}(\underline{\theta})$
consists of three terms
\[
\operatorname*{Res}_{\theta_{12}=i\eta}\operatorname*{Res}_{\theta_{23}=i\eta
}K_{1\dots n}(\underline{\theta})=R_{1\dots n}^{(1)}+R_{1\dots n}%
^{(2)}+R_{1\dots n}^{(3)}\,.
\]
This is because each pair of the $z$ integration contours will be
\textquotedblleft pinched\textquotedblright\ at three points. Due to symmetry
it is sufficient to determine the contribution from the $z_{1}$- and the
$z_{2}$-integrations and multiply the result by $m(m-1)$. The pinching points
are

\begin{itemize}
\item[(1)] $z_{1}=\theta_{2},\ z_{2}=\theta_{3},$

\item[(2)] $z_{1}=\theta_{1},\ z_{2}=\theta_{2},$

\item[(3)] $z_{1}=\theta_{2}-i\eta,\,z_{2}=\theta_{3}-i\eta.\,$
\end{itemize}

The contribution of (1) is given by $z_{1}$-and $z_{2}$-integrations
$\oint_{\theta_{2}}dz_{1}\oint_{\theta_{3}}dz_{2}\cdots$ along small circles
around $z_{1}=\theta_{2}$ and $z_{2}=\theta_{3}$ (see figure~\ref{f5.1}). The
S-matrices $\tilde{S}(\theta_{2}-z_{1})$ and $\tilde{S}(\theta_{3}-z_{2})$
yield the permutation operator $\tilde{S}(0)=\mathbf{P}$%
\begin{multline*}
\left(  \Omega\,\tilde{C}^{\beta_{m}}(\underline{\theta},z_{m})\cdots\tilde
{C}^{\beta_{2}}(\underline{\theta},\theta_{3})\tilde{C}^{\beta_{1}}%
(\underline{\theta},\theta_{2})\right)  _{\underline{\alpha}}=%
\begin{array}
[c]{c}%
\unitlength4mm\begin{picture}(12,8) \put(12,7){\oval(17,2)[lb]}
\put(12,7){\oval(20,6)[lb]} \put(12,7){\oval(10,8)[lb]}
\put(6,1){\oval(2,4)[rt]} \put(6,7){\oval(10,8)[lb]}
\put(3,7){\oval(6,10)[lb]} \put(3,1){\oval(6,2)[rt]} \put(5,1){\line(0,1){6}}
\put(8,1){\line(0,1){6}} \put(11,1){\line(0,1){6}}
\put(12,7){\oval(12,10)[lb]} \put(-.5,7.4){$\beta_1$} \put(.5,7.4){$\beta_2$}
\put(1.5,7.4){$\beta_3$} \put(3.2,7.4){$\beta_m$}
\multiput(4.8,7.3)(1,0){4}{1} \put(10.8,7.3){1} \put(4.4,0){$\alpha_1$}
\put(5.6,0){$\alpha_2$} \put(6.7,0){$\alpha_3$} \put(8,0){$\alpha_4$}
\put(10.8,0){$\alpha_n$} \multiput(12.2,1.6)(0,1.1){3}{1} \put(12.2,5.8){1}
\put(9,5){$\dots$} \put(11.5,4.5){$\vdots$} \end{picture}
\end{array}
\\
=\tilde{S}_{\alpha_{1}^{\prime}\alpha_{3}}^{\beta_{2}1}(\theta_{13})\tilde
{S}_{\alpha_{1}\alpha_{2}}^{\beta_{1}\alpha_{1}^{\prime}}(\theta_{12}%
)\prod_{j=3}^{m}\tilde{b}(\theta_{1}-z_{j})\left(  \Omega\tilde{C}^{\beta_{m}%
}(\underline{\check{\theta}},z_{m})\cdots\tilde{C}^{\beta_{3}}(\underline
{\check{\theta}},z_{3})\right)  _{\underline{\check{\alpha}}}\,.
\end{multline*}
We have used the fact that because of the $SU(N)$ ice rule only the amplitude
$\tilde{b}$ contributes to the S-matrices $\tilde{S}(\theta_{1}-z_{j})$ and
only $\tilde{a}=1$ to the S-matrices $\tilde{S}(\theta_{2}-z_{j}),\tilde
{S}(\theta_{3}-z_{j}),\tilde{S}(\theta_{i}-z_{1}),\tilde{S}(\theta_{i}-z_{2})$
after having applied Yang-Baxter relations. Further we use that for
$z_{12}\rightarrow\theta_{23}\rightarrow i\eta$ by assumption (iii)$^{(1)}$
(c.f. (\ref{3.26}))
\begin{align*}
L_{\underline{\beta}}(\underline{z})  &  \approx c_{1}\prod_{i=3}^{m}%
\tilde{\phi}(z_{i2})\tilde{S}_{\beta_{1}\beta_{2}}^{32}(z_{12})L_{\underline
{\check{\beta}}}(\underline{\check{z}})\\
c_{1}  &  =\tilde{\phi}(z_{12})=\tilde{\phi}(\theta_{23})=\tilde{\phi}(i\eta)
\end{align*}
and that because of (\ref{3.12})
\[
\operatorname*{Res}_{\theta_{12}=i\eta}\operatorname*{Res}_{\theta_{23}=i\eta
}\tilde{\phi}(\theta_{12})\tilde{\phi}(\theta_{13})\tilde{S}_{\alpha_{1}%
\alpha_{2}\alpha_{3}}^{321}(\theta_{1},\theta_{2},\theta_{3})=\tilde{\phi
}(i\eta)\tilde{\phi}(2i\eta)2\left(  i\eta\right)  ^{2}\epsilon_{\alpha
_{1}\alpha_{2}\alpha_{3}}\,.
\]
We combine this with the scalar functions $\tilde{h}$ and $p$ and after having
performed the remaining $z_{j}$-integrations we obtain
\[
R_{\underline{\alpha}}^{(1)}=c_{0}\prod_{i=4}^{n}\prod_{j=2}^{3}\tilde{\phi
}(\theta_{ij})\,\epsilon_{\underline{\hat{\alpha}}}K_{\underline{\check
{\alpha}}}(\underline{\check{\theta}})
\]
because of $\tilde{b}(\theta_{1}-z)\tilde{\phi}(\theta_{1}-z)\tilde{\phi
}(\theta_{2}-z)\tilde{\phi}(z-\theta_{3})\tilde{\phi}(\theta_{3}-z)\tau
(\theta_{2}-z)\tau(\theta_{3}-z)=-1$, the relation (iii')$_{3}$ of
(\ref{3.30}) for the p-functions $p(\underline{\theta},\theta_{2},\theta
_{3},\underline{\check{z}})|_{\theta_{12}=\theta_{23}=i\eta}=(-1)^{m}%
\,p(\underline{\check{\theta}},\underline{\check{z}})$ and the relation
(\ref{norm3}) for the normalization constants $N_{n}c_{1}2(i\eta)^{2}%
\tilde{\phi}(i\eta)\tilde{\phi}(2i\eta)=N_{n-3}c_{0}$.

The remaining contribution to (\ref{3.54}) is due to $R_{1\dots n}%
^{(2)}+R_{1\dots n}^{(3)}$. It is convenient to shift the particle with
momentum $\theta_{3}$ to the right by applying S-matrices using (i). Then we
have to prove that
\begin{equation}
\left(  \left(  R_{1\dots n}^{(2)}+R_{1\dots n}^{(3)}\right)  S_{43}\dots
S_{n3}+c_{0}\prod_{i=4}^{n}\prod_{j=2}^{3}\tilde{\phi}(\theta_{ij}%
)\,\varepsilon_{123}\,K_{4\dots n}(\underline{\check{\theta}})\sigma
_{3}\,\right)  _{\alpha_{1}\alpha_{2}\underline{\check{\alpha}}\alpha_{3}%
}=0\,. \label{3.56}%
\end{equation}
Note that because of (i) (c.f. (\ref{2.12}))
\[
K_{1\dots n}(\underline{\theta})S_{43}\dots S_{n3}=\prod_{i=4}^{n}%
a(\theta_{i3})K_{1\dots4\dots n3}(\theta_{1},\dots,\theta_{4},\dots,\theta
_{n},\theta_{3})\,.
\]

Due to Lemma \ref{l3} in appendix \ref{sc} it is sufficient to prove equation
(\ref{3.56}) only for $\alpha_{3}=1$ since the left hand side of this equation
is a highest weight co-vector. This is because Bethe ansatz vectors, in
particular also off-shell Bethe ansatz vectors are of highest weight (see
\cite{BKZ}). Therefore we consider this equation for the components with
$\alpha_{3}=1$. The contribution of pinching (2) is given by the $z_{1}$-,
$z_{2}$-integrations along the small circles around $z_{1}=\theta_{1}$ and
$z_{2}=\theta_{2}$ (see again figure~\ref{f5.1}). Now $\tilde{S}(\theta
_{1}-z_{1})$ and $\tilde{S}(\theta_{2}-z_{2})$ yield permutation operators
$\mathbf{P}$ and the co-vector part of this contribution for $\alpha_{3}=1$
is
\begin{multline}
\left(  \Omega\,\tilde{C}^{\beta_{m}}(\underline{\theta},z_{m})\cdots\tilde
{C}^{\beta_{2}}(\underline{\theta},\theta_{2})\tilde{C}^{\beta_{1}}%
(\underline{\theta},\theta_{1})P_{3}(1)\right)  _{\underline{\alpha}}\\=%
\begin{array}
[c]{c}%
\unitlength4mm\begin{picture} (12,8) \put(11,7){\oval(19,4)[lb]}
\put(11,4){\oval(12,2)[lb]} \put (11,3){\oval(14,2)[lb]}
\put(4,3){\line(0,1){4}} \put(5,4){\line(0,1){3}} \put(4,7){\oval(7,8)[lb]}
\put(3,7){\oval(6,10)[lb]} \put(10,1){\line(0,1){6}} \put(6,1){\line(0,1){6}}
\put(3,1){\oval(2,2)[rt]} \put(4,1){\oval(2,4)[rt]} \put(-.7,7.4){$\beta_1$}
\put(.2,7.4){$\beta_2$} \put(1.2,7.4){$\beta_m$} \put(3.8,7.3){1}
\put(4.8,7.3){1} \put(5.8,7.3){1} \put(9.8,7.3){1} \put(3.5,0){$\alpha_1$}
\put(4.7,0){$\alpha_2$} \put(5.9,0){$\alpha_4$} \put(9,0){$\alpha_3=1$}
\put(11.2,1.6){1} \put(11.2,2.7){1} \put(11.2,4.8){1} \put(7,4.3){$\dots$}
\put(10.5,3.9){$\vdots$} \end{picture}
\end{array}
\label{3.57}\\
=\delta_{\alpha_{1}}^{\beta_{1}}\delta_{\alpha_{2}}^{\beta_{2}}\left(
\Omega\tilde{C}^{\beta_{m}}(\underline{\check{\theta}},z_{m})\cdots\tilde
{C}^{\beta_{3}}(\underline{\check{\theta}},z_{3})\right)  _{\underline
{\check{\alpha}}}\delta_{\alpha_{3}}^{1}%
\end{multline}
where $P_{3}(1)$ projects onto the components with $\alpha_{3}=1$. We have
used the fact that because of the $SU(N)$ ice rule the amplitude $a$ only
contributes to the S-matrices $S(\theta_{1}-z_{j}),S(\theta_{2}-z_{j}%
),S(\theta_{3}-z_{j}),S(\theta_{i}-z_{1})$ after having applied Yang-Baxter
relations. One derives the formula%
\[
\operatorname*{Res}_{\theta_{23}=i\eta}\operatorname*{Res}_{\theta_{12}=i\eta
}\tilde{\phi}(\theta_{31})\tilde{\phi}(\theta_{32})\tilde{S}_{\alpha_{1}%
\alpha_{2}}^{32}(\theta_{12})=-\tilde{\phi}(i\eta)\tilde{\phi}(2i\eta
)2(i\eta)^{2}\epsilon_{\alpha_{1}\alpha_{2}1}%
\]
using (\ref{3.12}), $\epsilon^{321}=1$, $\tilde{S}_{1\alpha_{1}\alpha_{2}%
}^{321}(\theta_{3}+2\pi i,\theta_{1},\theta_{2})=b(\theta_{31}+2\pi
i)b(\theta_{32}+2\pi i)\tilde{S}_{\alpha_{1}\alpha_{2}}^{32}(\theta_{12})$
(because of the $SU(3)$ ice rule) and $\tilde{\phi}(\theta)=-\tilde{b}%
(\theta+2\pi i)\tilde{\phi}(\theta+2\pi i)$ (see (\ref{1.28})). We combine
this and (\ref{3.57}) with the scalar functions $\tilde{h}$ and $p$ taking
into account the property (iii)$^{(1)}$ of (\ref{3.12}) and after having
performed the remaining $z_{j}$-integrations we obtain%
\begin{equation}
R_{1\dots n}^{(2)}S_{43}\dots S_{n3}P_{3}(1)\,=-c_{0}\prod_{i=4}^{n}%
\prod_{j=2}^{3}\tilde{\phi}(\theta_{ij})\,\varepsilon_{123}\,K_{4\dots
n}(\underline{\check{\theta}})\sigma_{3}P_{3}(1)\,.
\end{equation}
We have used the following equations $a(\theta_{i3})\tilde{\phi}(\theta
_{i1})=\tilde{\phi}(\theta_{i3})$ (because of (\ref{1.28})), the definition
(\ref{2.19}) of $\tau(z)$ which implies $\tilde{\phi}(z-\theta_{2})\tilde
{\phi}(\theta_{1}-z)\tilde{\phi}(\theta_{2}-z)\tilde{\phi}(\theta_{3}%
-z)\tau(\theta_{1}-z)\tau(\theta_{2}-z)=1$, the second relation (iii')$_{3}$
of (\ref{3.30}) for the p-functions $p(\underline{\theta},\theta_{1}%
,\theta_{2},\underline{\check{z}})|_{\theta_{12}=\theta_{23}=i\eta}%
=\sigma\,p(\underline{\check{\theta}},\underline{\check{z}})$ and the relation
(\ref{norm3}) for the normalization constants $N_{n}c_{1}\tilde{\phi}%
(i\eta)\tilde{\phi}(2i\eta)2(i\eta)^{2}=N_{n-3}c_{0}$.

The contribution of pinching (3) is given by the $z_{1}$-, $z_{2}%
$-integrations along the small circles around $z_{1}=\theta_{2}-i\eta$ and
$z_{2}=\theta_{3}-i\eta$, (see again figure~\ref{f5.1}). Now $\tilde{S}%
(\theta_{3}-z_{2})$ yields $\tilde{S}_{\alpha\beta}^{\delta\gamma}(\theta
_{3}-z_{2})\rightarrow\Gamma_{(\rho\sigma)}^{\delta\gamma}\Gamma_{\alpha\beta
}^{(\rho\sigma)}$ and the residue of the Bethe ansatz state
\[
i\operatorname*{Res}_{z_{2}=\theta_{3}-i\eta}\left(  \Omega\,\tilde
{C}(\underline{\theta},z_{m})\cdots\tilde{C}(\underline{\theta},z_{1}%
)P_{n}(1)\,\right)  _{\underline{\alpha}}=%
\begin{array}
[c]{c}%
~~\unitlength4mm\begin{picture}(12,8) \put(11,7){\oval(19,4)[lb]}
\put(9,7){\oval(2,8)[rb]} \put(9.7,3.2){\makebox(0,0){$\bullet$}}
\put(10.3,2.7){\makebox(0,0){$\bullet$}}
\bezier{22}(9.7,3.2)(10,2.95)(10.3,2.7)
\put(11,1){\oval(2,4)[lt]}
\put(9,7){\oval(17,8)[lb]}
\put(4,1){\line(0,1){6}} \put(5,1){\line(0,1){6}} \put(6,1){\line(0,1){6}}
\put(11,7){\oval(23,10)[lb]} \put(-.7,7.4){$\beta_1$} \put(.2,7.4){$\beta_2$}
\put(1.2,7.4){$\beta_m$} \put(3.8,7.3){1} \put(4.8,7.3){1} \put(5.8,7.3){1}
\put(9.8,7.3){1} \put(3.5,0){$\alpha_1$} \put(4.7,0){$\alpha_2$}
\put(5.9,0){$\alpha_3$} \put(9,0){$\alpha_n=1$} \put(11.2,1.6){1}
\put(11.2,2.7){1} \put(11.2,4.8){1} \put(7,4.3){$\dots$}
\put(10.5,3.9){$\vdots$} \end{picture}
\end{array}
\]
vanishes for $\alpha_{3}=1$ because $\Gamma_{\alpha\beta}^{(\rho\sigma)}$ is
antisymmetric with respect to $\alpha,\beta$. Therefore equation (\ref{3.56})
is proved for $\alpha_{3}=1$ and because of Lemma \ref{l3} also in general.
\end{proof}

\paragraph{The higher level Bethe ansatz}

\begin{lemma}
\label{l6}Let the constant $R$ and the contour $\mathcal{C}_{\underline{z}}$
be defined as in the context of (\ref{2.16}). Then the higher level function
\begin{align*}
L_{\underline{\beta}}(\underline{z})  &  =\frac{1}{k!}\int_{\mathcal{C}%
_{\underline{z}}}\frac{du_{1}}{R}\cdots\int_{\mathcal{C}_{\underline{z}}}%
\frac{du_{k}}{R}\,\tilde{h}(\underline{z},\underline{u})p^{(1)}(\underline
{z},\underline{u})\,\tilde{\Psi}_{\underline{\beta}}^{(1)}(\underline
{z},\underline{u})\\
\tilde{\Psi}_{\underline{\beta}}^{(1)}(\underline{z},\underline{u})  &
=\left(  \Omega^{(1)}\,\tilde{C}^{(1)}(\underline{z},u_{k})\cdots\tilde
{C}^{(1)}(\underline{z},u_{1})\right)  _{\underline{\beta}}\\
\tilde{h}(\underline{z},\underline{u})  &  =\prod_{i=1}^{m}\prod_{j=1}%
^{k}\tilde{\phi}(z_{i}-u_{j})\prod_{1\leq i<j\leq k}\tau(u_{i}-u_{j})\,.
\end{align*}
satisfies $(\mathrm{i})^{(1)}$ - $(\mathrm{iii})^{(1)}$ of the equations
(\ref{3.22}) - (\ref{3.26}) if%
\begin{equation}%
\begin{array}
[c]{lll}%
(\mathrm{i}^{\prime})_{3}^{(1)} &  & p_{mk}^{(1)}(\underline{z},\underline
{u})~\text{is symmetric under }z_{i}\leftrightarrow z_{j}\\[2mm]%
(\mathrm{ii}^{\prime})_{3}^{(1)} &  & \left\{
\begin{array}
[c]{l}%
p_{mk}^{(1)}(z_{1}+2\pi i,z_{2},\dots,\underline{u})=(-1)^{k}p_{mk}%
^{(1)}(z_{1},z_{2},\dots,\underline{u})\\
p_{mk}^{(1)}(\underline{z},u_{1}+2\pi i,u_{2},\dots)=(-1)^{m}p_{mk}%
^{(1)}(\underline{z},u_{1},u_{2},\dots)
\end{array}
\right. \\[1mm]%
(\mathrm{iii}^{\prime})_{3}^{(1)} &  & p_{mk}^{(1)}(\underline{z}%
,\underline{u})|_{z_{12}=i\eta,u_{1}=z_{2}}=(-1)^{k-1}p_{m-2k-1}%
^{(1)}(\underline{\check{z}},\underline{\check{u}})\,.
\end{array}
\label{3.60}%
\end{equation}

\end{lemma}

\begin{proof}
The proofs of the second level equations (i)$^{(1)}$ and (ii)$^{(1)}$ are
similar to the ones for the first level. For the proof of (iii)$^{(1)}$ we
observe that for $z_{12}\rightarrow i\eta$ there is a pinching at $u_{j}%
=z_{2}$ which means that in $\tilde{\Psi}_{\underline{\beta}}^{(1)}%
(\underline{z},\underline{u})$ we may replace $\tilde{S}(z_{2}-u_{1}%
)\rightarrow\mathbf{P}$ and we have to consider (for $z_{12}\rightarrow i\eta$
and $u_{1}=z_{2})$%
\begin{align*}
\left(  \Omega^{(1)}\,\tilde{C}^{(1)}(\underline{z},u_{k})\cdots\tilde
{C}^{(1)}(\underline{z},z_{2})\right)  _{\underline{\beta}}  &  =%
\begin{array}
[c]{c}%
\unitlength6mm\begin{picture}(9,6.5)(.4,-.8) \put(1,2){\line(1,0){8}}
\put(1,4){\line(1,0){8}} \put(2,0){\line(0,1){5}} \put(4,0){\line(0,1){5}}
\put(8,0){\line(0,1){5}} \put(1,0){\oval(4,2)[rt]} \put(9,5){\oval(12,8)[lb]}
\put(1.3,0){$z_1$} \put(3.2,0){$z_2$} \put(4.2,0){$z_3$} \put(8.2,0){$z_m$}
\put(1.8,-.8){$\beta_1$} \put(2.8,-.8){$\beta_2$} \put(3.8,-.8){$\beta_3$}
\put(7.8,-.8){$\beta_m$} \put(8.4,2.3){$u_2$} \put(8.4,4.3){$u_k$}
\put(4.8,1.2){$u_1=z_2$} \put(5.5,3){$\dots$} \put(1.4,2.8){$\vdots$}
\put(.4,.8){3} \put(.4,1.8){3} \put(.4,3.8){3} \put(9.3,.8){2}
\put(9.3,1.8){2} \put(9.3,3.8){2} \put(1.9,5.3){2} \put(2.9,5.3){2}
\put(3.9,5.3){2} \put(7.9,5.3){2} \end{picture}
\end{array}
\\
&  =\tilde{S}_{\beta_{1}\beta_{2}}^{32}(z_{12})\prod_{j=2}^{k}\tilde{b}%
(z_{1}-u_{j})\check{\Psi}_{\underline{\check{\beta}}}(\underline{\check{z}%
},\underline{\check{u}})
\end{align*}
with $\underline{\check{\beta}}=(\beta_{3},\dots,\beta_{m}),\ \underline
{\check{z}}=(z_{3},\dots,z_{m}),\ \underline{\check{u}}=(u_{2},\dots,u_{k})$.
Therefore because of $\tilde{b}(z_{1}-u)\tilde{\phi}(z_{1}-u)\tilde{\phi
}(z_{2}-u)\tau(z_{2}-u)=-1$ and $(\mathrm{iii}^{\prime})_{3}^{(1)}$ for
$z_{12}\rightarrow i\eta$%
\[
L_{\underline{\beta}}(\underline{z})\approx\tilde{\phi}(z_{12})\prod_{i=3}%
^{m}\tilde{\phi}(z_{i2})\tilde{S}_{\beta_{1}\beta_{2}}^{23}(z_{12}%
)L_{\underline{\check{\beta}}}(\underline{\check{z}})
\]
which proves the claim (iii)$^{(1)}$ of (\ref{3.26}).
\end{proof}

\section{$SU(N)$ form factors}

In this section we perform the form factor program for the general $SU(N)$
S-matrix. For this purpose, we extend the procedures of the previous section,
i.e., the nested Bethe ansatz method now with $N-1$ levels combined with the
off-shell Bethe ansatz. Applications of the results to the $SU(N)$ Gross-Neveu
model \cite{GN} will be investigated in a separate article \cite{BFK3}.

\subsection{S-matrix}

The $SU(N)$ S-matrix is given by (\ref{1.2}) and (\ref{1.4}). Again, the
eigenvalue $S_{-}(\theta)$ has a pole at $\theta=i\eta=\frac{2}{N}i\pi$ which
means that there exist bound states of $r$ fundamental particles $\alpha
_{1}+\dots+\alpha_{r}\rightarrow(\rho_{1}\dots\rho_{r})\,$(with $\rho
_{1}<\dots<\rho_{r}$) which transform as the anti-symmetric $SU(N)$ tensor
representation of rank $r,\ (0<r<N)$. The masses of the bound states satisfy
$m_{r}=m_{N-r}$ which suggests Swieca's \cite{KuS,KS,KKS} picture that the
antiparticle of a particle of rank $r$ is to be identified with the particle
of rank $N-r$ (see also \cite{BK04,BFK}).

Iterating $N-1$ times the general bound state formula one obtains for the
scattering of a bound state $(\beta_{1}\beta_{2}\dots\beta_{N-1})$ with
another particle $\delta$ (analogously to (\ref{3.2}) for the $SU(3)$ case)
\begin{multline*}
S_{(\beta_{1}\beta_{2}\dots\beta_{N-1})\delta}^{\delta^{\prime}(\gamma
_{1}\gamma_{2}\dots\gamma_{N-1})}(\theta)\Gamma_{\alpha_{1}\alpha_{2}%
\dots\alpha_{N-1}}^{(\beta_{1}\beta_{2}\dots\beta_{N-1})}\\
=\Gamma_{\alpha_{1}^{\prime}\alpha_{2}^{\prime}\dots\alpha_{N-1}^{\prime}%
}^{(\gamma_{1}\gamma_{2}\dots\gamma_{N-1})}S_{\alpha_{1}\delta_{1}}%
^{\delta^{\prime}\alpha_{1}^{\prime}}(\theta+i\pi-i\eta)\dots S_{\alpha
_{N-1}\delta}^{\delta_{N-2}\alpha_{N-1}^{\prime}}(\theta-i\pi+i\eta)
\end{multline*}
with the total bound state fusion intertwiner%
\[
\Gamma_{\alpha_{1}\alpha_{2}\dots\alpha_{N-1}}^{(\beta_{1}\beta_{2}\dots
\beta_{N-1})}=\Gamma_{(\beta_{1}\beta_{2}\dots\beta_{N-2})\alpha_{N-1}%
}^{(\beta_{1}\beta_{2}\dots\beta_{N-1})}\dots\Gamma_{(\beta_{1}\beta
_{2})\alpha_{3}}^{(\beta_{1}\beta_{2}\beta_{3})}\Gamma_{\alpha_{1}\alpha_{2}%
}^{(\beta_{1}\beta_{2})}\,.
\]
Taking special cases for the external particles we obtain
\begin{align*}
S_{(1,2\dots,N-1)N}^{N(1,2\dots,N-1)}(\theta)  &  =b(\theta+i\pi-i\eta)\dots
b(\theta-i\pi+i\eta)=(-1)^{N-1}a(\pi i-\theta)\\
S_{(1,2\dots,N-1)N-1}^{N-1(1,2\dots,N-1)}(\theta)  &  =b(\theta+i\pi
-i\eta)\dots a(\theta-i\pi+i\eta)=(-1)^{N-1}b(\pi i-\theta)\\
S_{(2,3\dots,N)1}^{N(1,2\dots,N-1)}(\theta)  &  =-b(\theta+i\pi-i\eta)\dots
c(\theta-i\pi+i\eta)=(-1)^{N-1}c(\pi i-\theta)\,.
\end{align*}
which may be written as%
\begin{multline*}
S_{(\alpha_{1}\dots\alpha_{N-1})\alpha_{N}}^{\beta_{N}(\beta_{1}\dots
\beta_{N-1})}(\theta)\\
=(-1)^{N-1}\left(  \delta_{(\alpha_{1}\dots\alpha_{N-1})}^{(\beta_{1}%
\dots\beta_{N-1})}\delta_{\alpha_{N}}^{\beta_{N}}\,b(\pi i-\theta
)+\epsilon^{\beta_{N}\beta_{1}\dots\beta_{N-1}^{\prime}}\epsilon_{\alpha
_{1}\dots\alpha_{N-1}\alpha_{N}}\,c(\pi i-\theta)\right)
\end{multline*}
with the total anti-symmetric tensors $\epsilon_{\alpha_{1}\dots\alpha_{N}}$
and $\epsilon^{\alpha_{1}\dots\alpha_{N}}$ ($\epsilon_{1\dots N}%
=\epsilon^{N\dots1}=1$). These results may be interpreted as an unusual
crossing relation%
\begin{gather}
\epsilon_{\beta_{1}\beta_{2}\dots\beta_{N-1}\beta}\,S_{\alpha_{1}\delta_{1}%
}^{\delta^{\prime}\beta_{1}}(\theta_{1})\dots S_{\alpha_{N-1}\delta}%
^{\delta_{N-2}\beta_{N-1}}(\theta_{N-1})=(-1)^{N-1}\epsilon_{\alpha_{1}%
\alpha_{2}\dots\alpha_{N-1}\gamma}\,S_{\delta\beta}^{\gamma\delta^{\prime}%
}(i\pi-\theta)\label{4.4}\\%
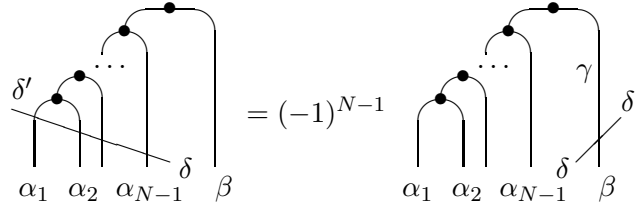
\begin{array}
[c]{c}%
\unitlength3mm\begin{picture}(28,8.5) \put(2,1){\oval(2,6)[t]}
\put(2,4){\makebox(0,0){$\bullet$}} \put(3,4){\oval(2,2)[t]}
\put(3,5){\makebox(0,0){$\bullet$}} \put(4,1){\line(0,1){3}}
\put(3.6,5.4){$\dots$} \put(5,6){\oval(2,2)[t]}
\put(5,7){\makebox(0,0){$\bullet$}} \put(6,1){\line(0,1){5}}
\put(7,7){\oval(4,2)[t]} \put(7,8){\makebox(0,0){$\bullet$}}
\put(9,1){\line(0,1){6}} \put(7,1.2){\line(-3,1){7}} \put(.3,-.3){$\alpha_1$}
\put(2.5,-.3){$\alpha_2$} \put(4.6,-.3){$\alpha_{N-1}$} \put(0,4){$\delta'$}
\put(7.4,.5){$\delta$} \put(9,-.3){$\beta$} \put(10.4,3){$=(-1)^{N-1}$}
\put(19,1){\oval(2,6)[t]} \put(19,4){\makebox(0,0){$\bullet$}}
\put(20,4){\oval(2,2)[t]} \put(20,5){\makebox(0,0){$\bullet$}}
\put(21,1){\line(0,1){3}} \put(20.6,5.4){$\dots$} \put(22,6){\oval(2,2)[t]}
\put(22,7){\makebox(0,0){$\bullet$}} \put(23,1){\line(0,1){5}}
\put(24,7){\oval(4,2)[t]} \put(24,8){\makebox(0,0){$\bullet$}}
\put(26,1){\line(0,1){6}} \put(25,1.2){\line(1,1){2}}
\put(17.3,-.3){$\alpha_1$} \put(19.5,-.3){$\alpha_2$}
\put(21.6,-.3){$\alpha_{N-1}$} \put(27,3.5){$\delta'$} \put(24.1,.5){$\delta$}
\put(26,-.3){$\beta$} \put(25,5){$\gamma$} \end{picture}
\end{array}
\nonumber
\end{gather}
with $\theta_{j}=\theta+i\pi-ji\eta$ if we write the charge conjugation
matrices as
\begin{align*}
\mathbf{C}_{(\alpha_{1}\dots\alpha_{N-1})\alpha_{N}}  & =\mathbf{C}%
_{\alpha_{1(}\alpha_{2}\dots\alpha_{N})}=\epsilon_{\alpha_{1}\dots\alpha_{N}%
}\,,\\
\mathbf{C}^{(\alpha_{1}\dots\alpha_{N-1})\alpha_{N}}  & =\mathbf{C}%
^{\alpha_{1(}\alpha_{2}\dots\alpha_{N})}=\epsilon^{\alpha_{1}\dots\alpha_{N}%
}\,.
\end{align*}
Therefore we have the relations (c.f. (\ref{2.6}) and (\ref{3.10}))%
\begin{gather}
\mathbf{C}_{\alpha(\alpha_{1}\dots\alpha_{N-1})}\mathbf{C}^{(\alpha_{1}%
\dots\alpha_{N-1})\beta}=\delta_{\alpha}^{\beta}\,,~\mathbf{C}_{\alpha
(\alpha_{1}\dots\alpha_{N-1})}A_{\beta}^{\alpha}\mathbf{C}^{\beta(\alpha
_{1}\dots\alpha_{N-1})}=(-1)^{N-1}\operatorname*{tr}A\nonumber\\
\text{and }\mathbf{C}_{(\beta_{1}\dots\beta_{N-1})\gamma}\Gamma_{\alpha
_{1}\dots\alpha_{N-1}}^{(\beta_{1}\dots\beta_{N-1})}=\epsilon_{\alpha_{1}%
\dots\alpha_{N-1}\gamma}\Gamma\label{4.4a}%
\end{gather}
where the constant $\Gamma$ is%
\[
\Gamma=i\sqrt{\frac{1}{2\pi}N\Gamma^{N}\left(  1-\frac{1}{N}\right)  }\,.
\]
These results are consistent with the picture (c.f. \cite{KuS}) that the bound
state of $N-1$ particles of rank 1 is to be identified with the anti-particle
of rank 1. As already remarked, the physical aspects of these facts will be
discussed elsewhere \cite{BFK3}.

For later convenience we consider as a generalization of (\ref{3.12}) the
total $N$-particle S-matrix (consisting of $N(N-1)/2$ factors) in terms of
$\tilde{S}$%
\begin{equation}
\tilde{S}_{12\dots N}=\left(  \tilde{S}_{12}\tilde{S}_{13}\dots\tilde{S}%
_{1N}\right)  \left(  \tilde{S}_{23}\dots\tilde{S}_{2N}\right)  \dots\tilde
{S}_{N-1N} \label{4.6a}%
\end{equation}
in the limit $\theta_{jj+1}\rightarrow i\eta~(j=1,\dots N-1)$ which behaves
as
\begin{equation}
\tilde{S}_{\alpha_{1}\alpha_{2}\dots\alpha_{N}}^{\beta_{N}\dots\beta_{2}%
\beta_{1}}\approx\left(  N-1\right)  !\frac{i\eta}{\theta_{12}-i\eta}%
\dots\frac{i\eta}{\theta_{N-1N}-i\eta}\,\epsilon^{\beta_{N}\dots\beta_{2}%
\beta_{1}}\epsilon_{\alpha_{1}\alpha_{2}\dots\alpha_{N}}\,. \label{4.6}%
\end{equation}
The algebraic structure of this relation follows because one can use the
Yang-Baxter equations to shift in (\ref{4.6a}) any factor $\tilde{S}%
_{ii+1}\sim\left(  1-\mathbf{P}\right)  _{ii+1}$ to the right or the left.
Therefore the expression is totally anti-symmetric with respect to the
$\alpha_{i}$ and the $\beta_{i}$. The factor follows from (\ref{1.5}).

\subsection{The general form factor formula}

In order to obtain a recursion relation where only form factors for the
fundamental particles of type $\alpha$ (which transform as the $SU(N)$ vector
representation) are involved, we have to apply iteratively the bound state
relation (iv) to get the $(N-1)$-bound state which is to be identified with
the anti-particle%
\begin{multline*}
\operatorname*{Res}_{\theta_{12}=i\eta}\dots\operatorname*{Res}_{\theta
_{N-2N-1}=i\eta}F_{1\dots n}^{\mathcal{O}}(\underline{\theta})\\
=F_{(1\dots N-1)N\dots n}^{\mathcal{O}}(\theta_{(1\dots N-1)},\theta_{N}%
,\dots,\theta_{n})\sqrt{2}^{N-2}\Gamma_{1\dots N-1}^{(1\dots N-1)}%
\end{multline*}
and finally the annihilation residue equation (iii)%
\begin{multline*}
\operatorname*{Res}_{\theta_{(1\dots N-1)N}=i\pi}F_{(1\dots N-1)N\dots
n}^{\mathcal{O}}(\theta_{(1\dots N-1)},\theta_{N},\dots,\theta_{n})\\
=2i\mathbf{C}_{(1\dots N-1)N}F_{N+1\dots n}^{\mathcal{O}}(\theta_{N+1}%
,\dots,\theta_{n})\left(  \mathbf{1}-\sigma_{N}^{\mathcal{O}}S_{Nn}\dots
S_{NN+1}\right)  .
\end{multline*}
Similar as for $N=3$ we obtain with (\ref{4.4a})%
\begin{multline}
\operatorname*{Res}_{\theta_{N-1N}=i\eta}\dots\operatorname*{Res}_{\theta
_{12}=i\eta}F_{123\dots n}^{\mathcal{O}}(\theta_{1},\dots,\theta
_{n})\label{4.7}\\
=2i\varepsilon_{1\dots N}\sqrt{2}^{N-2}\Gamma F_{N+1\dots n}^{\mathcal{O}%
}(\theta_{N+1},\dots,\theta_{n})\left(  \mathbf{1}-\sigma_{N}^{\mathcal{O}%
}S_{Nn}\dots S_{NN+1}\right)  .
\end{multline}

\paragraph{Ansatz for form factors:}

\label{an}

We propose the $n$-particle form factors of an operator $\mathcal{O}(x)$ as
given by the same formula (\ref{2.10}) as for $SU(2)$ and $SU(3)$ in terms of
the K-function and the minimal form factor function $F\left(  \theta\right)  $
given by (\ref{1.22}) which belongs to the highest weight $w=(2,0,\dots,0)$.
The form factor equations (i) and (ii) for K-function write again as
(\ref{2.12}) and (\ref{2.14}). Consistency of (\ref{4.7}), (\ref{2.14}) and
the crossing relation (\ref{4.4}) means that the statistics factors are of the
form $\sigma_{\alpha}^{\mathcal{O}}=\sigma^{\mathcal{O}}(r_{\alpha})$ if the
particle of type $\alpha$ belongs to an $SU(N)$ representation of rank
$r_{\alpha}=1,\dots,N-1$ and%
\begin{equation}
\sigma^{\mathcal{O}}(r)=e^{i\pi(1-1/N)rQ_{\mathcal{O}}}~~~\text{for
~}Q_{\mathcal{O}}=n\operatorname{mod}N \label{4.7b}%
\end{equation}
as an extension of (\ref{2.15a}) and (\ref{3.15}). For the K-function
$K_{\underline{\alpha}}^{\mathcal{O}}(\underline{\theta})$ we propose again
the ansatz in form of the integral representation (\ref{2.16}) with
(\ref{2.18}) and (\ref{2.19}). The Bethe ansatz co-vector $\tilde{\Psi
}_{\underline{\alpha}}(\underline{\theta},\underline{z})$ is again of the form
(\ref{3.16}) and for the function $L_{\underline{\beta}}(\underline{z})$ one
makes again an analogous ansatz as for the K-function (\ref{2.16}) where the
indices run over sets with one element less. For the $SU(N)$ case we have to
iterate this $N-2$ times
\begin{align}
\tilde{\Psi}_{\underline{\alpha}}^{(l-1)}(\underline{\theta},\underline{z})
&  =L_{\underline{\beta}}^{(l)}(\underline{z}){\tilde{\Phi}}_{\underline
{\alpha}}^{(l-1)\underline{\beta}}(\underline{\theta},\underline
{z})\,\nonumber\\
L_{\underline{\beta}}^{(l)}(\underline{z})  &  =\frac{1}{k!}\int
_{\mathcal{C}_{\underline{z}}}\frac{du_{1}}{R}\cdots\int_{\mathcal{C}%
_{\underline{z}}}\frac{du_{k}}{R}\,\tilde{h}(\underline{z},\underline
{u})p^{(l)}(\underline{z},\underline{u})\,\tilde{\Psi}_{\underline{\beta}%
}^{(l)}(\underline{z},\underline{u})\label{3.7d}\\
\tilde{h}(\underline{z},\underline{u})  &  =\prod_{i=1}^{m}\prod_{j=1}%
^{k}\tilde{\phi}(\theta_{i}-z_{j})\prod_{1\leq i<j\leq k}\tau(z_{i}%
-z_{j})\nonumber
\end{align}
for $l=1,\dots,N-2,~\alpha_{i}=l,\dots,N,~\beta_{i}=l+1,\dots,N,~(\tilde{\Psi
}^{(0)}=\tilde{\Psi})$. By this procedure one obtains the nested Bethe
ansatz\footnote{In \cite{BKZ2} the nested off-shell Bethe ansatz was
formulated in terms of \textquotedblleft Jackson-type integrals" instead of
contour integrals.}.

As for the $Z(N)$ and $A(N-1)$ models \cite{BK04,BFK} the `Jost-function'
$\phi\left(  \theta\right)  =\tilde{\phi}(\theta)/a\left(  \theta\right)
=\tilde{\phi}(-\theta)$ is a solution of the equation%
\begin{equation}
\prod_{k=0}^{N-2}\phi\left(  \theta+ki\eta\right)  \prod_{k=0}^{N-1}F\left(
\theta+ki\eta\right)  =1 \label{4.8}%
\end{equation}
which is typical for models where the bound state of $N-1$ particles is the
anti-particle. The solution is%
\[
\phi(z)=\Gamma\left(  \frac{z}{2\pi i}\right)  \Gamma\left(  1-\frac{1}%
{N}-\frac{z}{2\pi i}\right)
\]
and again we define $\tau(z)=1/(\phi(z)\phi(-z))$.

\paragraph{The higher level Bethe ansatz:}

In order that the form factor $F^{\mathcal{O}}\left(  \underline{\theta
}\right)  $ given by (\ref{2.10}) and (\ref{3.7d}) satisfies the form factor
equations (i) - (iii) the higher level L-functions $L_{\beta_{1},\beta
_{2},\dots,\beta_{m}}^{(l)}(z_{1},z_{2},\dots,z_{m}),$ $(l<\beta_{i}\leq
N,\ m=n_{l})$ have to satisfy:

\begin{description}
\item[(i)$^{(l)}$] Watson's equations%
\[
L_{\dots ij\dots}^{(l)}(\dots,z_{i},z_{j},\dots)=L_{\dots ji\dots}^{(l)}%
(\dots,z_{j},z_{i},\dots)\tilde{S}_{ij}(z_{ij})\,,
\]

\item[(ii)$^{(l)}$] the crossing relation%
\[
L_{\beta_{1},\beta_{2},\dots,\beta_{m}}^{(l)}(z_{1},z_{2},\dots,z_{m}%
)=L_{\beta_{2},\dots,\beta_{m},\beta_{1}}^{(l)}(z_{2},\dots,z_{m},z_{1}-2\pi
i)\,
\]
and

\item[(iii)$^{(l)}$] the function $L_{\underline{\beta}}^{(l)}(\underline{z})$
has to possess simple poles at $z_{12},\dots,z_{N-l-1,N-l}=i\eta$ and it has
to factorize in the neighborhood of these poles as%
\begin{align}
L_{\underline{\beta}}^{(l)}(\underline{z})~  &  \approx~c_{l}\prod
_{i=N-l+1}^{m}\prod_{j=2}^{N-l}\tilde{\phi}(z_{ij})\,\tilde{S}_{\underline
{\hat{\beta}}}^{N\dots l+1}(\underline{\hat{z}})\,L_{\underline{\check{\beta}%
}}(\underline{\check{z}})\label{4.10}\\%
\begin{array}
[c]{c}%
\unitlength4mm\begin{picture}(4,4) \put(2,3){\oval(4,2)}
\put(2,3){\makebox(0,0){$L^{(l)}$}} \put(1,1){\line(0,1){1}}
\put(3,1){\line(0,1){1}} \put(1.6,.2){$\underline{\beta}$}
\put(1.5,1.3){$\dots$} \end{picture}
\end{array}
~  &  \approx~c_{l}\prod_{i=N-l+1}^{mn_{l}}\prod_{j=2}^{N-l}\tilde{\phi
}(z_{ij})~%
\begin{array}
[c]{c}%
\unitlength4mm\begin{picture}(2,5) \put(1,2.5){\oval(2,2)}
\put(1,2.5){\makebox(0,0){$\tilde S$}} \put(.5,1){\line(0,1){.6}}
\put(1.5,1){\line(0,1){.6}} \put(.5,4){\line(0,-1){.6}}
\put(1.5,4){\line(0,-1){.6}} \put(.7,0){$\underline{\hat{\beta}}$}
\put(.2,4.5){$_{N~l+1}$} \put(.7,1.3){$_{\dots}$} \put(.7,3.9){$_{\dots}$}
\end{picture}~~~\unitlength4mm\begin{picture}(4,4) \put(2,3){\oval(4,2)}
\put(2,3){\makebox(0,0){$L^{(l)}$}} \put(1,1){\line(0,1){1}}
\put(3,1){\line(0,1){1}} \put(1.7,0){$\underline{\check{\beta}}$}
\put(1.5,1.3){$\dots$} \end{picture}
\end{array}
\nonumber
\end{align}
where we have used the short notations $\underline{\hat{\beta}}=(\beta
_{1},\dots,\beta_{N-l}),$\\$\underline{\check{\beta}}=(\beta_{N-l+1},\dots
,\beta_{m})$ and $\underline{\hat{z}}=(z_{1},\dots,z_{N-l}),$~$\underline
{\check{z}}=(z_{N-l+1},\dots,z_{m}).$ The S-matrix $\tilde{S}_{\underline
{\hat{\beta}}}^{N\dots l+1}(\underline{\hat{z}})$ describes the total
scattering of the $N-l$ particles with rapidities $\underline{\hat{z}}$ for
the initial quantum numbers $\underline{\hat{\beta}}$ and the final
configuration of quantum numbers $(N,\dots,l+1)$.
\end{description}

\noindent The constants satisfy the recursion relation%
\begin{equation}
c_{l}=c_{l+1}\prod_{j=1}^{N-l-1}\tilde{\phi}(ji\eta) \label{4.12}%
\end{equation}
with the solution
\begin{equation}
c_{1}=\tilde{\phi}^{N-2}(i\eta)\tilde{\phi}^{N-3}(2i\eta)\cdots\tilde{\phi
}((N-2)i\eta) \label{4.13}%
\end{equation}
if $c_{N-1}=1$ (see appendix \ref{se}).

\begin{lemma}
\label{l2}The higher level function $L_{\underline{\beta}}^{(l)}(\underline
{z})$ satisfies $(\mathrm{i})^{(l)}$ - $(\mathrm{iii})^{(l)}$, if the higher
level p-function in the integral representation (\ref{3.7d}) satisfies%
\begin{equation}%
\begin{array}
[c]{ll}%
(\mathrm{i}^{\prime})^{(l)} & p^{(l)}(\underline{z},\underline{u})~\text{is
symmetric under }z_{i}\leftrightarrow z_{j}\\[2mm]%
(\mathrm{ii}^{\prime})^{(l)} & \left\{
\begin{array}
[c]{l}%
p_{mk}^{(l)}(z_{1}+2\pi i,z_{2},\dots,\underline{u})=(-1)^{k}p_{mk}%
^{(l)}(z_{1},z_{2},\dots,\underline{u})\\
p_{mk}^{(l)}(\underline{z},u_{1}+2\pi i,u_{2},\dots)=(-1)^{m}p_{mk}%
^{(l)}(\underline{z},u_{1},u_{2},\dots)
\end{array}
\right.  \\[1mm]%
(\mathrm{iii}^{\prime})^{(l)} & p_{mk}^{(l)}(\underline{z},z_{2},\dots
z_{N-l},\underline{\check{u}})|_{z_{12}=\dots=z_{z_{N-l-1,N-l}}=i\eta
}=(-1)^{k-N-l+1}\,p_{\check{m}\check{k}}^{(l)}(\underline{\check{z}%
},\underline{\check{u}})
\end{array}
\label{4.9}%
\end{equation}
with $\check{m}=M-N+l,~\check{k}=M-N+l-1$.
\end{lemma}

\noindent This lemma is proved in appendix \ref{se}.

\paragraph{The p-function:}

In order that the form factor $F^{\mathcal{O}}\left(  \theta\right)  $ given
by (\ref{2.10}) and (\ref{2.16}) - (\ref{2.19}) satisfies the form factor
equations (i) - (iii) the p-function $p^{\mathcal{O}}(\underline{\theta
},\underline{z})$ in (\ref{2.16}) which depends on the explicit nature of the
local operator $\mathcal{O}$ is assumed to satisfy%
\begin{equation}%
\begin{array}
[c]{ll}%
(\mathrm{i}^{\prime})  & p_{nm}(\underline{\theta},\underline{z})~\text{is
symmetric under }\theta_{i}\leftrightarrow\theta_{j}\\[2mm]%
(\mathrm{ii}^{\prime})  & \left\{
\begin{array}
[c]{l}%
\sigma p_{nm}(\theta_{1}+2\pi i,\theta_{2},\dots,\underline{z})=(-1)^{m+N-1}%
p_{nm}(\theta_{1},\theta_{2},\dots,\underline{z})\\
p_{nm}(\underline{\theta},z_{1}+2\pi i,z_{2},\dots)=(-1)^{n}p_{nm}%
(\underline{\theta},z_{1},z_{2},\dots)
\end{array}
\right. \\[1mm]%
(\mathrm{iii}^{\prime})  & p_{nm}(\underline{\theta},\theta_{2},\dots
,\theta_{N},\underline{\check{z}})|_{\theta_{12}=\dots=\theta_{N-1N}=i\eta
}=(-1)^{m-N+1}\,p_{n-Nm-N+1}(\underline{\check{\theta}},\underline{\check{z}%
})\\
& p_{nm}(\underline{\theta},\theta_{1},\dots,\theta_{N-1},\underline
{\check{z}})|_{\theta_{12}=\dots=\theta_{N-1N}=i\eta}=\sigma\,p_{n-Nm-N+1}%
(\underline{\check{\theta}},\underline{\check{z}})
\end{array}
\label{4.25}%
\end{equation}
where $\underline{\theta}=(\theta_{1},\dots,\theta_{n}),\,\underline
{\check{\theta}}=(\theta_{N+1},\dots,\theta_{n}),\,\underline{z}=(z_{1}%
,\dots,z_{m})$ and\\
$\underline{\check{z}}=(z_{N},\dots,z_{m})$. In order to
simplify the notation, we have in these equations suppressed the dependence of
the p-function $p^{\mathcal{O}}$ and the statistics factor $\sigma
^{\mathcal{O}}$ on the operator $\mathcal{O}(x)$.

\begin{theorem}
\label{TN}The co-vector valued function $F_{\underline{\alpha}}(\underline
{\theta})$ given by the ansatz (\ref{2.10}) and the integral representation
(\ref{2.16}) satisfies the form factor equations $(\mathrm{i}),(\mathrm{ii})$
and $(\mathrm{iii})$ of (\ref{1.10}) -- (\ref{1.14}), in particular
(\ref{4.7}) if

\begin{enumerate}
\item $L_{\underline{\beta}}(\underline{z})$ satisfies the equation
$(\mathrm{i})^{(1)}$, $(\mathrm{ii})^{(1)}$, $(\mathrm{iii})^{(1)}$ and
(\ref{4.13}) of lemma \ref{l2},

\item $p^{\mathcal{O}}(\underline{\theta},\underline{z})$ satisfies the
equation $(\mathrm{i}^{\prime})$, $(\mathrm{ii}^{\prime})$ and $(\mathrm{iii}%
^{\prime})$ of (\ref{4.25}) and

\item the normalization constants satisfy%
\begin{equation}
(N-1)!(i\eta)^{N-1}\left(  \prod_{j=1}^{N-1}\left(  \tilde{\phi}%
(ji\eta)F(ji\eta)\right)  ^{N-j}\right)  N_{n}=2i\sqrt{2}^{N-2}\Gamma
N_{n-N}\,. \label{norm}%
\end{equation}

\end{enumerate}
\end{theorem}

The proof of this theorem can be found in appendix \ref{sd}.

\subsection{Examples}

To illustrate our general results we present some simple examples.

\paragraph{The energy momentum tensor:}

For the local operator $\mathcal{O}(x)=T^{\rho\sigma}(x)$ (where $\rho
,\sigma=\pm$ denote the light cone components) the p-function is, as for the
sine-Gordon model in \cite{BK}%
\[
p^{T^{\rho\sigma}}(\underline{\theta},\underline{z})=\sum\limits_{i=1}%
^{n}e^{\rho\theta_{i}}\sum\limits_{i=1}^{m}e^{\sigma z_{i}}\,.
\]
For the $n=N$ particle form factor there are $n_{l}=N-l$ integrations in the
$l$-th level of the off-shell Bethe ansatz. The $SU(N)$ weights are (see
\cite{BKZ2})
\[
w=\left(  n-n_{1},n_{1}-n_{2},\dots,n_{N-2}-n_{N-1},n_{N-1}\right)  =\left(
1,1,\dots,1,1\right)  \,.
\]
We calculate the form factor of the particle $\alpha$ and the bound state
$(\underline{\beta})=(\beta_{1},\dots,\beta_{N-1})$ of $N-1$ particles. In
each level all integrations up to one may be performed iteratively using the
bound state relation (iv) (similar as in the proof of theorem \ref{TN}). Then
all remaining integrations in the higher levels can be done by means of the
formula%
\[
\int_{-i\infty}^{i\infty}ds\Gamma(a+s)\Gamma(b+s)\Gamma(c-s)\Gamma(d-s)=2\pi
i\frac{\Gamma(a+c)\Gamma(a+d)\Gamma(b+c)\Gamma(b+d)}{\Gamma(a+b+c+d)}\,.
\]
The result for the form factor of the particle $\alpha$ and the bound state
$(\underline{\beta})$ writes as%
\begin{align}
F_{\alpha(\underline{\beta})}^{T^{\rho\sigma}}(\theta_{1},\theta_{2})  &
=K_{\alpha(\underline{\beta})}^{T^{\rho\sigma}}(\theta_{1},\theta
_{2})\,G(\theta_{12})\nonumber\\
K_{\alpha(\underline{\beta})}^{T^{\rho\sigma}}(\theta_{1},\theta_{2})  &
=N_{2}^{T^{\rho\sigma}}\left(  e^{\rho\theta_{1}}+e^{\rho\theta_{2}}\right)
\int_{\mathcal{C}_{\underline{\theta}}}\frac{dz}{R}\tilde{\phi}(\theta
_{1}-z)e^{\sigma z}L(\theta_{2}-z)\label{4.18}\\
&  ~~~\times\epsilon_{\delta\underline{\gamma}}\tilde{S}_{\alpha\epsilon
}^{\delta1}(\theta_{1}-z)\tilde{S}_{(\underline{\beta})1}^{\epsilon
(\underline{\gamma})}(\theta_{2}-z)\nonumber
\end{align}
where the summation is over $\underline{\gamma}$ and $\delta>1$ and
$G(\theta)$ is the minimal form factor function of two particles of rank $r=1$
and $r=N-1$. The functions $G(\theta)$ and $L(\theta)$ are given by%
\begin{gather*}
G(i\pi-\theta)F(\theta)\phi(\theta)=1\\
L(\theta)=\Gamma\left(  \frac{1}{2}+\frac{\theta}{2\pi i}\right)
\Gamma\left(  -\frac{1}{2}+\frac{1}{N}-\frac{\theta}{2\pi i}\right)  .
\end{gather*}
The remaining integral in(\ref{4.18}) may be performed (similar as in
\cite{BK}) with the result\footnote{In \cite{BKZ1,BKZ2} this result has been
obtained using Jackson type integrals.}%
\[
\langle\,0\,|\,T^{\rho\sigma}(0)\,|\,\theta_{1},\theta_{2}\,\rangle
_{\alpha(\underline{\beta})}^{in}=4M^{2}\epsilon_{\alpha\underline{\beta}%
}e^{\frac{1}{2}(\rho+\sigma)\left(  \theta_{1}+\theta_{2}+i\pi\right)  }%
\frac{\sinh\tfrac{1}{2}\left(  \theta_{12}-i\pi\right)  }{\theta_{12}-i\pi
}G(\theta_{12})\,.
\]
Similar as in \cite{BK} one can prove the eigenvalue equation
\[
\left(  \int dxT^{\pm0}(x)-\sum_{i=1}^{n}p_{i}^{\pm}\right)  |\,\theta
_{1},\dots,\theta_{n}\rangle_{\underline{\alpha}}^{in}=0
\]
for arbitrary states.

\paragraph{The fields $\psi_{\alpha}(x)$:}

Because the Bethe ansatz yields highest weight states we obtain the matrix
elements of the spinor field $\psi(x)=\psi_{1}(x)$. The p-function for the
local operator $\psi^{(\pm)}(x)$ is (see also \cite{BFKZ})%
\[
p^{\psi^{(\pm)}}(\underline{\theta},\underline{z})=\exp\pm\frac{1}{2}\left(
\sum\limits_{i=1}^{m}z_{i}-\left(  1-\frac{1}{N}\right)  \sum\limits_{i=1}%
^{n}\theta_{i}\right)  \,.
\]
For example the 1-particle form factor is
\[
\langle\,0\,|\,\psi^{(\pm)}(0)\,|\,\theta\,\rangle_{\alpha}=\delta_{\alpha
1}\,e^{\mp\frac{1}{2}\left(  1-\frac{1}{N}\right)  \theta}\,.
\]
The last two formulae are consistent with the proposal of Swieca et al.
\cite{KuS,KKS} that the statistics of the fundamental particles in the chiral
$SU(N)$ Gross-Neveu model should be $\sigma=\exp\left(  2\pi is\right)  $,
where $s=\frac{1}{2}\left(  1-\frac{1}{N}\right)  $ is the spin (see also
(\ref{4.7b})). For the $n=N+1$ particle form factor there are again
$n_{l}=N-l$ integrations in the $l$-th level of the off-shell Bethe ansatz and
the $SU(N)$ weights are $w=\left(  2,1,\dots,1,1\right)  $. Similar as above
one obtains the 2-particle and 1-bound state form factor%
\begin{align*}
F_{\alpha\beta(\underline{\gamma})}^{\psi^{(\pm)}}(\theta_{1},\theta
_{2},\theta_{3})  &  =K_{\alpha\beta(\underline{\gamma})}^{\psi^{(\pm)}%
}(\theta_{1},\theta_{2},\theta_{3})F(\theta_{12})G(\theta_{13})G(\theta
_{23})\\
K_{\alpha\beta(\underline{\gamma})}^{\psi^{(\pm)}}  &  =N^{\psi}e^{\mp\frac
{1}{2}\left(  1-\frac{1}{N}\right)  \sum\theta_{i}}\int_{\mathcal{C}%
_{\underline{\theta}}}\frac{dz}{R}\tilde{\phi}(\theta_{1}-z)\tilde{\phi
}(\theta_{2}-z)L(\theta_{3}-z)e^{\pm\frac{1}{2}z}\\
&  ~~~\times\epsilon_{\delta\underline{\gamma}}\tilde{S}_{\alpha_{1}\epsilon
}^{\delta1}(\theta_{1}-z)\tilde{S}_{\alpha_{2}\zeta}^{\epsilon1}(\theta
_{2}-z)\tilde{S}_{(\underline{\beta})1}^{\zeta(\underline{\gamma})}(\theta
_{3}-z)\,.
\end{align*}
We were not able to perform this integration. In \cite{BFK3} we will discuss
the $1/N$ expansion. There we will also discuss the physical interpretation of
the results for the chiral Gross-Neveu model.

\subsection*{Acknowledgments}

We thank A. Fring, R. Schrader, B. Schroer, and A. Zapletal for useful
discussions. H.B.
thanks A. Belavin, A. Nersesyan, and A. Tsvelik for interesting discussions.
M.K. thanks E. Seiler and P. Weisz for discussions and hospitality at the
Max-Planck Institut f\"{u}r Physik (M\"{u}nchen), where parts of this work have been
performed. H.B. thanks the condensed matter group of ICTP (Trieste) for
hospitality, where part of this work was done. H.B. was supported partially by
the grant Volkswagenstiftung within in the project "Nonperturbative aspects of
quantum field theory in various space-time dimensions". A.F. acknowledges
support from PRONEX under contract CNPq 66.2002/1998-99 and CNPq (Conselho
Nacional de Desenvolvimento Cient\'{\i}fico e Tecnol\'{o}gico). This work is
also supported by the EU network EUCLID, 'Integrable models and applications:
from strings to condensed matter', HPRN-CT-2002-00325.

\appendix

\section*{Appendix}

\section{A Lemma\label{sc}}

\begin{lemma}
\label{l3}Let $v^{1\dots n}\in V^{1\dots n}$ be a highest weight vector, i.e.
$E_{1\dots n}v^{1\dots n}=0$. Then $v^{1\dots n}$ vanishes if the components
$v^{\alpha_{1}\alpha_{2}\dots\alpha_{n}}$ for $\alpha_{1}=1$ vanish.
\end{lemma}

\begin{proof}
By definition $E$ acts on the basis vectors $e_{\underline{\alpha}}^{1\dots
n}=e_{\alpha_{1}}\otimes e_{\alpha_{2}}\otimes\cdots\otimes e_{\alpha_{n}}$
as
\begin{align*}
E_{1\dots n}e_{\underline{\alpha}}^{1\dots n}  &  =\sum_{i=1}^{n}e_{\alpha
_{1}}\otimes\cdots\otimes Ee_{\alpha_{i}}\otimes\cdots\otimes e_{\alpha_{n}}\\
Ee_{\alpha}  &  =e_{\alpha-1}~~\text{where }e_{0}=0\,.
\end{align*}
Therefore we may write%
\begin{align*}
0  &  =E_{1\dots n}v^{1\dots n}=\sum_{\underline{\alpha},\alpha_{1}%
>1}v^{\alpha_{1}\alpha_{2}\dots\alpha_{n}}E_{1\dots n}e_{\underline{\alpha}%
}^{1\dots n}\\
&  =\sum_{\underline{\alpha},\alpha_{1}>1}v^{\alpha_{1}\alpha_{2}\dots
\alpha_{n}}\sum_{i=1}^{n}e_{\alpha_{1}}\otimes\cdots\otimes e_{\alpha_{i}%
-1}\otimes\cdots\otimes e_{\alpha_{n}}\\
&  =\sum_{\underline{\alpha},\alpha_{1}=2}v^{2\alpha_{2}\dots\alpha_{n}}%
e_{1}\otimes\cdots\otimes e_{\alpha_{n}}+\sum_{\underline{\alpha},\alpha
_{1}>2}w^{\alpha_{1}\alpha_{2}\dots\alpha_{n}}e_{\alpha_{1}}\otimes
\cdots\otimes e_{\alpha_{n}}%
\end{align*}
for some vector $w^{1\dots n}$. Because the second term $w^{1\dots n}$ is
orthogonal to the first one all components $v^{2\alpha_{2}\dots\alpha_{n}}$
vanish. Iterating this procedure proves the claim.
\end{proof}

\section{Proof of Theorem\label{sd}
\protect
\ref{TN}}

The proof of the main theorem of this article for $SU(N)$ is a straightforward
extension of the one for $SU(3)$ above.

\begin{proof}
The form factor equations (i) and (ii) may be proved quite analogously as for
$SU(3)$. The proof of (iii) is similar to that for the $Z(N)$ model in
\cite{BFK}. We use the short hand notations%
\begin{align*}
\underline{\theta}  &  =(\theta_{1},\dots,\theta_{n}),~\underline{\hat{\theta
}}=(\theta_{1},\dots,\theta_{N}),~\underline{\check{\theta}}=(\theta
_{N+1},\dots,\theta_{n}),\\
\underline{\alpha}  &  =(\alpha_{1},\dots,\alpha_{n}),~\underline{\hat{\alpha
}}=(\alpha_{1},\dots,\alpha_{N}),~\underline{\check{\alpha}}=(\alpha
_{N+1},\dots,\alpha_{n}),\\
\underline{z}  &  =(z_{1},\dots,z_{m}),~\underline{\hat{z}}=(z_{1}%
,\dots,z_{N-1}),~\underline{\check{z}}=(z_{N},\dots,z_{m}),\\
\underline{\beta}  &  =(\beta_{1},\dots,\beta_{m}),~\underline{\hat{\beta}%
}=(\beta_{1},\dots,\beta_{N-1}),~\underline{\check{\beta}}=(\beta_{N}%
,\dots,\beta_{m})\,.
\end{align*}
That $F_{1\dots n}^{\mathcal{O}}(\underline{\theta})$ given by (\ref{2.10}),
(\ref{2.16}), (\ref{2.18}) and (\ref{3.7d}) satisfies (iii) in the form of
(\ref{4.7}) is equivalent to that $K_{1\dots n}(\underline{\theta})$ satisfies%
\begin{multline}
\operatorname*{Res}_{\theta_{N-1N}=i\eta}\dots\operatorname*{Res}_{\theta
_{12}=i\eta}K_{1\dots n}(\underline{\theta})=c_{0}\,\prod_{i=N+1}^{n}%
\prod_{j=2}^{N}\tilde{\phi}(\theta_{ij})\,\varepsilon_{1\dots N}K_{N+1\dots
n}(\underline{\check{\theta}})\label{4.7a}\\
\times\left(  \mathbf{1}-\sigma_{N}S_{Nn}\dots S_{NN+1}\right)
\end{multline}%
\[
c_{0}=2i\sqrt{2}^{N-2}\Gamma\prod_{j=1}^{N-1}F^{-(N-j)}(ji\eta)
\]
where the relation of $F(z)$ and $\phi(z)$ given by (\ref{4.8}) has been used.
The residues of $K_{1\dots n}(\underline{\theta})$ consists again of three
terms
\[
\operatorname*{Res}_{\theta_{N-1N}=i\eta}\dots\operatorname*{Res}_{\theta
_{12}=i\eta}K_{1\dots n}(\underline{\theta})=R_{1\dots n}^{(1)}+R_{1\dots
n}^{(2)}+R_{1\dots n}^{(3)}%
\]
because $N-1$ of the $z$ integration contours will be \textquotedblleft
pinched\textquotedblright\ at three points. Again due to symmetry it is
sufficient to determine the contribution from the $z_{1},\dots,$ $z_{N-1}%
$-integrations and multiply the result by $m\dots(m-N+2)$. The pinching points
are

\begin{itemize}
\item[(1)] $z_{1}=\theta_{2},\dots,\,z_{N-1}=\theta_{N}$,

\item[(2)] $z_{1}=\theta_{1},\dots,\,z_{N-1}=\theta_{N-1}$,

\item[(3)] $z_{1}=\theta_{2}-i\eta,\dots,\,z_{N-1}=\theta_{N}-i\eta$,
\end{itemize}

The contribution of (1) is given by $N-1$ integrations along small circles
around $z_{1}=\theta_{2},z_{2}=\theta_{3},\dots,z_{N-1}=\theta_{N}$ (see
figure~\ref{f5.1}). The S-matrices $\tilde{S}(\theta_{2}-z_{1}),\dots
,\,\tilde{S}(\theta_{N}-z_{N-1})$ yield the permutation operator $\tilde
{S}(0)=\mathbf{P}$. Therefore for $\theta_{12},\dots,\theta_{N-2N-1}%
,\theta_{N-1N}\rightarrow i\eta$%
\begin{multline}
\left(  \Omega\,\tilde{C}^{\beta_{m}}(\underline{\theta},z_{m})\cdots\tilde
{C}^{\beta_{N}}(\underline{\theta},z_{N})\tilde{C}^{\beta_{N-1}}%
(\underline{\theta},\theta_{N})\cdots\tilde{C}^{\beta_{1}}(\underline{\theta
},\theta_{2})\right)  _{\underline{\alpha}}\label{4.71}\\
=%
\begin{array}
[c]{c}%
\unitlength4mm\begin{picture}(15,9.5)(13.5,0) \put(20,8){\oval(12,10)[lb]}
\put(19.5,5.6){$\vdots$} \put(19.5,4.2){$:$} \put(22.4,2){$\dots$}
\put(20,1){\line(0,1){7}} \put(20,1){\oval(2,4)[rt]}
\put(21,1){\oval(2,6)[rt]} \put(23,1){\oval(2,8)[rt]}
\put(23,8){\oval(14,6)[lb]} \put(21,8){\oval(12,8)[lb]}
\put(25,1){\line(0,1){7}} \put(26,2){$\dots$} \put(28,1){\line(0,1){7}}
\put(29,8){\oval(16,10)[lb]} \put(29,8){\oval(14,8)[lb]}
\put(29,8){\oval(10,6)[lb]} \put(29,8){\oval(22,2)[lb]} \put(19,1){$\theta_1$}
\put(25.2,1){$\theta_N$} \put(28.2,1){$\theta_n$} \put(19.7,0){$\alpha_1$}
\put(24.7,0){$\alpha_N$} \put(27.8,0){$\alpha_n$} \put(13.6,8.8){$\beta_1$}
\put(15.3,8.8){$\beta_{N-1}$} \put(18,8.8){$\beta_m$} \put(15,2.2){$z_1$}
\put(16.8,5.4){$z_{N-1}$} \put(18.6,7.3){$z_m$} \put(29.2,2.6){1}
\put(29.2,3.6){1} \put(29.2,4.6){1} \put(29.2,6.6){1} \put(19.9,8.1){1}
\put(20.8,8.1){1} \put(21.8,8.1){1} \put(23.8,8.1){1} \put(24.8,8.1){1}
\put(27.8,8.1){1} \end{picture}
\end{array}
\\
=\prod_{j=N}^{m}\tilde{b}(\theta_{1}-z_{j})\left(  \tilde{S}_{1N}\dots
\tilde{S}_{12}\right)  _{\underline{\hat{\alpha}}}^{\underline{\hat{\beta}}%
,1}\left(  \Omega\tilde{C}^{\beta_{m}}(\underline{\check{\theta}},z_{m}%
)\cdots\tilde{C}^{\beta_{N}}(\underline{\check{\theta}},z_{N})\right)
_{\underline{\check{\alpha}}}\,.
\end{multline}
It has been used that due to the $SU(N)$ ice rule only the amplitude
$b(\cdot)$ contributes to the S-matrices $S(\theta_{1}-z_{j})$ and $a(\cdot)$
to the S-matrices $S(\theta_{2}-z_{j}),\dots,S(\theta_{N}-z_{j}),S(\theta
_{i}-z_{1}),\dots,S(\theta_{i}-z_{N-2})$ after having applied Yang-Baxter
relations. One observes that the product of S-matrices in (\ref{4.10})
together with the one in (\ref{4.71}) yields the total $N-1$ S-matrix
\[
\tilde{S}_{\underline{\hat{\beta}}}^{N\dots2}(\underline{\hat{z}})\left(
\tilde{S}_{1N}\dots\tilde{S}_{12}\right)  _{\underline{\hat{\alpha}}%
}^{\underline{\hat{\beta}},1}=\tilde{S}_{\underline{\hat{\alpha}}}^{N\dots
21}(\underline{\hat{\theta}})
\]
for which the residue formula (\ref{4.6}) applies%
\[
\operatorname*{Res}_{\theta_{12}=i\eta}\dots\operatorname*{Res}_{\theta
_{N-1N}=i\eta}\tilde{S}_{\alpha_{1}\dots\alpha_{N}}^{N\dots21}=(N-1)!(i\eta
)^{N-1}\epsilon^{N\dots21}\epsilon_{\alpha_{1}\dots\alpha_{N}}\,.
\]
We combine (\ref{4.71}) with the function $L_{\underline{\beta}}(\underline
{z})$ with the property (\ref{4.10}) for ($l=1$) and the scalar functions
$\tilde{h}$ and $p$ and after having performed the remaining $z_{j}%
$-integrations we obtain%
\[
R_{1\dots n}^{(1)}=c_{0}\prod_{i=N+1}^{n}\prod_{j=2}^{N}\tilde{\phi}%
(\theta_{ij})\,\varepsilon_{1\dots N}(\underline{\hat{\theta}})K_{N+1\dots
n}(\underline{\check{\theta}})\,.
\]
We have used the following equations $\tilde{b}(\theta_{1}-z)\tilde{\phi
}(\theta_{1}-z)=-\tilde{\phi}(z-\theta_{2})$ (for $\theta_{12}=i\eta$)
together with the definition (\ref{2.23}) of $\tau(z)$, the relation (iii') of
(\ref{4.25}) for the p-function $p(\underline{\theta},\theta_{2},\dots
,\theta_{N},\underline{\check{z}})=(-1)^{m-N+1}\,p(\underline{\check{\theta}%
},\underline{\check{z}})$ (for $\theta_{12}=\dots=\theta_{N-1N}=i\eta$), the
relation (\ref{norm}) for the normalization constants $N_{n}c_{1}%
(N-1)!(i\eta)^{N-1}\prod_{j=1}^{N-1}\tilde{\phi}(ji\eta)=N_{n-N}c_{0}$ and the
recursion relation (\ref{4.12}) for the constants $c_{l}$ with the solution
(\ref{D.5}).

The remaining contribution to (\ref{4.7a}) is due to $R_{2}$ and $R_{3}$%
\[
R_{1\dots n}^{(2)}+R_{1\dots n}^{(3)}=-c_{0}\prod_{i=N+1}^{n}\prod_{j=2}%
^{N}\tilde{\phi}(\theta_{ij})\,\varepsilon_{1\dots N}K_{N+1\dots n}%
(\underline{\check{\theta}})\sigma_{N}S_{Nn}\dots S_{NN+1}%
\]
It is convenient to shift the particle with momentum $\theta_{N}$ to the right
by applying S-matrices and write
\begin{equation}
\left(  R_{1\dots n}^{(2)}+R_{1\dots n}^{(3)}\right)  S_{N+1N}\dots
S_{nN}+c_{0}\prod_{i=N+1}^{n}\prod_{j=2}^{N}\tilde{\phi}(\theta_{ij}%
)\,\varepsilon_{1\dots N}K_{N+1\dots n}(\underline{\check{\theta}})\sigma
_{N}=0 \label{7.44}%
\end{equation}
where the components of this co-vector are now denoted by $v_{\alpha_{1}%
\dots\alpha_{N-1N+1}\dots\alpha_{n}\alpha_{N}}$. Note that because of (i)
(\ref{2.12})
\[
K_{1\dots n}(\underline{\theta})S_{N+1N}\dots S_{nN}=\prod_{i=N+1}^{n}%
a(\theta_{iN})K_{1\dots N+1\dots nN}(\theta_{1},\dots,\theta_{N+1}%
,\dots,\theta_{n},\theta_{N})\,.
\]

Again due to Lemma \ref{l3} in appendix \ref{sc} it is sufficient to prove
equation (\ref{7.44}) only for $\alpha_{N}=1$ since the left hand side of this
equation is a highest weight co-vector. This is because Bethe ansatz vectors,
in particular also off-shell Bethe ansatz vectors are of highest weight (see
\cite{BKZ}). Therefore we consider this equation for the components with
$\alpha_{N}=1$ only. The contribution of $R_{1\dots n}^{(2)}$ is given by the
$z_{1},\dots,z_{N-1}$-integrations along the small circles around
$z_{1}=\theta_{1},\dots,z_{N-1}=\theta_{N-1}$ (see again figure~\ref{f5.1}).
Now $\tilde{S}(\theta_{1}-z_{1})_{1},\dots,\tilde{S}(\theta_{N-1}-z_{N-1})$
yield permutation operators $\mathbf{P}$ and the co-vector part of this
contribution for $\alpha_{N}=1$ is
\begin{multline}
\left(  \Omega\,\tilde{C}^{\beta_{m}}(\underline{\theta},z_{m})\cdots\tilde
{C}^{\beta_{N-1}}(\underline{\theta},\theta_{N-1})\cdots\tilde{C}^{\beta_{1}%
}(\underline{\theta},\theta_{1})P_{N}(1)\right)  _{\alpha_{1}\dots
\alpha_{N-1N+1}\dots\alpha_{n}\alpha_{N}}\label{4.73}\\
=%
\begin{array}
[c]{c}%
\unitlength4mm\begin{picture}(17,10)(13,0) \put(19.5,5.6){$\vdots$}
\put(19.5,4.2){$:$} \put(21.4,2){$\dots$} \put(19,1){\oval(2,4)[rt]}
\put(20,1){\oval(2,6)[rt]} \put(22,1){\oval(2,8)[rt]}
\put(19,8){\oval(10,10)[lb]} \put(20,8){\oval(10,8)[lb]}
\put(24,1){\line(0,1){7}} \put(25.5,2){$\dots$} \put(28,1){\line(0,1){7}}
\put(22,8){\oval(12,6)[lb]} \put(29,8){\oval(18,10)[lb]}
\put(29,8){\oval(16,8)[lb]} \put(29,8){\oval(12,6)[lb]}
\put(29,8){\oval(22,2)[lb]} \put(19,1){$\theta_1$}
\put(24.2,1){$\theta_{N+1}$} \put(28.2,1){$\theta_N$} \put(19.7,0){$\alpha_1$}
\put(23.7,0){$\alpha_{N+1}$} \put(27.,0){$\alpha_N=1$}
\put(13.6,8.8){$\beta_1$} \put(15.3,8.8){$\beta_{N-1}$}
\put(18,8.8){$\beta_m$} \put(15,2.2){$z_1$} \put(16.8,5.4){$z_{N-1}$}
\put(18.6,7.3){$z_m$} \put(29.2,2.6){1} \put(29.2,3.6){1} \put(29.2,4.6){1}
\put(29.2,6.6){1} \put(19.9,8.1){1} \put(20.8,8.1){1} \put(22.8,8.1){1}
\put(23.8,8.1){1} \put(27.8,8.1){1} \end{picture}
\end{array}
\\
=\delta_{\alpha_{1}}^{\beta_{1}}\cdots\delta_{\alpha_{N-1}}^{\beta_{N-1}%
}\left(  \Omega\tilde{C}^{\beta_{m}}(\underline{\check{\theta}},z_{m}%
)\cdots\tilde{C}^{\beta_{N}}(\underline{\check{\theta}},z_{N})\right)
_{\underline{\check{\alpha}}}\delta_{\alpha_{N}}^{1}%
\end{multline}
where $P_{N}(1)$ projects onto the components with $\alpha_{N}=1$. We have
used the fact that because of the $SU(N)$ ice rule the amplitude $a(\cdot)$
only contributes to the S-matrices $S(\theta_{1}-z_{j}),S(\theta_{2}%
-z_{j}),S(\theta_{N}-z_{j}),S(\theta_{i}-z_{1})$ after having applied
Yang-Baxter relations. We use $\tilde{\phi}(\theta)=-\tilde{b}(\theta+2\pi
i)\tilde{\phi}(\theta+2\pi i)$ to replace for $i=1,\dots,N-1$ and $\beta\neq1$%

\[
\tilde{\phi}(\theta_{Ni})=\tilde{S}_{11}^{11}(\theta_{Ni})\tilde{\phi}%
(\theta_{Ni})=-\tilde{S}_{1\beta}^{\beta1}(\theta_{Ni}+2\pi i)\tilde{\phi
}(\theta_{Ni}+2\pi i)
\]
therefore using again (\ref{4.6}) we obtain
\begin{align*}
&  \operatorname*{Res}_{\theta_{N-1N}=i\eta}\cdots\operatorname*{Res}%
_{\theta_{12}=i\eta}\tilde{\phi}(\theta_{N1})\cdots\tilde{\phi}(\theta
_{NN-1})\tilde{S}_{\alpha_{1}\dots\alpha_{N-1}}^{N\dots2}(\theta_{1}%
,\cdots,\theta_{N-1})\\
&  =(-1)^{N-1}\tilde{\phi}(i\eta)\cdots\tilde{\phi}((N-1)i\eta)\\
&  \times\operatorname*{Res}_{\theta_{N-1N}=i\eta}\cdots\operatorname*{Res}%
_{\theta_{12}=i\eta}\tilde{S}_{1\alpha_{1}\cdots\alpha_{N-1}}^{N\cdots
21}(\theta_{N}+2\pi i,\theta_{1},\cdots,\theta_{N-1})\\
&  =-\tilde{\phi}(i\eta)\cdots\tilde{\phi}((N-1)i\eta)(N-1)!(i\eta
)^{N-1}\epsilon_{\alpha_{1}\cdots\alpha_{N-1}1}\,.
\end{align*}
We combine (\ref{4.73}) with the function $L_{\underline{\beta}}(\underline
{z})$ with the property (\ref{4.10}) (for $l=1$) and the scalar functions
$\tilde{h}$ and $p$ and after having performed the remaining $z_{j}%
$-integrations we obtain%
\[
R_{1\dots n}^{(2)}S_{N+1N}\dots S_{nN}P_{N}(1)=-c_{0}\prod_{i=N+1}^{n}%
\prod_{j=2}^{N}\tilde{\phi}(\theta_{ij})\,\varepsilon_{1\dots N}%
(\underline{\hat{\theta}})K_{N+1\dots n}(\underline{\check{\theta}})\sigma
_{N}P_{N}(1)\,.
\]
We have used the following equations: $a(\theta_{iN})\tilde{\phi}(\theta
_{i1})=\tilde{\phi}(\theta_{iN})$ (because of (\ref{1.28})), the definition
(\ref{2.23}) of $\tau(z)$, the relation (iii') of (\ref{4.25}) for the
p-function $p(\underline{\theta},\theta_{1},\dots,\theta_{N-1},\underline
{\check{z}})=\,\sigma p(\underline{\check{\theta}},\underline{\check{z}})$
(for $\theta_{12}=\dots=\theta_{N-1N}=i\eta$), the relation (\ref{norm}) for
the normalization constants $N_{n}c_{1}(N-1)!(i\eta)^{N-1}\prod_{j=1}%
^{N-1}\tilde{\phi}(ji\eta)=N_{n-N}c_{0}$ and the recursion relation
(\ref{4.12}) for the constants $c_{l}$.

The contribution of pinching (3) is given by the $z_{1},\dots,z_{N-1}%
$-integrations along the small circles around $z_{1}=\theta_{2}-i\eta
,\dots,\,z_{N-2}=\theta_{N-1}-i\eta,\,z_{N-1}=\theta_{N}-i\eta$, (see again
figure~\ref{f5.1}). As for $N=3$ the S-matrix $\tilde{S}_{\alpha\beta}%
^{\delta\gamma}(\theta_{N}-z_{N-1})$ yields $\Gamma_{(\rho\sigma)}%
^{\delta\gamma}\Gamma_{\alpha\beta}^{(\rho\sigma)}$ and the residue of the
Bethe ansatz state vanishes,
\[
\,\operatorname*{Res}_{z_{N-1}=\theta_{N}-i\eta}\left(  \Omega\,\tilde
{C}(\underline{\theta},z_{m})\cdots\tilde{C}(\underline{\theta},z_{1}%
)P_{N}(1)\,\right)  _{\underline{\alpha}}=0\,,
\]
because $\Gamma_{\alpha\beta}^{(\rho\sigma)}$ is antisymmetric with respect to
$\alpha,\beta$. Therefore equation (\ref{7.44}) is proved for $\alpha_{N}=1$
and because of Lemma \ref{l3} also in general.
\end{proof}

\section{The higher level Bethe ansatz\label{se}}

\paragraph{Proof of lemma
\protect
\ref{l6}:}

For the higher level functions $L_{\underline{\beta}}^{(l)}(\underline{z})$
one may verify the equations (i)$^{(l)}$ and (ii)$^{(l)}$ quite analogously to
the corresponding ones for the main theorem (e.g. for $N=3$).

We prove (iii)$^{(l)}$ by induction and assume:%
\begin{equation}
L_{\underline{\gamma}}^{(l+1)}(\underline{u})\approx c_{l+1}\tilde
{S}_{\underline{\hat{\gamma}}}^{N\dots l+2}(\underline{\hat{u}})\,\prod
_{i=N-l}^{k}\prod_{j=2}^{N-l-1}\tilde{\phi}(u_{ij})L_{\underline{\check
{\gamma}}}^{(l+1)}(\underline{\check{u}}) \label{D.2}%
\end{equation}
for $u_{12},\dots,u_{N-l-2,N-2}\rightarrow i\eta$. In the integral
representation (\ref{3.7d}) of $L_{\underline{\beta}}^{(l)}(\underline{z})$
there are pinchings at $u_{1}=z_{2},\dots,u_{N-l-1}=z_{N-l}$ if $z_{12}%
,\dots,z_{N-l-1,N-l}\rightarrow i\eta$. Therefore in $\tilde{\Psi}%
_{\underline{\beta}}^{(l)}(\underline{z},\underline{u})$ the S-matrices
$\tilde{S}(z_{2}-u_{1}),\dots,\tilde{S}(z_{N-l}-u_{N-l-1})$ yield the
permutation operator $\mathbf{P}$ and we have to consider
\begin{multline*}
{\tilde{\Phi}}_{\underline{\beta}}^{(l)\underline{\gamma}}(\underline
{z},\underline{u})\,=\left(  \Omega^{(l)}\,\tilde{C}^{(l)\gamma_{k}%
}(\underline{z},u_{k})\cdots\tilde{C}^{(l)\gamma_{N-l-1}}(\underline
{z},z_{N-l})\cdots\tilde{C}^{(l)\gamma_{1}}(\underline{z},z_{2})\right)
_{\underline{\beta}}\\
=~~%
\begin{array}
[c]{c}%
\unitlength6mm\begin{picture}(12,8) \put(1,4){\line(1,0){10}}
\put(1,6){\line(1,0){10}} \put(2,1){\line(0,1){6}} \put(6,1){\line(0,1){6}}
\put(10,1){\line(0,1){6}} \put(1,1){\oval(4,2)[rt]}
\put(1,1){\oval(6.5,4.5)[rt]} \put(11,7){\oval(16,10)[lb]}
\put(11,7){\oval(13.5,7.5)[lb]} \put(1.3,1){$z_1$} \put(2.3,1){$z_2$}
\put(4.4,1){$z_{N-l}$} \put(10.2,1){$z_m$} \put(6.6,2.7){$u_{N-l-1}$}
\put(8,5.4){$u_k$} \put(6.8,1.5){$u_1=z_2$} \put(7.5,4.8){$\dots$}
\put(3.2,1.3){${\dots}$} \put(1.4,4.8){$\vdots$} \put(1.4,2.3){${\vdots}$}
\put(.2,1.8){$\gamma_1$} \put(-.4,3){$\gamma_{N-l}$} \put(.2,5.8){$\gamma_k$}
\put(11.4,1.8){l+1} \put(11.3,3){l+1} \put(11.3,5.8){l+1} \put(1.6,7.3){l+1}
\put(3.9,7.3){l+1} \put(9.6,7.3){l+1} \put(1.6,.3){$\beta_1$}
\put(2.6,.3){$\beta_2$} \put(4,.3){$\beta_{N-l}$} \put(9.6,.3){$\beta_m$}
\end{picture}
\end{array}
\\
=\left(  \tilde{S}_{12}(z_{12})\dots\tilde{S}_{1N-l}(z_{1N-l})\right)
_{\beta_{1}\dots\beta_{N-l}}^{\gamma_{1}\dots\gamma_{N-l-1}l+1}\prod
_{j=N-l}^{k}\tilde{b}(z_{1}-u_{j}){\tilde{\Phi}}_{\underline{\check{\beta}}%
}^{(l)\underline{\check{\gamma}}}(\underline{\check{z}},\underline{\check{u}})
\end{multline*}
where $\underline{\check{\beta}}=(\beta_{N-l+1},\dots,\beta_{m}),\ \underline
{\check{z}}=(z_{N-l+1},\dots,z_{m}),\ \underline{\check{u}}=(u_{N-l}%
,\dots,u_{k})$. We may write for $\underline{\hat{u}}=(u_{1},\dots
,u_{N-l-1})=(z_{2},\dots,z_{N-l})$%
\[
\tilde{S}_{\underline{\hat{\gamma}}}^{N\dots l+2}(\underline{\hat{u}})\left(
\tilde{S}_{12}(z_{12})\dots\tilde{S}_{1N-l}(z_{1N-l})\right)  _{\beta_{1}%
\dots\beta_{N-l}}^{\gamma_{1}\dots\gamma_{N-l-1}l+1}=\tilde{S}_{\underline
{\hat{\beta}}}^{N\dots l+1}(\underline{\hat{z}})
\]
with the notation $\underline{\hat{z}}=(z_{1},\dots,z_{N-l})$. Therefore using
the assumption (\ref{D.2}) we obtain when $z_{12},\dots,z_{N-l-1,N-l}%
\rightarrow i\eta$
\begin{multline*}
L_{\underline{\beta}}^{(l)}(\underline{z})\approx\frac{1}{\check{k}!}%
\oint_{z_{2}}\frac{du_{1}}{R}\cdots\oint_{z_{N-l}}\frac{du_{N-l-1}}{R}%
\int_{\mathcal{C}_{\underline{z}}}\frac{du_{N-l}}{R}\cdots\int_{\mathcal{C}%
_{\underline{z}}}\frac{du_{k}}{R}\,\tilde{h}(\underline{z},\underline{u})\\
\times L_{\underline{\gamma}}^{(l+1)}(\underline{u})\,{\tilde{\Phi}%
}_{\underline{\beta}}^{(l)\underline{\gamma}}(\underline{z},\underline
{u})=c_{l}\tilde{S}_{\underline{\hat{\beta}}}^{N\dots l+1}(\underline{\hat{z}%
})\prod_{i=N-l+1}^{m}\prod_{j=2}^{N-l}\tilde{\phi}(z_{ij})L_{\underline
{\check{\beta}}}^{(l)}(\underline{\check{z}})
\end{multline*}
where $\check{k}=k-N+l+1$. The following formulae have been used: $\tilde
{b}(z_{1}-u_{j})\prod_{i=3}^{N-l}\tilde{\phi}(u_{j}-z_{i})\prod_{i=1}%
^{N-l}\tilde{\phi}(z_{i}-u_{j})\prod_{i=2}^{N-l}\tau(z_{i}-u_{j})=-1,$ the
relation $\mathrm{(iii}^{\prime}\mathrm{)}^{(l)}$ of (\ref{4.9}) for the
p-function $p_{mk}^{(l)}(\underline{z},z_{2},\dots z_{N-l},\underline
{\check{u}})=(-1)^{k-N-l+1}\,p_{\check{m}\check{k}}^{(l)}(\underline{\check
{z}},\underline{\check{u}})$ (if $z_{12}=\dots=z_{z_{N-l-1,N-l}}=i\eta$) and
the recursion relation $c_{l}=c_{l+1}\prod_{j=1}^{N-l-1}\tilde{\phi}(ji\eta)$.
The solution to this recursion relation with $c_{N-1}=1$ is
\begin{equation}
c_{1}=\tilde{\phi}^{N-2}(i\eta)\tilde{\phi}^{N-3}(2i\eta)\cdots\tilde{\phi
}((N-2)i\eta) \label{D.5}%
\end{equation}
\endproof

\end{document}